\documentclass[11pt]{article}
\usepackage[letterpaper,margin=1.05in]{geometry}
\usepackage[T1]{fontenc}
\usepackage{lmodern,microtype}
\usepackage{amsmath,amssymb,amsthm,mathtools}
\usepackage{booktabs,array,tabularx,enumitem,needspace,float}
\usepackage{xcolor}
\usepackage{tikz}
\usetikzlibrary{arrows.meta}
\usepackage[colorlinks=true,linkcolor=blue!45!black,citecolor=blue!45!black,urlcolor=blue!45!black]{hyperref}
\hypersetup{
 pdftitle={Late-time asymptotics for scalar perturbations from hypersurfaces with a static end},
 pdfauthor={Stefanos Aretakis},
 pdfsubject={},
 pdfkeywords={wave equation, initial data with a static end, all real powers, spatial tails, Newman-Penrose charge}
}
\setlist{itemsep=3pt,topsep=5pt}
\numberwithin{equation}{section}
\newtheorem{theorem}{Theorem}[section]
\newtheorem{proposition}[theorem]{Proposition}
\newtheorem{lemma}[theorem]{Lemma}

\theoremstyle{definition}

\newtheorem{remark}[theorem]{Remark}
\newcommand{\R}{\mathbb R}
\newcommand{\HH}{\mathcal H^+}
\newcommand{\scri}{\mathcal I^+}
\newcommand{\dd}{\,\mathrm d}
\newcommand{\Os}[1][\infty]{O_{\mathrm{sym}}^{(#1)}}
\newcommand{\calI}{\mathcal I[\psi]}

\newcommand{\EN}[1]{\mathcal E_N[#1]}

\newcommand{\Boxg}{\Box_g}
\newcommand{\norm}[1]{\left\lVert#1\right\rVert}

\title{\vspace{-1.4em}Late-time asymptotics for scalar perturbations\\
from hypersurfaces with a static end}
\author{Stefanos Aretakis\thanks{Department of Mathematics, University of Toronto,
40 St.\ George Street, Toronto, ON M5S 2E4, Canada.
Email: \href{mailto:aretakis@math.toronto.edu}{aretakis@math.toronto.edu}.}}
\date{27 November 2026}

\begin{document}
\maketitle
\thispagestyle{empty}
\begin{abstract}
We study spherically symmetric solutions of the scalar wave equation arising from
noncompact initial data on stationary, spherically symmetric,
asymptotically flat black hole exteriors with a nondegenerate horizon.
The spacelike initial hypersurface $\mathcal S$ extends from the bifurcation sphere or a regular future-horizon cross-section to
spatial infinity and has a static end: it agrees exactly with $t=0$ for sufficiently large $r$. The initial data are allowed to have any real power $p$ of decay or growth:
\[
 \psi|_{\mathcal S}=a r^{-p}+O(r^{-p-\delta}),\qquad
 T\psi|_{\mathcal S}=b r^{-p-1}+O(r^{-p-1-\delta}),\qquad \delta>0.
\]
We obtain global late-time asymptotics for solutions arising from such data. We show that for $p=3$, the spatial tail contributes at the same order as the curvature generated tail. We also derive logarithmic divergence of $r\psi$ towards null infinity for generic $p=1$ data, the constant limit $a+b$ at $p=0$, and
polynomial growth for generic $p<0$ data.

\end{abstract}

{\fontsize{9}{10.5}\selectfont\setlength{\parskip}{0pt}\tableofcontents}

\section{Introduction}
\label{sec:introduction}

\subsection{Basic background}

The late-time behavior of solutions to the linear wave equation
\begin{equation}
 \Box_g\psi=0
 \label{wave}
\end{equation}
on a black hole exterior $(\mathcal M,g)$ depends both on scattering
by the geometry and on the asymptotics of the initial data prescribed
at infinity. Compactly supported data remove the second source of
leading behavior. In this paper, we provide precise global late-time
asymptotics for spherically symmetric solutions to \eqref{wave} whose initial data on a hypersurface with a static end have any real power of decay or growth.

We consider backgrounds which are stationary, spherically symmetric and asymptotically
flat, with a nondegenerate horizon. They satisfy the geometric
conditions and the energy boundedness and integrated local decay
assumptions of Section~\ref{sec:setting}; these analytic assumptions
are part of the background hypotheses. The class includes Schwarzschild
and subextremal Reissner--Nordstr\"om. We denote by $M>0$ the mass,
by $r$ the area radius, so that a symmetry sphere has area $4\pi r^2$,
and by $r_h$ the horizon radius. The stationary Killing vector field
$T$ generates time translations and is normalized to have unit
timelike length at infinity. We write
$D(r)=-g(T,T)$; in particular,
$D(r)=1-2M/r+O(r^{-2})$ at infinity.

We prescribe initial data on a spacelike hypersurface with a
\emph{static end}: it agrees exactly with $t=0$ for all sufficiently
large $r$ and reaches spatial infinity $i^0$. We distinguish the two
principal configurations by their notation:
\begin{itemize}
 \item The static hypersurface $\{t=0\}$ meets the
 \emph{bifurcation sphere} $\mathcal B$, where the past and future
 event horizons intersect.
 \item The hypersurface $\Sigma_0$ meets a smooth spherically symmetric
 cross-section of the future event horizon $\HH$ strictly to the
 future of $\mathcal B$.
\end{itemize}
We also allow a compact deformation of $\{t=0\}$ with the same
horizon endpoint $\mathcal B$, denoted by $\Sigma_{\mathcal B}$.
The data are smooth in coordinates regular at the relevant horizon
endpoint. In statements and formulas covering these choices together,
$\mathcal S$ denotes the chosen initial hypersurface.

\begin{figure}[H]
\centering
\begin{tikzpicture}[x=.96cm,y=.96cm,
 every node/.style={font=\small},
 boundary/.style={line width=.8pt},
 initial/.style={draw=blue!65!black,line width=1.25pt},
 future/.style={draw=red!65!black,line width=1.25pt}]
 \begin{scope}
  \coordinate (B) at (0,0);
  \coordinate (IZ) at (6,0);
  \coordinate (IP) at (3,3);
  \coordinate (G) at (.85,.85);
  \coordinate (J) at (2.55,1.19);
  \coordinate (F) at (3.68,2.32);
  \fill[orange!7] (B)--(IP)--(IZ)--cycle;
  \fill[blue!7] (B)--(IP)
     .. controls (2.75,2.7) and (2.62,1.8) .. (J)
     .. controls (2.5,.8) and (2.48,.4) .. (2.5,0)--cycle;
  \draw[boundary] (B)--(IP)--(IZ);
  \draw[black!45,line width=.65pt] (2.5,0)
     .. controls (2.48,.4) and (2.5,.8) .. (J)
     .. controls (2.62,1.8) and (2.75,2.7) .. (IP);
  \draw[initial] (B)--(IZ);
  \draw[green!45!black,line width=1pt] (B)
     .. controls (.7,.12) and (1.4,.24) .. (2,.24)
     .. controls (2.8,.24) and (3.5,0) .. (4,0);
  \draw[future] (G)
     .. controls (1.3,.8) and (2.10,.74) .. (J)--(F);
  \fill (B) circle (1.3pt);
  \fill (IZ) circle (1.3pt);
  \fill (IP) circle (1.3pt);
  \fill[red!65!black] (G) circle (1.5pt);
  \fill[red!65!black] (J) circle (1.4pt);
  \node[below left] at (B) {$\mathcal B$};
  \node[right] at (IZ) {$i^0$};
  \node[above] at (IP) {$i^+$};
  \node[rotate=45] at (1.48,1.90) {$\HH$};
  \node[rotate=-45] at (4.70,1.68) {$\scri$};
  \node[blue!65!black,below] at (2.9,0) {$\{t=0\}$};
  \node[green!45!black] at (.95,.40) {$\Sigma_{\mathcal B}$};
  \node[left,red!65!black] at (G) {$\Gamma_\tau$};
  \node[red!65!black] at (1.70,1.09) {$\mathcal N_\tau$};
  \node[red!65!black] at (1.70,.70) {$\mathcal C_\tau$};
  \node[red!65!black,rotate=45,font=\footnotesize]
     at (3.05,1.87) {$u=\tau$};
  \node[font=\footnotesize] at (2.08,1.50) {$r\le R$};
  \node[font=\footnotesize] at (4.12,.91) {$r\ge R$};
  \node[anchor=west,font=\footnotesize,black!65] at (2.65,.47) {$r=R$};
  \node[font=\footnotesize] at (3,-.75)
     {(a) Initial endpoint at $\mathcal B$};
 \end{scope}
 \begin{scope}[xshift=7.4cm]
  \coordinate (B) at (0,0);
  \coordinate (IZ) at (6,0);
  \coordinate (IP) at (3,3);
  \coordinate (S) at (.45,.45);
  \coordinate (G) at (.95,.95);
  \coordinate (J) at (2.55,1.19);
  \coordinate (F) at (3.68,2.32);
  \begin{scope}
   \clip (S) .. controls (1.3,.05) and (2.8,0) .. (4,0)
       --(IZ)--(IP)--cycle;
   \fill[orange!7] (B)--(IP)--(IZ)--cycle;
   \fill[blue!7] (B)--(IP)
       .. controls (2.75,2.7) and (2.62,1.8) .. (J)
       .. controls (2.5,.8) and (2.48,.4) .. (2.5,0)--cycle;
  \end{scope}
  \draw[boundary] (B)--(IP)--(IZ);
  \draw[black!25,line width=.6pt] (B)--(IZ);
  \draw[black!45,line width=.65pt] (2.5,0)
     .. controls (2.48,.4) and (2.5,.8) .. (J)
     .. controls (2.62,1.8) and (2.75,2.7) .. (IP);
  \draw[initial] (S) .. controls (1.3,.05) and (2.8,0) .. (4,0)--(IZ);
  \draw[future] (G)
     .. controls (1.4,.87) and (2.10,.74) .. (J)--(F);
  \fill (B) circle (1.3pt);
  \fill (IZ) circle (1.3pt);
  \fill (IP) circle (1.3pt);
  \fill[blue!65!black] (S) circle (1.5pt);
  \fill[red!65!black] (G) circle (1.5pt);
  \fill[red!65!black] (J) circle (1.4pt);
  \node[below left] at (B) {$\mathcal B$};
  \node[right] at (IZ) {$i^0$};
  \node[above] at (IP) {$i^+$};
  \node[rotate=45] at (1.48,1.90) {$\HH$};
  \node[rotate=-45] at (4.70,1.68) {$\scri$};
  \node[left,blue!65!black,font=\footnotesize] at (S) {$S_{\mathcal H^+}$};
  \node[blue!65!black] at (1.65,.34) {$\Sigma_0$};
  \node[blue!65!black,below] at (4.8,0) {$t=0$};
  \node[left,red!65!black] at (G) {$\Gamma_\tau$};
  \node[red!65!black] at (1.75,1.13) {$\mathcal N_\tau$};
  \node[red!65!black] at (1.75,.72) {$\mathcal C_\tau$};
  \node[red!65!black,rotate=45,font=\footnotesize]
     at (3.05,1.87) {$u=\tau$};
  \node[font=\footnotesize] at (2.08,1.50) {$r\le R$};
  \node[font=\footnotesize] at (4.12,.91) {$r\ge R$};
  \node[anchor=west,font=\footnotesize,black!65] at (2.65,.47) {$r=R$};
  \node[font=\footnotesize] at (3,-.75)
     {(b) Initial endpoint to the future of $\mathcal B$};
 \end{scope}
\end{tikzpicture}
\caption{The initial hypersurfaces and future hypersurfaces. }
\label{fig:intro-hypersurfaces}
\end{figure}
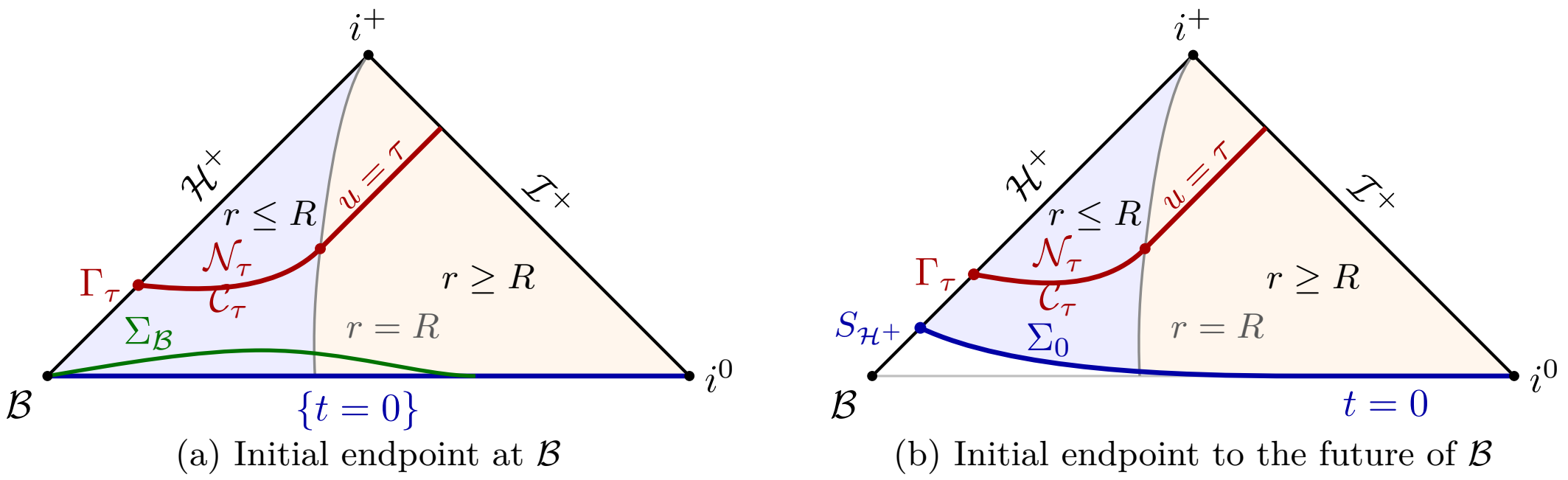

\paragraph{The future hypersurfaces and the time coordinate.}
We use the single time coordinate $\tau$ associated with the future
hypersurfaces $\mathcal N_\tau$. Each $\mathcal N_\tau$ consists of
a compact spacelike portion $\mathcal C_\tau$, from the horizon
sphere $\Gamma_\tau$ to a fixed radius $R>r_h$, and an outgoing null
end. We write $u=t-r_*(r)$, where $\dd r_*/\dd r=D^{-1}$;
on the null end, $\tau=u$ exactly. The coordinate $\tau$ is regular
at the future horizon, satisfies $T\tau=1$, and labels the cut of
future null infinity $\scri$ reached by $\mathcal N_\tau$.
At each fixed exterior radius it differs from $t$ by a constant.
We write the evolved field as $\psi(\tau,r)$, including at $r=r_h$.
Section~\ref{sec:hypersurfaces} gives the precise construction.
Figure~\ref{fig:intro-hypersurfaces} shows the initial hypersurfaces,
the future family, and the bounded-radius and exterior regions.

\subsection{Previous work}
\label{sec:literature}

Price predicted inverse-power late-time tails for scalar perturbations
of black holes \cite{Price}. Dafermos--Rodnianski established decay on
Schwarzschild using the red-shift effect \cite{DRRedshift} and developed
the physical-space method of $r$-weighted energy estimates
\cite{DRrp}. Dafermos--Rodnianski--Shlapentokh-Rothman proved boundedness
and decay throughout the subextremal Kerr family \cite{DRSDecay} and
constructed a scattering theory relating radiation fields on the
horizons and null infinities \cite{DRSScattering}.
Tataru proved sharp pointwise
decay bounds on stationary asymptotically flat backgrounds under energy
boundedness and local energy decay assumptions \cite{Tataru}.

Precise leading-order asymptotics on stationary, spherically symmetric
backgrounds, including Schwarzschild and subextremal
Reissner--Nordstr\"om, were obtained in \cite{AAG1}, together with
formulas relating their coefficients to conserved asymptotic charges.
The weighted hypotheses in that work permit suitable noncompact data. Subsequent work
expressed the coefficient directly in terms of compactly supported
Schwarzschild data on hypersurfaces meeting the bifurcation sphere
\cite{AAG2}. We use those asymptotic estimates and the interpretation
of the coefficient as a flux of the initial data. The additional issue
here is to determine how prescribed asymptotics at spatial infinity
contribute to the late-time field.

Hintz proved sharp asymptotics on a broader class of stationary
asymptotically flat backgrounds satisfying spectral assumptions,
including subextremal Kerr \cite{Hintz}. On the common admissible class
of spherically symmetric data, his leading coefficient agrees with ours for
sufficiently rapid decay. His stated extension to noncompact data
requires stronger decay assumptions than the $p=3$ spatial class
treated here.

Looi--Xiong obtained precise expansions for semilinear waves and
identified explicit contributions from asymptotic coefficients at
infinity \cite{LX}. Their treatment of noncompact data uses
hypersurfaces approaching future null infinity. Thus contributions
from noncompact data at infinity have precedents; the problem here is
to derive the complete coefficient from data prescribed on a
hypersurface ending at spatial infinity.

Gajic derived precise late-time asymptotics for wave equations with
inverse-square potentials on Schwarzschild \cite{GajicInvSquare}.
His work on extremal Kerr identifies asymptotic instabilities
associated with non-axisymmetric azimuthal modes
\cite{GajicExtremalKerr}, and his analysis of charged scalar fields
on near-extremal Reissner--Nordstr\"om establishes oscillating late-time tails
\cite{GajicCharged}.
Gajic--Kehrberger explained how asymptotic charges connect the behavior
of a field near null infinity to its late-time tails \cite{GK22}.
Their later work derives global asymptotics on dynamical black hole
backgrounds settling to Schwarzschild \cite{GK}.

In Minkowski space, the contribution of spatial power tails follows
directly from the classical radial d'Alembert formula \cite{John}.
Related flat-space formulas also appear in the frequency-domain
analysis of Bhattacharjee--Chakrabarty--Chow--Paul--Virmani \cite{BCCPV}.
Our Minkowski section recalls this mechanism and fixes the normalizations.

The present paper gives a quantitative global classification from the
initial data on $\mathcal S$ for every real power of decay or growth.
It includes the spatial contribution to the coefficient at $p=3$ and
treats initial hypersurfaces meeting either the bifurcation sphere or
a later cross-section of the future event horizon.

\subsection{Classification of late-time asymptotics for all \texorpdfstring{$p$}{p}}
\label{sec:intro-classification}

\paragraph{The initial data.}
We denote by $n_{\mathcal S}$ the future-directed unit normal to the initial hypersurface $\mathcal S$. The initial data are the field and its
normal derivative:
\[
 f=\psi|_{\mathcal S},\qquad \pi=n_{\mathcal S}\psi|_{\mathcal S} ,\qquad h=T\psi|_{\mathcal S}.
\]
On the static outer part of $\mathcal S$,
$h=\sqrt D\,\pi$. We assume
\begin{equation}
 f=a r^{-p}+\Os[m+1](r^{-p-\delta}),\qquad
 h=b r^{-p-1}+\Os[m](r^{-p-1-\delta}),
 \qquad p\in\R,\quad\delta>0.
 \label{eq:intro-spatial-data}
\end{equation}
Here $m=m(p)=30+2\lceil\max\{0,-p\}\rceil$ is a sufficient finite
derivative count for the main theorem. For an integer $N\ge0$, we write
\begin{equation}
 z=\Os[N](r^{-q})
 \quad\Longleftrightarrow\quad
 |\partial_r^j z(r)|\le C_jr^{-q-j}
 \quad\text{for }j=0,1,\ldots,N\text{ and }r\ge R_0.
 \label{eq:symbol-notation}
\end{equation}
Thus $j$ is the number of radial derivatives, and the constants $C_j$
are independent of $r$. The superscript $\infty$ means that these
bounds hold for every nonnegative integer $j$. When the displayed
variable is $r_*$ or $v$, the derivatives in this notation are taken
with respect to that variable.

\paragraph{The radiation field.}
At infinity the quantity of interest is the rescaled field $r\psi$.
Its \emph{radiation field} is
\begin{equation}
 F(\tau)=\lim_{r\to\infty}r\psi(\tau,r),
 \label{eq:intro-radiation-field}
\end{equation}
when this limit is finite. For generic data in
\eqref{eq:intro-spatial-data}, it is finite for $p>1$, diverges
logarithmically at $p=1$, and diverges as a power of $r$ for $p<1$.

\paragraph{The Newman--Penrose charges.}
We denote by $L_{\rm out}$ differentiation along an outgoing radial null
ray parametrized by the area radius, so that $L_{\rm out}r=1$.
The ordinary Newman--Penrose charge is
\begin{equation}
 I_0[\psi]=\lim_{r\to\infty}r^2L_{\rm out}(r\psi).
 \label{eq:intro-NP-definition}
\end{equation}
This limit is taken along an outgoing null ray, and is independent
of the ray when finite in the classes considered here. 

For $p>1$, the time integral normalized to vanish at late time is
\begin{equation}
 \psi^{(1)}(\tau,r)=-\int_\tau^\infty\psi(s,r)\,\dd s,
 \qquad T\psi^{(1)}=\psi.
 \label{eq:intro-time-primitive}
\end{equation}
The integration is at fixed $r$. This time integral is another solution
of the wave equation on the future exterior. Its Newman--Penrose
charge, when finite, is the \emph{time-inverted charge}
\begin{equation}
 I_0^{(1)}[\psi]=I_0[\psi^{(1)}].
 \label{eq:intro-inverse-NP-definition}
\end{equation}
The charge $I_0^{(1)}$ is finite for the classes $p\ge3$ used below.

Both charges can be read from the initial data on $\mathcal S$. We write
$S_{\mathcal H^+}=\mathcal S\cap\HH$ for the actual horizon endpoint and, for
$p\ge3$, we define
\begin{equation}
 B_{\mathcal S}=\frac1{4\pi}\left(
       \int_{S_{\mathcal H^+}}f\,\dd S+
       \int_{\mathcal S}\pi\,\dd\mu_{\mathcal S}\right).
 \label{eq:intro-geometric-B}
\end{equation}
Here $\dd S$ is the area measure on the horizon sphere and
$\dd\mu_{\mathcal S}$ is the induced volume measure on $\mathcal S$.
The integrals converge in the stated range. The last two columns of
Table~\ref{tab:intro-black-hole} give the charges explicitly in terms
of $a,b$ and $B_{\mathcal S}$. Section~\ref{sec:intro-critical} explains the additional contribution
at $p=3$.

\paragraph{The classification.}
Table~\ref{tab:intro-black-hole} summarizes the black hole results
in the same format as the Minkowski classification in
Section~\ref{sec:minkowski-classification}. To shorten its entries,
we write
\[
 C_p=b+(1-p)a,\qquad
 A_p(r)=\frac{2^{-p}C_p}{1-p}r^{1-p}\quad(p<1).
\]
The bounded-radius column holds uniformly on every interval
$r_h\le r\le R$, so it includes the future event horizon.
The two columns concerning null infinity distinguish the limit
$r\to\infty$ at fixed $\tau$ from the subsequent late-time behavior
of the finite radiation field $F(\tau)$.
All asymptotic equivalences assume that the displayed coefficient
is nonzero. The cancellation cases are explained below the table.

The energy column concerns the energy of first derivatives of the
initial field. Its convergence at the static end of $\mathcal S$ is equivalent
to convergence, for sufficiently large $R_0$, of
\[
 \int_{R_0}^\infty r^2\bigl(|\partial_r f|^2+|h|^2\bigr)\,\dd r.
\]
 If $p^2a^2+b^2>0$, this
integral is finite precisely for $p>1/2$.
\begin{table}[H]
\centering
\small
\setlength{\tabcolsep}{2.5pt}
\renewcommand{\arraystretch}{1.30}
\begin{tabularx}{\textwidth}{@{}
 >{\raggedright\arraybackslash}p{.70in}
 >{\raggedright\arraybackslash}p{1.00in}
 >{\raggedright\arraybackslash}p{.98in}
 >{\raggedright\arraybackslash}X
 >{\raggedright\arraybackslash}p{.50in}
 >{\raggedright\arraybackslash}p{.40in}
 >{\raggedright\arraybackslash}p{.80in}@{}}
\toprule
Power $p$
 & Field $\psi$ at bounded radius, including $\HH$\newline
   $\tau\to\infty$
 & Approach to null infinity\newline
   $r\psi(\tau,r)$:\newline
   $r\to\infty$, $\tau$ fixed
 & Late time along null infinity\newline
   $F(\tau)$ as $\tau\to\infty$
 & Finite initial energy?
 & $I_0$
 & $I_0^{(1)}$\\
\midrule
$p<0$
 & $\sim C_p\tau^{-p}$ (growth)
 & $\sim A_p(r)$
 & $\infty$
 & No & $\infty$ & Undefined\\
$p=0$
 & $\to a+b$
 & $\sim(a+b)r$
 & $\infty$
 & Yes/No & $\infty$ & Undefined\\
$0<p<\tfrac12$
 & $\sim C_p\tau^{-p}$
 & $\sim A_p(r)$
 & $\infty$
 & No & $\infty$ & Undefined\\
$p=\tfrac12$
 & $\sim C_p\tau^{-1/2}$
 & $\sim A_p(r)$
 & $\infty$
 & No (log) & $\infty$ & Undefined\\
$\tfrac12<p<1$
 & $\sim C_p\tau^{-p}$
 & $\sim A_p(r)$
 & $\infty$
 & Yes & $\infty$ & Undefined\\
$p=1$
 & $\sim b\tau^{-1}$
 & $\sim\tfrac b2\log(2r/\ell)$
 & $\infty$
 & Yes & $\infty$ & Undefined\\
$1<p<2$
 & $\sim C_p\tau^{-p}$
 & $\to F(\tau)$
 & $\sim\dfrac{C_p}{2(p-1)}\tau^{1-p}$
 & Yes & $\infty$ & $\infty$\\
$p=2$
 & $\sim4I_0\tau^{-2}$
 & $\to F(\tau)$
 & $\sim2I_0\tau^{-1}$
 & Yes & $\frac{b-a}{4}$ & $\infty$\\
$2<p<3$
 & $\sim C_p\tau^{-p}$
 & $\to F(\tau)$
 & $\sim\dfrac{C_p}{2(p-1)}\tau^{1-p}$
 & Yes & $0$ & $\infty$\\
$p=3$
 & $\sim-8I_0^{(1)}\tau^{-3}$
 & $\to F(\tau)$
 & $\sim-2I_0^{(1)}\tau^{-2}$
 & Yes & $0$ & $MB_{\mathcal S}$\newline $+\frac{2a-b}{8}$\\
$p>3$
 & $\sim-8I_0^{(1)}\tau^{-3}$
 & $\to F(\tau)$
 & $\sim-2I_0^{(1)}\tau^{-2}$
 & Yes & $0$ & $MB_{\mathcal S}$\\
\midrule
Compactly supported data
 & $\sim-8MB_{\mathcal S}\tau^{-3}$
 & $\to F(\tau)$
 & $\sim-2MB_{\mathcal S}\tau^{-2}$
 & Yes & $0$ & $MB_{\mathcal S}$\\
\bottomrule
\end{tabularx}
\caption{Black hole asymptotics from the initial data on $\mathcal S$.}
\label{tab:intro-black-hole}
\end{table}

\paragraph{The logarithmic case and the energy exception.}
At $p=1$, subtracting the explicit logarithmic divergence produces a
finite renormalized radiation field. With the same fixed reference
length $\ell>0$ in both logarithms,
\begin{equation}
 \begin{aligned}
 F_{\rm ren}(\tau)
 &=\lim_{r\to\infty}\left[r\psi(\tau,r)
                       -\frac b2\log(2r/\ell)\right],\\
 F_{\rm ren}(\tau)
 &=-\frac b2\log(\tau/\ell)+O(\tau^{-\eta}),
 \qquad \eta>0.
 \end{aligned}
 \label{eq:intro-renormalized-radiation}
\end{equation}
If $b=0$, the ordinary radiation field is finite and has the
remainder bound in \eqref{eq:intro-renormalized-radiation}.

At $p=0$, the energy is infinite if $b\ne0$. If $b=0$, its
finiteness depends on the remainders: $\delta>1/2$ guarantees finite
energy, and exact constant position data with zero velocity outside
a compact set also have finite derivative energy. Thus finite initial
energy and a finite radiation field are different requirements;
in particular, $1/2<p\le1$ can have finite energy while $r\psi$
diverges at infinity.

\paragraph{Remainders and vanishing coefficients.}
Theorem~\ref{thm:global-classification} gives a uniform global estimate
which connects the regions displayed in the table. Each bounded-radius
entry $A\tau^{-q}$ has a quantitative remainder
$O_R(\tau^{-q-\eta})$ for some $\eta>0$, uniformly for
$r_h\le r\le R$. Each finite radiation-field power has the analogous
positive improvement in its time exponent. For $p<1$, the fixed-time
error in the null-infinity entry is $O_\tau(r^{1-p-\gamma})$ for a
positive $\gamma$. All constants are controlled by finitely many
weighted norms of the initial data. These statements concern late $\tau$.

If a displayed leading coefficient vanishes, the same estimates give
improved upper bounds. For compactly supported data, the last row
records the curvature-generated tail. The corresponding Minkowski
solutions vanish in the late region described in
Section~\ref{sec:minkowski-classification}.

\subsection{The spatial infinity contribution at \texorpdfstring{$p=3$}{p = 3}}
\label{sec:intro-critical}

At $p=3$, the signal inherited from the spatial tail and the signal
generated by scattering off the black hole geometry both have order
$\tau^{-3}$ at bounded radius and on the future horizon. Their
radiation fields both have order $\tau^{-2}$ along future null
infinity. Thus both mechanisms enter the leading coefficient at
$p=3$. In the notation used in the later main theorem,
$\mathcal I=I_0^{(1)}[\psi]$, and
\begin{equation}
 \boxed{\quad\mathcal I=M B_{\mathcal S}+\frac{2a-b}{8}.\quad}
 \label{eq:introcharge}
\end{equation}
Thus the $p=3$ row of Table~\ref{tab:intro-black-hole} has
the explicit coefficients
\begin{align*}
 -8\mathcal I&=b-2a-8MB_{\mathcal S}
 &&\hbox{at bounded radius and on }\HH,\\
 -2\mathcal I&=\frac{b-2a}{4}-2MB_{\mathcal S}
 &&\hbox{at }\scri.
\end{align*}
The first coefficient multiplies $\tau^{-3}$ and the second
multiplies $\tau^{-2}$.

On the static hypersurface $\{t=0\}$ ending at the
bifurcation sphere, the geometric integral \eqref{eq:intro-geometric-B} becomes
\begin{equation}
 B_{\{t=0\}}=r_h^2 f(r_h)
              +\int_{r_h}^\infty\frac{r^2}{D(r)}h(r)\,\dd r.
 \label{eq:introB}
\end{equation}
For data on $\Sigma_0$, we use \eqref{eq:intro-geometric-B} with
$\mathcal S=\Sigma_0$, horizon sphere
$S_{\mathcal H^+}=\Sigma_0\cap\HH$, and normal derivative
$\pi=n_{\Sigma_0}\psi$. The same formula with
$\mathcal S=\Sigma_{\mathcal B}$ applies to the compactly deformed
bifurcation-ended hypersurface. In each configuration the contribution
from spatial infinity is $(2a-b)/8$.

This term records an incoming component present at arbitrarily large
radii on the initial hypersurface. Such data continue to supply a
signal at increasingly late times. Its leading effect has a classical
flat-space counterpart. The black hole calculation determines how
it combines with $MB_{\mathcal S}$ and gives a quantitative remainder.
Section~\ref{sec:interpretation} explains this interpretation
and gives explicit examples. In particular,
Section~\ref{sec:cutoff-example} shows how the cutoff and late-time
limits yield different asymptotic coefficients.

\subsection{The strategy of the proof}

For every real $p$, we first propagate the initial data through
the far-exterior region $\mathcal W_A$ to the outgoing hypersurface
$u=u_0=-A$, and then across a fixed interval of retarded time to
$\mathcal N_0$. A single weighted bootstrap treats all spatial
powers, including the logarithm at $p=1$. The characteristic
integral formula identifies the asymptotic coefficients and bounds
the remainders on these future hypersurfaces. For $p\ge3$, a
conserved current then evaluates the time-inverted charge in terms
of the initial data. The current flux through a
long incoming null segment is precisely where the contribution from
spatial infinity enters. Weighted estimates and finite-time propagation
in coordinates regular at the horizon verify the hypotheses of the
asymptotic theorem in \cite[Theorem~1.1]{AAG1}, with constants controlled by the initial data. We construct the time integral on $\mathcal N_{u_0}$ and its
future domain, using its compact spacelike portion at the horizon.

For $2<p<3$, the transferred data are decomposed into annuli whose
radii successively double. Applying two weighted decay estimates to
these pieces and summing them gives the intermediate decay powers.
Applying this decay argument to $T\psi$ and integrating in time
treats $1<p<2$, with $p=2$ supplied by the ordinary-charge estimate.
For all $1<p<3$, we use these bounds to propagate the coefficient of
$Q=\partial_v(r\psi)$ and integrate to arbitrary radii, obtaining
the global leading profile and its quantitative remainder. One
further time integration gives the logarithmic case $p=1$.

For $0<p<1$, the already established estimate applies to $T\psi$,
whose spatial exponent is $p+1$. Its time integral converges. The
limit $L(r)=\lim_{\tau\to\infty}\psi(\tau,r)$ is independent of $r$
by horizon regularity, and the behavior at infinity fixes its value. It is zero for
$0<p<1$ and equals $a+b$ for $p=0$. Finitely many further time
integrations handle every negative power. The field on the reference
hypersurface $\mathcal N_0$ fixes the integration constants. This procedure allows
growing initial data while applying finite-energy estimates only to
the time derivatives for which those estimates are justified.

Section~\ref{sec:minkowski} gives the Minkowski preparation.
Section~\ref{sec:setting} specifies the geometric setting, the initial data and the energy assumptions. Section~\ref{sec:global-classification}
states the main theorem. Section~\ref{sec:exterior} gives the common
propagation to $\mathcal N_{u_0}$ and $\mathcal N_0$ for every real
$p$, together with the expansions towards null infinity.
Section~\ref{sec:commuted} computes the initial data of $T\psi$
for every real $p$, and Section~\ref{sec:frac-spatial-transfer}
proves the radial derivative estimates for $p>1$. For $p\ge3$,
Section~\ref{sec:flux} computes the time-inverted Newman--Penrose
constant on the initial hypersurface, and
Section~\ref{sec:application} proves the global leading-order
expansion for $p\ge3$, with quantitative remainders. Section~\ref{sec:interpretation} then explains the
coefficient for $p=3$ and explains why compactly supported
approximations can lose the contribution from spatial infinity.
Sections~\ref{sec:frac-global},
\ref{sec:slow-spatial-endpoints}, and~\ref{sec:slow-global} establish
the global leading-order expansions for $1<p<3$, $p=1$, and $p<1$,
respectively. For $p<1$, the values of $\psi$ on $\mathcal N_0$
provided by Section~\ref{sec:exterior} fix the constants in the
time integrations of Section~\ref{sec:slow-global}.

\section{The Minkowski case}
\label{sec:minkowski}

This section records elementary consequences of the classical radial
d'Alembert formula in $(3+1)$-dimensional Minkowski space; see
\cite[Chapter~2, Section~4]{John}. These classical calculations identify
the relevant decay exponents and fix the normalizations used later.
We prescribe the initial data on the hypersurface
$\{t=0\}$, which reaches spatial infinity. We assume that
the solution is smooth at the centre $r=0$.

\subsection{The incoming combination of the initial data}
\label{sec:minkowski-formula}

We use the standard time coordinate $t\in\R$ and area radius $r\ge0$, and
write
\[
 f(r)=\psi(0,r),\qquad h(r)=T\psi(0,r),\qquad T=\partial_t,
 \qquad \phi=r\psi,\quad u=t-r,\quad v=t+r.
\]
The derivative $T$ is taken at fixed $r$. The coordinates $u,v$ are
respectively constant on outgoing and incoming radial null rays.
Here $f,h$ extend to smooth radial functions on $\R^3$. The reduced
field satisfies
\[
 (\partial_t^2-\partial_r^2)\phi=0,\qquad \phi(t,0)=0.
\]

For completeness, we define the odd extensions
\[
 f_{\rm odd}(s)=s f(|s|),\qquad
 h_{\rm odd}(s)=s h(|s|),\qquad s\in\R.
\]
The one-dimensional Cauchy formula gives
\[
 \phi(t,r)
 =\frac12\bigl[f_{\rm odd}(r+t)+f_{\rm odd}(r-t)\bigr]
   +\frac12\int_{r-t}^{r+t}h_{\rm odd}(s)\,\dd s.
\]
In the region $u=t-r>0$, the lower integration endpoint is $-u$.
Oddness gives the exact cancellation
\[
 f_{\rm odd}(-u)=-u f(u),\qquad
 \int_{-u}^{v}h_{\rm odd}(s)\,\dd s
 =\underbrace{\int_{-u}^{u}h_{\rm odd}(s)\,\dd s}_{=0}
   +\int_u^v s h(s)\,\dd s.
\]
Consequently,
\begin{align}
 \phi(t,r)
 &=\frac12\bigl[vf(v)-uf(u)\bigr]
      +\frac12\int_u^v s h(s)\,\dd s \notag\\
 &=\frac12\int_u^v k(s)\,\dd s,
 & k(s)&=(sf(s))'+s h(s).
 \label{eq:minkowski-integral}
\end{align}
The function $k$ is twice the incoming characteristic derivative of
the reduced data: $k(r)=2\partial_v\phi(0,r)$, where
$\partial_v=\tfrac12(\partial_t+\partial_r)$. In particular,
\begin{equation}
 \psi(t,0)=k(t),\qquad
 \left.\partial_r\phi\right|_u=k(v),\qquad
 T\phi=\frac12\bigl[k(v)-k(u)\bigr].
 \label{eq:minkowski-three-identities}
\end{equation}
The first identity follows by taking the regular limit $r\downarrow0$.
These formulas apply to arbitrary smooth radial data, including data
with infinite total derivative energy.

We fix $R_0>0$ and first assume that the exterior data are exact powers:
\begin{equation}
 f(r)=a r^{-p},\qquad h(r)=b r^{-p-1}
 \quad(r\ge R_0),\qquad p\in\R.
 \label{eq:minkowski-power-data}
\end{equation}
The smooth interior data are otherwise arbitrary. We set
\begin{equation}
 C_p=b+(1-p)a,\qquad k(r)=C_p r^{-p}\quad(r\ge R_0).
 \label{eq:minkowski-Cp}
\end{equation}
Every real $p$ is admissible for this smooth Cauchy problem: the powers
are imposed outside a compact set, and finite propagation speed
makes polynomial growth at infinity compatible with a smooth classical
solution at each finite time.

\paragraph{The region determined by the exact exterior powers.}
At a point $P=(t,r)$ with $u>R_0$, the interval in
\eqref{eq:minkowski-integral} satisfies
\[
 [u,v]=[t-r,t+r]\subset(R_0,\infty).
\]
Thus the solution at $P$ is determined by the initial values $f(s)$
and $h(s)$ at radii $t-r\le s\le t+r$ on $\{t=0\}$.
Every radius in this interval exceeds $R_0$, so the entire interval
lies in the part of the initial hypersurface where the prescribed
power formulas hold. The full causal past of $P$ also contains the central
part of $\{t=0\}$. The strong Huygens property of the
$(3+1)$-dimensional wave equation localizes the Cauchy representation
to the intersection of the past light cone with $\{t=0\}$; the radii
on that sphere range from $t-r$ to $t+r$. For radial data, the
odd-extension cancellation above gives this localization explicitly.

\begin{figure}[htbp]
\centering
\begin{tikzpicture}[x=.86cm,y=.86cm,
  every node/.style={font=\small},
  boundary/.style={line width=.8pt},
  threshold/.style={draw=green!45!black,line width=1.1pt},
  characteristic/.style={draw=orange!70!black,line width=.9pt},
  future/.style={draw=red!65!black,line width=1.3pt}]
 \coordinate (O) at (0,0);
 \coordinate (IZ) at (6,0);
 \coordinate (IP) at (0,6);
 \coordinate (R) at (1,0);
 \coordinate (C) at (0,1);
 \coordinate (N) at (2.5,3.5);
 \coordinate (P) at (1.4,3.3);
 \coordinate (CP) at (0,1.9);
 \coordinate (A) at (1.9,0);
 \coordinate (B) at (4.7,0);
 \fill[green!9] (C)--(N)--(IP)--cycle;
 \draw[boundary] (O)--(IP)--(IZ)--cycle;
 \draw[blue!60!black,line width=1.7pt] (R)--(IZ);
 \draw[threshold] (C)--(N)
   node[pos=.53,below,sloped,font=\footnotesize,
        inner sep=1.2pt] {$u=R_0$};
 \draw[green!45!black,line width=.6pt] (R)--(C);
 \draw[characteristic] (A)--(CP)--(P)--(B);
 \draw[orange!70!black,line width=2.2pt] (A)--(B);
 \draw[future] (0,3.4) .. controls (.25,3.4) and (.50,3.35) ..
   (.65,3.5) -- (1.575,4.425)
   node[pos=.58,above,sloped,font=\footnotesize,inner sep=1.2pt]
   {$u=\tau$};
 \node[left,red!65!black] at (0,3.4) {$\mathcal N_\tau$};
 \fill (O) circle (1.2pt);
 \fill (IZ) circle (1.2pt);
 \fill (IP) circle (1.2pt);
 \fill[green!45!black] (R) circle (1.5pt);
 \fill[green!45!black] (C) circle (1.5pt);
 \fill[orange!70!black] (A) circle (1.5pt);
 \fill[orange!70!black] (B) circle (1.5pt);
 \fill[orange!70!black] (P) circle (1.7pt);
 \node[above right] at (P) {$P$};
 \node[above] at (IP) {$i^+$};
 \node[right] at (IZ) {$i^0$};
 \node[left] at (0,4.8) {$r=0$};
 \node[left] at (C) {$t=R_0$};
 \node[below] at (R) {$R_0$};
 \node[below] at (A) {$u(P)$};
 \node[below] at (B) {$v(P)$};
 \node[rotate=-45] at (4.25,2.10) {$\scri$};
 \node[green!35!black] at (.57,4.65) {$u>R_0$};
 \node[below] at (3,-.62) {$\{t=0\}$};
\end{tikzpicture}
\caption{The relevant Minkowsi regions and hypersurfaces.}
\label{fig:minkowski-spatial-data}
\end{figure}
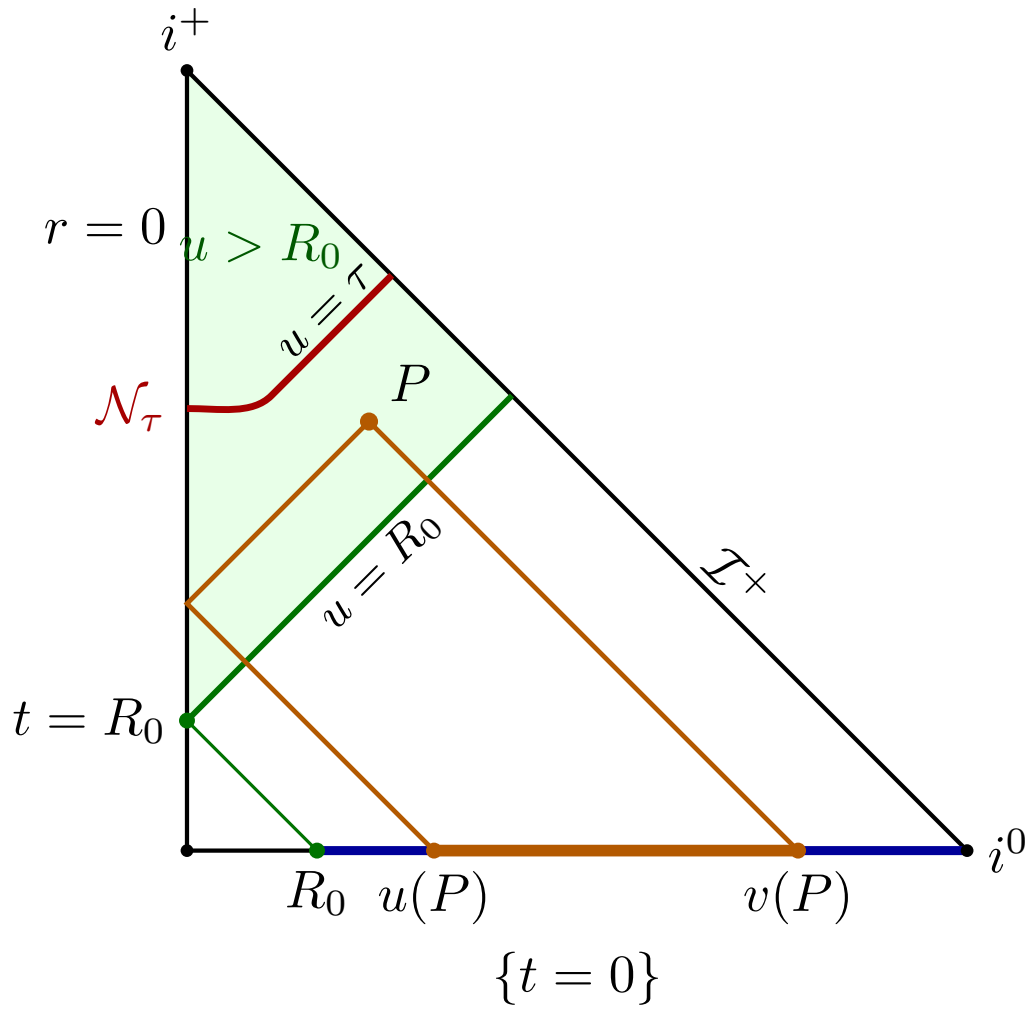

Figure~\ref{fig:minkowski-spatial-data} shows the relevant region.
It uses the compactified null coordinates
\[
 U_{\rm c}=\arctan(u/\ell),\qquad
 V_{\rm c}=\arctan(v/\ell),\qquad \ell>0,
\]
with horizontal coordinate proportional to $V_{\rm c}-U_{\rm c}$
and vertical coordinate proportional to $V_{\rm c}+U_{\rm c}$.
At fixed radius, $u\to+\infty$ approaches future timelike infinity
$i^+$. At fixed $u$, the limit $r\to\infty$ approaches future null
infinity $\scri$. On the initial slice $t=0$, spatial infinity $i^0$
is described by $u=-r\to-\infty$ and $v=r\to+\infty$.

Substitution of the exact powers on $[u,v]$ now gives the complete
solution throughout $u>R_0$:
\begin{equation}
 \boxed{\quad
 \phi(t,r)=
 \begin{cases}
 \displaystyle\frac{C_p}{2(1-p)}
       \bigl(v^{1-p}-u^{1-p}\bigr),&p\ne1,\\[6pt]
 \displaystyle\frac b2\log\frac vu,&p=1.
 \end{cases}\quad}
 \label{eq:flat-general-power}
\end{equation}

We introduce an antiderivative of the incoming derivative $k/2$.
For any reference length $\ell>0$, we define
\begin{equation}
 G_p(s)=
 \begin{cases}
 \displaystyle\frac{C_p}{2(1-p)}s^{1-p},&p\ne1,\\[5pt]
 \displaystyle\frac b2\log(s/\ell),&p=1.
 \end{cases}
 \qquad G_p'(s)=\frac12 k(s).
 \label{eq:minkowski-incoming-primitive}
\end{equation}
Then
\begin{equation}
 \phi(t,r)=G_p(v)-G_p(u),\qquad u>R_0.
 \label{eq:flat-profile}
\end{equation}
The prime denotes differentiation with respect to the argument $s$:
differentiating the power and logarithmic expressions gives
$G_p'(s)=\frac{d}{ds}G_p(s)=\frac{C_p}{2}s^{-p}=k(s)/2$,
including $p=1$, for which $C_1=b$.
The additive constant in the antiderivative cancels. A finite base point
defines the antiderivative for every real $p$. The relative minus sign
in \eqref{eq:flat-profile} follows from regular passage through the
centre and the resulting odd extension of the reduced field.

\subsection{Timelike asymptotics for every real exponent}

For $0\le r\le R_1$ and $t>R_0+R_1$, formula
\eqref{eq:minkowski-integral} is an average of $k$:
\begin{equation}
 \psi(t,r)=\frac1{2r}\int_{t-r}^{t+r}k(s)\,\dd s
          =\frac1r\int_{t-r}^{t+r}G_p'(s)\,\dd s.
 \label{eq:flat-derivative}
\end{equation}
Here $G_p'=k/2$ is the derivative computed above; at $r=0$ we take
the continuous limit. We keep $R_1$ fixed and let $t\to\infty$.
After setting $s=t+z$, we have $|z|\le r\le R_1$, so the ratio
$|z|/t\le R_1/t$ tends uniformly to zero. We take
$t>\max\{R_0+R_1,2R_1\}$ and use the finite Taylor formula
with remainder
\[
 k(t+z)=k(t)+k'(t)z+\frac{k''(t)}2z^2+\frac{k'''(t)}6z^3
       +E_4(t,z).
\]
Here $E_4(t,z)=z^4k^{(4)}(\xi)/24$ for some $\xi$ between
$t$ and $t+z$. Since $t/2\le\xi\le3t/2$ and
\[
 k^{(4)}(\xi)=C_pp(p+1)(p+2)(p+3)\xi^{-p-4},
\]
we have $|E_4(t,z)|\le C_{p,C_p}|z|^4t^{-p-4}$ for every real $p$.
This bound gives the explicit error after averaging:
\[
 \left|\frac1{2r}\int_{-r}^r E_4(t,z)\,\dd z\right|
 \le \frac{C_{p,C_p}}{2r}t^{-p-4}\int_{-r}^r|z|^4\,\dd z
 =\frac{C_{p,C_p}}5r^4t^{-p-4}
 \le C_{p,C_p,R_1}t^{-p-4}.
\]
The odd terms integrate to zero on $[-r,r]$, and
$\frac1{2r}\int_{-r}^{r}z^2\,\dd z=r^2/3$.
Since $k(t)=C_pt^{-p}$ and $k''(t)=C_pp(p+1)t^{-p-2}$,
we obtain, for every $p\in\R$, the uniform expansion
\begin{equation}
 \psi(t,r)
 =C_p t^{-p}
  +\frac{C_p p(p+1)}6\,r^2t^{-p-2}
  +O_{p,C_p,R_1}(t^{-p-4}).
 \label{eq:minkowski-timelike}
\end{equation}
Consequently $C_p\ne0$ gives decay for $p>0$, a constant late field
for $p=0$, and polynomial growth for $p<0$. At the centre the leading
expression is exact: $\psi(t,0)=C_p t^{-p}$ for $t>R_0$.

\paragraph{The logarithmic formula at bounded radius.}
At $p=1$, the identity $(s(a/s))'=0$ gives $k(s)=b/s$ and $C_1=b$.
To expand the logarithmic formula, we write $\zeta=r/t$. Factoring $t$ out
of both $t+r$ and $t-r$ cancels the common $\log t$ term. The
remaining logarithm has the convergent odd power series
\[
 \log\frac{t+r}{t-r}
 =\log(1+\zeta)-\log(1-\zeta)
 =2\sum_{j=0}^{\infty}\frac{\zeta^{2j+1}}{2j+1}
 =2\zeta+\frac23\zeta^3+O(\zeta^5),\qquad |\zeta|\le\tfrac12.
\]
For $0\le r\le R_1$ and $t>\max\{R_0+R_1,2R_1\}$, this expansion
is uniform. Dividing by $2r$ gives
\begin{equation}
 \begin{aligned}
 \psi(t,r)
 &=\frac b{2r}\log\frac{t+r}{t-r}\\
 &=\frac b t+\frac{br^2}{3t^3}
      +O\bigl(|b|r^4t^{-5}\bigr)
  =\frac b t+\frac{br^2}{3t^3}+O_{b,R_1}(t^{-5}).
 \end{aligned}
 \label{eq:minkowski-pone}
\end{equation}
The expressions at $r=0$ are their continuous limits, yielding
$\psi(t,0)=b/t$. Thus the logarithm of the quotient produces the
pure leading power $b/t$ at bounded radius. Along an outgoing null
ray, $u$ is fixed while $v=u+2r$ grows; that limit gives the
logarithmic radial behavior computed in the next subsection.

At $p=0$, formula \eqref{eq:flat-general-power} gives the exact identity
\begin{equation}
 \psi(t,r)=a+b\qquad(u>R_0).
 \label{eq:minkowski-pzero}
\end{equation}
For negative integer $p$, the exact expression is a polynomial in
$t,r$; the expansion \eqref{eq:minkowski-timelike} then terminates.

\subsection{Radiation fields and time derivatives}
\label{sec:minkowski-radiation}

The ordinary radiation field of $\psi$ is the rescaled limit
\[
 F(u)=\lim_{r\to\infty}r\psi(u+r,r).
\]
Here $u$ is held fixed while the point $(t,r)=(u+r,r)$ approaches
future null infinity. We also consider the radiation field of the
stationary time derivative $T\psi$, namely
\[
 \mathcal R(u)=\lim_{r\to\infty}r(T\psi)(u+r,r),
 \qquad T\psi=\left.\partial_t\psi\right|_r.
\]
Thus $T$ differentiates the solution at fixed radius before the limit
to null infinity is taken. The energy flux through null infinity is
expressed in terms of $|\mathcal R(u)|^2$; see the linear discussion
in \cite[Sections~1--2]{BSB}. For the exact power data with $p>1$
considered below, $\mathcal R(u)=\partial_uF(u)$.

For arbitrary radial data, \eqref{eq:minkowski-integral} gives
\begin{equation}
 F(u)=\frac12\int_u^\infty k(s)\,\dd s,\qquad u>0,
 \label{eq:minkowski-radiation-criterion}
\end{equation}
with a finite radiation field precisely when the displayed improper
integral converges.

For the exact powers with $C_p\ne0$, there are three cases.
If $p>1$, then
\begin{equation}
 F(u)=F_p(u):=\frac{C_p}{2(p-1)}u^{1-p},\qquad u>R_0.
 \label{eq:minkowski-Fp}
\end{equation}
If $p=1$, the rescaled field has logarithmic growth:
\begin{equation}
 \phi(u,r)=\frac b2\log\frac{u+2r}{u}
          =\frac b2\log(r/\ell)+O_u(1).
 \label{eq:minkowski-null-log}
\end{equation}
If $p<1$, including $p=0$ and all negative exponents, then
\begin{equation}
 \phi(u,r)\sim A_p(r):=
       \frac{C_p}{2^p(1-p)}r^{1-p}
       \quad(r\to\infty,\ u>R_0\ \hbox{fixed}).
 \label{eq:minkowski-null-power}
\end{equation}
For $0<p<1$, this gives decay of $\psi$ and power growth of $r\psi$
along each fixed outgoing ray. At fixed radius both $u$ and $v$ tend
to infinity together, and the difference in
\eqref{eq:flat-general-power} has the extra cancellation responsible
for \eqref{eq:minkowski-timelike}.

For the radiation field of $T\psi$, we assume that $k(s)$ has a finite
limit $L$ at infinity. The exact identity
\eqref{eq:minkowski-three-identities} then gives
\begin{equation}
 \mathcal R(u)=\lim_{r\to\infty}rT\psi(u+r,r)
              =\frac12\bigl[L-k(u)\bigr].
 \label{eq:minkowski-derivative-radiation}
\end{equation}
For the exact powers with $C_p\ne0$ and $u>R_0$, the results are
\begin{equation}
 \begin{aligned}
 \mathcal R(u)&=
 \begin{cases}
 -\dfrac{C_p}{2}u^{-p},&p>0,\\[4pt]
 0,&p=0,
 \end{cases}\\[3pt]
 rT\psi(u+r,r)&\sim 2^{-p-1}C_p r^{-p}
 \quad(p<0,\ r\to\infty,\ u\ \hbox{fixed}).
 \end{aligned}
 \label{eq:minkowski-derivative-cases}
\end{equation}
In particular, for $0<p\le1$ the time derivative has the finite
radiation field $\mathcal R(u)=-C_pu^{-p}/2$, while the rescaled
field $r\psi$ has the growth given above.

\paragraph{A renormalized radiation field.}
For exact powers, the formula valid for every real $p$ permits the
subtraction
\begin{equation}
 F_{\mathrm{ren}}(u)
 :=\lim_{r\to\infty}\bigl[\phi(u,r)-G_p(u+2r)\bigr]
 =-G_p(u),\qquad u>R_0.
 \label{eq:minkowski-renormalized}
\end{equation}
For $p>1$ this agrees with $F$. For $p\le1$ it subtracts the
explicitly growing incoming antiderivative $G_p$. At $p=1$ the choice of
$\ell$ fixes the additive constant. Using the full antiderivative
$G_p(u+2r)$ fixes both the growing terms and the finite term in
this normalization. Its derivative is
\[
 \partial_uF_{\mathrm{ren}}(u)=-\frac12 k(u).
\]
Whenever $L=\lim_{s\to\infty}k(s)$ is finite, the radiation field
$\mathcal R$ of $T\psi$ satisfies
\[
 \mathcal R(u)=\partial_uF_{\mathrm{ren}}(u)+\frac12 L.
\]
For the exact powers, $L=0$ when $p>0$ and $L=C_0$ when $p=0$.
The initial energy is evaluated from $(f,h)$, and the ordinary NP
charge is the outgoing null limit of a derivative of $r\psi$.

\subsection{Energy and the constant-background exception}
\label{sec:minkowski-energy}

The standard derivative energy of the data on $\{t=0\}$ is
\begin{equation}
 E[f,h]=2\pi\int_0^\infty
       r^2\bigl(h(r)^2+f'(r)^2\bigr)\,\dd r.
 \label{eq:minkowski-energy}
\end{equation}
Its exterior part for \eqref{eq:minkowski-power-data} is exactly
\begin{equation}
 E_{\mathrm{ext}}
 =2\pi\bigl(b^2+p^2a^2\bigr)
            \int_{R_0}^\infty r^{-2p}\,\dd r.
 \label{eq:minkowski-exterior-energy}
\end{equation}
Whenever $b^2+p^2a^2>0$, the energy is finite precisely for
$p>\tfrac12$. The divergence at $p=\tfrac12$ is logarithmic; below
that exponent it is a power divergence.

The coefficient in \eqref{eq:minkowski-exterior-energy} vanishes only
for zero exterior data, or for $p=0,b=0$. In the latter case $f=a$
outside a compact set, so the derivative energy is finite even when
$a\ne0$. The constant background has zero derivative energy; the
decaying representative $f-a$ lies in the usual energy space.

For $\tfrac12<p\le1$, the data can have finite initial energy and a
divergent rescaled field at null infinity. Their late radiation flux,
measured by the stationary time derivative, is finite:
\begin{equation}
 \int_{R_0}^\infty|\mathcal R(u)|^2\,\dd u
 =\frac{C_p^2}{4}\int_{R_0}^\infty u^{-2p}\,\dd u<\infty.
 \label{eq:minkowski-late-flux}
\end{equation}
This integral measures the flux on $u>R_0$. When $C_p=0$, an
infinite-energy outgoing tail can be carried towards arbitrarily
negative retarded times, while the flux on $u>R_0$ is zero.

\subsection{Ordinary and time-inverted NP limits}
\label{sec:minkowski-charges}

For spherically symmetric solutions, we use the normalization
\[
 I_0[\chi]=\lim_{r\to\infty}
         r^2\left.\partial_r(r\chi)\right|_u.
\]
For general radial data the exact identity
\eqref{eq:minkowski-three-identities} gives, whenever the limit exists,
\begin{equation}
 I_0[\psi]=\frac14\lim_{s\to\infty}s^2k(s).
 \label{eq:minkowski-NP-criterion}
\end{equation}
For exact power data with $C_p\ne0$,
\begin{equation}
 r^2\left.\partial_r\phi\right|_u
 =C_p r^2(u+2r)^{-p}
 \sim 2^{-p}C_p r^{2-p}.
 \label{eq:minkowski-NP-growth}
\end{equation}
Thus $I_0$ diverges for $p<2$, equals $C_2/4=(b-a)/4$ at $p=2$,
and vanishes for $p>2$. At every real exponent the weighted coefficient
$\lim r^p\partial_r\phi|_u=2^{-p}C_p$ remains well defined.

For $p>1$ there is a normalized pointwise time integral,
\begin{equation}
 \psi^{(1)}(t,r)=-\int_t^\infty\psi(s,r)\,\dd s,
 \qquad T\psi^{(1)}=\psi,\qquad
 \psi^{(1)}(t,r)\longrightarrow0\quad(t\to\infty).
 \label{eq:minkowski-time-primitive}
\end{equation}
The integral and its required local derivatives converge; integration
of the wave equation, or the characteristic formula, verifies that
$\psi^{(1)}$ is again a smooth radial wave. Its incoming combination
on the relevant tail is
\begin{equation}
 k_1(s)=-\int_s^\infty k(y)\,\dd y
       =-\frac{C_p}{p-1}s^{1-p}.
 \label{eq:minkowski-primitive-k}
\end{equation}
We therefore obtain
\begin{equation}
 r^2\left.\partial_r(r\psi^{(1)})\right|_u
 \sim-\frac{C_p}{2^{p-1}(p-1)}r^{3-p}.
 \label{eq:minkowski-inverse-NP-growth}
\end{equation}
In particular, for $C_p\ne0$,
\begin{equation}
 I_0^{(1)}[\psi]:=I_0[\psi^{(1)}]
 =
 \begin{cases}
 -C_3/8=(2a-b)/8,&p=3,\\
 0,&p>3,
 \end{cases}
 \label{eq:minkowski-inverse-charge}
\end{equation}
whereas the limit diverges for $1<p<3$.
For $p\le1$ the future integral in
\eqref{eq:minkowski-time-primitive} diverges when $C_p\ne0$.
For $C_p\ne0$ these are distinct conditions: a pointwise time integral exists for
$p>1$, its ordinary radiation field exists for $p>2$,
and its NP limit is finite for $p\ge3$.
In particular, for $2<p<3$ the ordinary charge is zero and the
defining limit for $I_0^{(1)}$ diverges.

\paragraph{Higher time integrals.}
The same calculation describes arbitrarily large $p$. For an integer
$n\ge1$ and $p>n$, we define
\begin{equation}
 \psi^{(n)}(t,r)
 =\frac{(-1)^n}{(n-1)!}
    \int_t^\infty(s-t)^{n-1}\psi(s,r)\,\dd s.
 \label{eq:minkowski-higher-primitives}
\end{equation}
Then $T\psi^{(n)}=\psi^{(n-1)}$, with $\psi^{(0)}=\psi$, and
\[
 k_n(s)=
 \frac{(-1)^n C_p}{\prod_{j=1}^n(p-j)}s^{n-p}
 \qquad(s>R_0).
\]
Consequently, within the convergence range $p>n$,
\begin{equation}
 I_0[\psi^{(n)}]=
 \begin{cases}
 0,&p>n+2,\\[2pt]
 \displaystyle\frac{(-1)^n C_{n+2}}{4(n+1)!},&p=n+2,
 \end{cases}
 \label{eq:minkowski-higher-charges}
\end{equation}
and the limit diverges for $n<p<n+2$ if $C_p\ne0$.
At $p=n+2$ all lower-order charges vanish.
For example, $p=4$ gives $I_0=I_0^{(1)}=0$ and
$I_0[\psi^{(2)}]=C_4/24$. If $C_p\ne0$ and $p\le n$, the defining
future integral for $\psi^{(n)}$ diverges. The cancelling
cases are treated separately below.

\subsection{Classification and exact cancellations}
\label{sec:minkowski-classification}

Table~\ref{tab:minkowski-all-p} covers every $p\in\R$ and every pair
$(a,b)$ for the exact exterior class
\eqref{eq:minkowski-power-data}. The first eleven rows assume
$C_p\ne0$; the last two rows cover its complement. The function
$A_p$ is defined in \eqref{eq:minkowski-null-power}.

To compare directly with Table~\ref{tab:intro-black-hole}, we use
one family of future hypersurfaces with outgoing null ends. We choose
a smooth radial function $H_{\mathrm M}$, even near $r=0$, with
\[
 H_{\mathrm M}(r)\ge r,\qquad
 0\le H_{\mathrm M}'(r)<1\quad(0\le r<R),\qquad
 H_{\mathrm M}(r)=r\quad(r\ge R),
\]
and a smooth transition at a fixed $R>0$. We define
\[
 \tau=t-H_{\mathrm M}(r),\qquad
 \mathcal N_\tau=\{t=\tau+H_{\mathrm M}(r)\}.
\]
Thus $\mathcal N_\tau$ has a spacelike interior portion and the
exact outgoing null end $u=\tau$; Figure~\ref{fig:minkowski-spatial-data}
shows one such hypersurface. At fixed radius, $t-\tau=H_{\mathrm M}(r)$
is a fixed shift, and $F(u)$ becomes $F(\tau)$ on null infinity.
The initial data are prescribed on $\{t=0\}$.
The approach column takes $r\to\infty$ at fixed $\tau>R_0$;
the next column gives the behavior of $F(\tau)$ as $\tau\to\infty$.
Bounded-radius asymptotics are uniform on every $0\le r\le R_1$.
Since $u=\tau+H_{\mathrm M}(r)-r\ge\tau$, the exact identities
for $u>R_0$ apply throughout the region of the table.

\begin{table}[H]
\centering
\small
\setlength{\tabcolsep}{2.5pt}
\renewcommand{\arraystretch}{1.30}
\begin{tabularx}{\textwidth}{@{}
 >{\raggedright\arraybackslash}p{.70in}
 >{\raggedright\arraybackslash}p{1.00in}
 >{\raggedright\arraybackslash}p{.98in}
 >{\raggedright\arraybackslash}X
 >{\raggedright\arraybackslash}p{.50in}
 >{\raggedright\arraybackslash}p{.40in}
 >{\raggedright\arraybackslash}p{.80in}@{}}
\toprule
Power $p$
 & Field $\psi$ at bounded radius\newline
   $\tau\to\infty$
 & Approach to null infinity\newline
   $r\psi(\tau,r)$:\newline
   $r\to\infty$, $\tau$ fixed
 & Late time along null infinity\newline
   $F(\tau)$ as $\tau\to\infty$
 & Finite initial energy?
 & $I_0$
 & $I_0^{(1)}$\\
\midrule
$p<0$
 & $\sim C_p\tau^{-p}$ (growth)
 & $\sim A_p(r)$
 & $\infty$
 & No & $\infty$ & Undefined\\
$p=0$
 & $a+b$ exactly
 & $(a+b)r$ exactly
 & $\infty$
 & Yes/No & $\infty$ & Undefined\\
$0<p<\tfrac12$
 & $\sim C_p\tau^{-p}$
 & $\sim A_p(r)$
 & $\infty$
 & No & $\infty$ & Undefined\\
$p=\tfrac12$
 & $\sim C_p\tau^{-1/2}$
 & $\sim A_p(r)$
 & $\infty$
 & No (log) & $\infty$ & Undefined\\
$\tfrac12<p<1$
 & $\sim C_p\tau^{-p}$
 & $\sim A_p(r)$
 & $\infty$
 & Yes & $\infty$ & Undefined\\
$p=1$
 & $\sim b\tau^{-1}$
 & $\sim\tfrac b2\log(2r/\ell)$
 & $\infty$
 & Yes & $\infty$ & Undefined\\
$1<p<2$
 & $\sim C_p\tau^{-p}$
 & $\to F(\tau)$
 & $\sim\dfrac{C_p}{2(p-1)}\tau^{1-p}$
 & Yes & $\infty$ & $\infty$\\
$p=2$
 & $\sim4I_0\tau^{-2}$
 & $\to F(\tau)$
 & $\sim2I_0\tau^{-1}$
 & Yes & $\frac{b-a}{4}$ & $\infty$\\
$2<p<3$
 & $\sim C_p\tau^{-p}$
 & $\to F(\tau)$
 & $\sim\dfrac{C_p}{2(p-1)}\tau^{1-p}$
 & Yes & $0$ & $\infty$\\
$p=3$
 & $\sim-8I_0^{(1)}\tau^{-3}$
 & $\to F(\tau)$
 & $\sim-2I_0^{(1)}\tau^{-2}$
 & Yes & $0$ & $\frac{2a-b}{8}$\\
$p>3$
 & $\sim C_p\tau^{-p}$
 & $\to F(\tau)$
 & $\sim\dfrac{C_p}{2(p-1)}\tau^{1-p}$
 & Yes & $0$ & $0$\\
\midrule
Any $p$:\newline $C_p=0$,\newline $a\ne0$
 & $0$ for $\tau>R_0$
 & $\to0$
 & $0$ for $\tau>R_0$
 & Iff $p>\tfrac12$ & $0$ & $0$\\
Compactly supported data
 & $0$ for $\tau>R_0$
 & $\to0$
 & $0$ for $\tau>R_0$
 & Yes & $0$ & $0$\\
\bottomrule
\end{tabularx}
\caption{Minkowski asymptotics from initial
hypersurface $\{t=0\}$.}
\label{tab:minkowski-all-p}
\end{table}

Indeed, $C_p=0$ is exactly the relation
\begin{equation}
 b=(p-1)a.
 \label{eq:minkowski-cancellation}
\end{equation}
For exact tails it makes $k=0$ on $[R_0,\infty)$, and
\eqref{eq:minkowski-integral} gives $\psi=0$ throughout $u>R_0$.
All normalized future time integrals then exist at every fixed radius,
for every order, and vanish in the same region. Their NP limits
are zero. This conclusion includes noncompact, infinite-energy
outgoing data. 

The normalization here is decay at future timelike infinity, as in
\eqref{eq:minkowski-time-primitive}. For example, when $f=a/r$ and $h=0$ outside a
compact set, the future-normalized time integral has initial limit $-a$
at spatial infinity. Adding the constant $a$ makes that initial
limit zero, but changes its late value to $a$. The zero inverse
charges in the cancelling rows of the table refer to the explicit
future normalization. For the first time integral of the $p\ge3$ data used in the black
hole comparison, the spatial and future normalizations agree.

At $p=1$ the relation is simply $b=0$. Thus a smooth position datum
equal to $a/r$ outside a compact set, with zero exterior velocity,
gives $\psi=0$ throughout $u>R_0$.
At $p=0,b=0$, by contrast, the solution equals the constant $a$
in the late region and has finite derivative energy. This latter
case belongs to the $C_0\ne0$ row when $a\ne0$.
The energy and the incoming amplitude are therefore independent
tests, as reflected in the table.

\subsection{Asymptotic powers, remainders, and the case \texorpdfstring{$p=3$}{p = 3}}
\label{sec:minkowski-remainders}

The exact exterior assumption is convenient for the complete table.
For a quantitative asymptotic version, we fix an integer $N\ge0$
and assume instead that
\begin{equation}
 f(r)=a r^{-p}+\Os[N+1](r^{-p-\delta}),\qquad
 h(r)=b r^{-p-1}+\Os[N](r^{-p-1-\delta}),
 \quad p\in\R,\quad\delta>0.
 \label{eq:minkowski-symbol-data}
\end{equation}
The superscripts specify the derivative counts in
\eqref{eq:symbol-notation}. Since $k=(rf)'+rh$ uses one radial
derivative of $f$, these assumptions give
\begin{equation}
 k(r)=C_p r^{-p}+\rho(r),\qquad
 \rho=\Os[N](r^{-p-\delta}),
 \label{eq:minkowski-k-remainder}
\end{equation}
and \eqref{eq:minkowski-integral} immediately gives
\begin{equation}
 \psi(t,r)
 =C_p t^{-p}
  +O_{R_1}(t^{-p-2}+t^{-p-\delta})
 \quad(0\le r\le R_1).
 \label{eq:minkowski-quantitative-remainder}
\end{equation}
For $0\le j\le N$, differentiating the integral formula gives the
corresponding estimate for $T^j\psi=\partial_t^j\psi$, with each
power of $t$ decreased by $j$. When $C_p=0$, the same identity gives
the sharper bound $\psi=O_{R_1}(t^{-p-\delta})$.
For $p>1$,
\begin{equation}
 F(u)=\frac{C_p}{2(p-1)}u^{1-p}
          +O(u^{1-p-\delta}).
 \label{eq:minkowski-radiation-remainder}
\end{equation}
For $p<1$ and $C_p\ne0$, the leading power divergence
\eqref{eq:minkowski-null-power} is unchanged. At $p=1,b\ne0$,
the leading logarithmic divergence is unchanged.

If $C_p=0$, only the leading incoming term has been removed.
The first nonzero term of $k$ controls the remaining timelike signal
and the charge limits.
The exact criteria
\eqref{eq:minkowski-radiation-criterion} and
\eqref{eq:minkowski-NP-criterion} still apply.
For instance, a residual bound $k=O(r^{-p-\delta})$ ensures
convergence of $F$ when $p+\delta>1$. If $k(r)=c r^{-2}+o(r^{-2})$, then $I_0=c/4$;
if $k=o(r^{-2})$, then $I_0=0$.
Energy must still be checked on $f,h$ separately. In particular,
for data tending to a constant with $p=0,b=0$, a remainder exponent
$\delta>\tfrac12$ is sufficient for finite energy.

As a useful example with a slower position tail, we consider
\begin{equation}
 f(r)=\frac{a_1}{r}+\frac{a_2}{r^2}+\Os[N+1](r^{-2-\delta}),
 \qquad
 h(r)=\frac{b_2}{r^3}+\Os[N](r^{-3-\delta}).
 \label{eq:minkowski-coulomb-example}
\end{equation}
Then $k=(b_2-a_2)r^{-2}+\Os[N](r^{-2-\delta})$ and
\[
 I_0=\frac{b_2-a_2}{4}.
\]
More generally, finite sums of exterior powers can be treated term
by term in $k$. For a renormalized radiation field, we subtract the antiderivatives
of all nonintegrable terms of $k$. Subtracting just the
leading antiderivative in \eqref{eq:minkowski-renormalized} is sufficient
if $\rho\in L^1([R_0,\infty))$, in particular if $p+\delta>1$.

\paragraph{The Minkowski coefficient for $p=3$.}
\label{sec:flat-interpretation}
We now set $p=3$ in the Minkowski formulas above. This calculation
identifies $\mathcal I_\infty=(2a-b)/8$, which appears as the
contribution from spatial infinity in the black hole coefficient
\eqref{eq:introcharge}. For this power, the initial tail produces
the same decay rates as the leading black hole asymptotics.
We write $C_{\mathrm{in}}=(2a-b)/4=-C_3/4$. For the exact exterior
data $f=ar^{-3}$ and $h=br^{-4}$, \eqref{eq:flat-general-power} reduces to
\begin{equation}
 \psi(t,r)
 =\frac{C_{\mathrm{in}}}{r}
    \left[\frac1{(t+r)^2}-\frac1{(t-r)^2}\right]
 =(b-2a)\frac{t}{(t^2-r^2)^2},\qquad u>R_0.
 \label{eq:flat}
\end{equation}
Hence
\begin{equation}
 \lim_{t\to\infty}t^3\psi(t,r)=b-2a=-4C_{\mathrm{in}},
 \qquad
 \lim_{u\to\infty}u^2F(u)=\frac{b-2a}{4}=-C_{\mathrm{in}}.
 \label{eq:flat-limits}
\end{equation}
Together with \eqref{eq:minkowski-inverse-charge}, this is exactly
$-8\mathcal I_\infty$ and $-2\mathcal I_\infty$, where
$\mathcal I_\infty=(2a-b)/8$.
Related flat-space formulas and their interpretation as signals
inherited from noncompact data are discussed in
\cite[Sections~3.1--3.2]{BCCPV}.

The four relevant thresholds have different meanings:
$p=\tfrac12$ separates finite from infinite energy, while $p=1$
separates generic finite radiation fields from divergent rescaled
fields at null infinity.
The ordinary NP limit becomes finite at $p=2$, and the first
time-inverted NP limit becomes finite at $p=3$. The last is also the exponent
at which the flat data-induced signal has the same $t^{-3}$ power
as the spherically symmetric black hole tail. In flat space the compact interior
data have left the late region; on a black hole background they
can contribute through curvature scattering. The additional bulk
term in the black hole coefficient is established in the subsequent
sections.

\section{The geometric setting}
\label{sec:setting}

We denote by $r_h>0$ the horizon radius. The open exterior is
\[
 \mathcal M_{\mathrm{ext}}
 =\mathbb R_t\times(r_h,\infty)_r\times\mathbb S^2.
\]
We use coordinates $(t,r,\theta,\varphi)$, where $t\in\mathbb R$ is
the stationary time with Killing field $T=\partial_t$,
$r\in(r_h,\infty)$ is the area radius, and
$\omega=(\theta,\varphi)$ denotes the angular variables on the unit
sphere. In the usual angular coordinate chart,
$0<\theta<\pi$ and $\varphi\in\mathbb R/(2\pi\mathbb Z)$, with
\[
 \dd\omega^2=\dd\theta^2+\sin^2\theta\,\dd\varphi^2.
\]
Smooth angular charts cover the poles. The horizon endpoints are
included using the regular coordinates described below.

\subsection{The background and coordinates}

We consider the exterior metric
\begin{equation}
 g=-D(r)\dd t^2+D(r)^{-1}\dd r^2+r^2\dd\omega^2,
 \qquad r>r_h.
 \label{eq:metric}
\end{equation}
We assume $D\in C^\infty([r_h,\infty))$,
\begin{equation}
 D(r_h)=0,\qquad D'(r_h)>0,\qquad D(r)>0\quad(r>r_h),
 \label{eq:horizon}
\end{equation}
and, for $M>0$, $d_2\in\R$, and $\eta>0$,
\begin{equation}
 D(r)=1-\frac{2M}{r}+\frac{d_2}{r^2}
                 +\Os[\infty](r^{-2-\eta}).
 \label{eq:D}
\end{equation}
Here the superscript $\infty$ has the meaning in
\eqref{eq:symbol-notation}: the metric remainder obeys the weighted
radial bounds at every nonnegative integer derivative order.
The metric has a smooth regular extension across the nondegenerate future
horizon, and across the bifurcation sphere when that endpoint is used.
These metric assumptions define the geometric background class.

We fix the tortoise coordinate $r_*$ and double-null coordinates by
\[
 \frac{\dd r_*}{\dd r}=D^{-1},\qquad u=t-r_*,\qquad v=t+r_*.
\]
We write $r=r(r_*)$ for the inverse of this strictly increasing map.
The tortoise coordinate ranges over $\mathbb R$, so the exterior
double-null domain is $(u,v)\in\mathbb R^2$. Thus
\begin{equation}
 \partial_u r=-\frac D2,\quad \partial_v r=\frac D2,
 \quad T=\partial_u+\partial_v,
 \quad \left.\partial_r\right|_u=\frac2D\partial_v.
 \label{eq:coords}
\end{equation}
The derivatives $\partial_u,\partial_v$ denote differentiation in
the double-null chart. In the ingoing coordinates $(v,r,\omega)$,
\[
 g=-D(r)\dd v^2+2\dd v\dd r+r^2\dd\omega^2,
 \qquad T=\left.\partial_v\right|_r.
\]
This chart is regular at the future horizon, with
$v\in\mathbb R$ and $r\in[r_h,\infty)$, and the metric extends
smoothly to $r<r_h$ near each finite-$v$ horizon point. To include
the bifurcation sphere, we set
\[
 \kappa=\frac{D'(r_h)}2,\qquad
 \mathsf U=-e^{-\kappa u},\qquad
 \mathsf V=e^{\kappa v}.
\]
The exterior has $\mathsf U<0$ and $\mathsf V>0$; its future horizon
has $\mathsf U=0,\mathsf V>0$, and the bifurcation sphere has
$\mathsf U=\mathsf V=0$. The assumed extension is smooth in these
regular horizon coordinates.

\subsection{Hypersurfaces with static and outgoing null ends}
\label{sec:hypersurfaces}

\paragraph{Initial hypersurfaces with a static end.}
The static initial hypersurface $\{t=0\}$ meets the bifurcation
sphere $\mathcal B$ in its closure. We write $\Sigma_0$ for an initial
hypersurface meeting a spherically symmetric cross-section of $\HH$ strictly
to the future of $\mathcal B$. We also allow compact deformations of
$\{t=0\}$ that still meet $\mathcal B$, and denote such a hypersurface
by $\Sigma_{\mathcal B}$.

In statements common to these configurations, we choose one initial
hypersurface and denote it by $\mathcal S$. We assume that it is
rotation-invariant, spacelike, and Cauchy for its future exterior
development, and smooth up to its horizon endpoint in regular coordinates.
We write $S_{\mathcal H^+}=\mathcal S\cap\HH$ for its horizon endpoint.
This sphere equals $\mathcal B$ when $\mathcal S=\{t=0\}$ or
$\mathcal S=\Sigma_{\mathcal B}$, and is the later horizon
cross-section when $\mathcal S=\Sigma_0$. Each initial
hypersurface has a \emph{static end}: after a constant translation of
$t$, we have
\[
 \mathcal S\cap\{r\ge R_0\}=\{t=0,\ r\ge R_0\}
\]
for some fixed $R_0>r_h$. Its outer end reaches spatial infinity $i^0$.

\paragraph{The future hypersurfaces $\mathcal N_\tau$.}
We use one stationary family to describe the evolved solution and to
measure its energy. We choose a smooth function $H(r)$, regular at
$r=r_h$, and a fixed $R>r_h$ such that
\[
 H'(r)>0,\qquad D(r)H'(r)<2\quad(r_h\le r<R),
 \qquad H(r)=2r_*(r)\quad(r\ge R).
\]
We choose the transition smoothly and define
\begin{equation}
 \begin{aligned}
 \tau&=v-H(r),&
 \mathcal N_\tau&=\{v=\tau+H(r)\},\\
 \mathcal C_\tau&=\mathcal N_\tau\cap\{r\le R\},&
 \Gamma_\tau&=\mathcal N_\tau\cap\HH.
 \end{aligned}
 \label{eq:future-hypersurfaces}
\end{equation}
The compact portion $\mathcal C_\tau$ is spacelike for $r<R$ and meets
the future horizon at the sphere $\Gamma_\tau$. Its outgoing null end is
\[
 \mathcal N_\tau\cap\{r\ge R\}=\{u=\tau,\ r\ge R\}.
\]
The compact portion and the outgoing null portion meet on the
two-sphere $\mathcal N_\tau\cap\{r=R\}$.
Indeed, the radial part of the metric induced on $\mathcal N_\tau$ is
$H'(2-DH')\,\dd r^2$, which is positive on the spacelike portion and
zero on the null portion. The coordinates $(\tau,r,\omega)$ are
regular at the future horizon. We use the stationary coordinate derivatives
\begin{equation}
 T=\left.\partial_\tau\right|_r,\qquad
 Y=\left.\partial_r\right|_\tau
   =\left.\partial_r\right|_v+H'(r)T,\qquad [T,Y]=0.
 \label{eq:future-derivatives}
\end{equation}
On the outgoing null portion, $\tau=u$ and
$Y=\partial_r|_u$ exactly. At each fixed exterior radius, $\tau$ differs
from $t$ by a constant, and on $\HH$ it equals $v-H(r_h)$.

We choose the relevant members of this family in the future development
of $\mathcal S$. The member $\mathcal N_{u_0}$, with $u_0<0$, is used to
transfer the initial data to an outgoing null end;
Lemma~\ref{lem:compact-transfer} gives the quantitative choice of its
compact portion. We fix the additive constants in $r_*$ and $\tau$
consistently so that $\mathcal N_0$ is a sufficiently late reference
hypersurface, while $\tau=u$ on every outgoing null portion. The initial data are prescribed on $\mathcal S$; values on $\mathcal N_{u_0}$ and
$\mathcal N_0$ are restrictions of the evolved solution.

\paragraph{Spacelike hypersurfaces used to compare the cited estimates.}
\label{par:spacelike-comparison}
The derivative estimates in \cite[Section~7]{AAG1} use spacelike
hypersurfaces approaching null infinity. For that comparison, we choose
a smooth height function $\widehat H$, regular at the horizon, with
$\widehat H'>0$, $D\widehat H'<2$, and
$\widehat H'=2/D-r^{-2}$ for all sufficiently large $r$.
On the compact part we choose it as in \cite[Section~2.2]{AAG1}.
We define
\[
 \widehat\tau=v-\widehat H(r),\qquad
 \widehat{\mathcal N}_s=\{\widehat\tau=s\},\qquad
 \widehat Y=\left.\partial_r\right|_{\widehat\tau}.
\]
Writing $\Delta H=H-\widehat H$ and $\zeta=H'-\widehat H'$, we have
\[
 \widehat\tau=\tau+\Delta H(r),\qquad
 Y=\widehat Y+\zeta(r)T,\qquad
 \zeta(r)=r^{-2}\quad\hbox{for large }r.
\]
Thus $\Delta H$ is bounded. We fix the additive constant in
$\widehat H$ so that the relevant members lie in the future
development of the initial hypersurface.
These spacelike hypersurfaces are used only in the comparison
argument of Section~\ref{sec:frac-decay-23}; the conclusions of
our theorems are stated on $\mathcal N_\tau$.

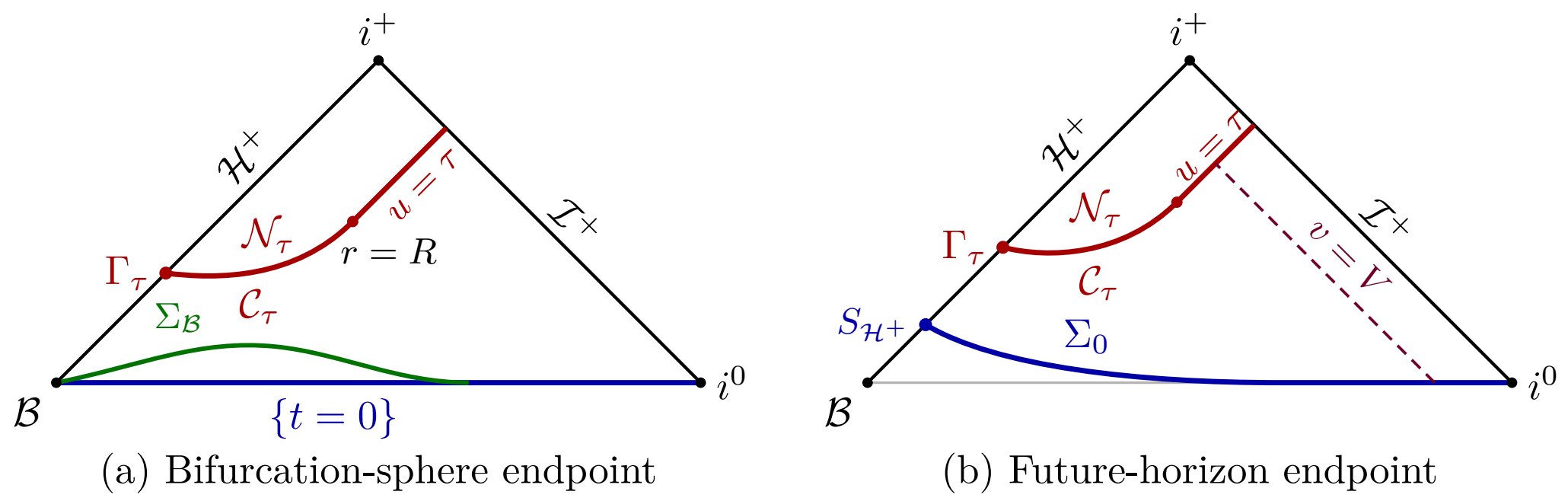
\begin{figure}[H]
\centering
\begin{tikzpicture}[x=1.12cm,y=1.12cm,
 every node/.style={font=\small},
 boundary/.style={line width=.8pt},
 initial/.style={draw=blue!65!black,line width=1.3pt},
 future/.style={draw=red!65!black,line width=1.3pt}]
 \begin{scope}
  \coordinate (B) at (0,0);
  \coordinate (IZ) at (5,0);
  \coordinate (IP) at (2.5,2.5);
  \coordinate (G) at (.85,.85);
  \coordinate (J) at (2.3,1.25);
  \coordinate (F) at (3.025,1.975);
  \draw[boundary] (B)--(IP)--(IZ);
  \draw[initial] (B)--(IZ);
  \draw[green!45!black,line width=1.1pt]
     (B) .. controls (.65,.13) and (1.1,.34) .. (1.7,.28)
     .. controls (2.2,.23) and (2.65,0) .. (3.2,0);
  \draw[future] (G) .. controls (1.25,.8) and (1.85,.8) .. (J)
     -- (F);
  \fill (B) circle (1.3pt);
  \fill (IZ) circle (1.3pt);
  \fill (IP) circle (1.3pt);
  \fill[red!65!black] (G) circle (1.6pt);
  \fill[red!65!black] (J) circle (1.4pt);
  \node[below left] at (B) {$\mathcal B$};
  \node[right] at (IZ) {$i^0$};
  \node[above] at (IP) {$i^+$};
  \node[rotate=45] at (1.45,1.77) {$\HH$};
  \node[rotate=-45] at (4.04,1.25) {$\scri$};
  \node[blue!65!black,below] at (2.15,0) {$\{t=0\}$};
  \node[green!45!black,above,font=\footnotesize] at (.95,.27)
      {$\Sigma_{\mathcal B}$};
  \node[left,red!65!black] at (G) {$\Gamma_\tau$};
  \node[red!65!black] at (1.65,1.14) {$\mathcal N_\tau$};
  \node[red!65!black] at (1.58,.59) {$\mathcal C_\tau$};
  \node[red!65!black,rotate=45,font=\footnotesize]
      at (2.86,1.53) {$u=\tau$};
  \node[font=\footnotesize] at (2.58,1.02) {$r=R$};
  \node[align=center] at (2.5,-.70)
      {(a) Bifurcation-sphere endpoint};
 \end{scope}
 \begin{scope}[xshift=7.05cm]
  \coordinate (B) at (0,0);
  \coordinate (IZ) at (5,0);
  \coordinate (IP) at (2.5,2.5);
  \coordinate (S) at (.45,.45);
  \coordinate (G) at (1.05,1.05);
  \coordinate (J) at (2.4,1.4);
  \coordinate (F) at (3,2);
  \draw[boundary] (B)--(IP)--(IZ);
  \draw[black!30,line width=.6pt] (B)--(IZ);
  \draw[initial] (S) .. controls (1.0,.12) and (2.1,0) .. (3.2,0)
      -- (IZ);
  \draw[future] (G) .. controls (1.4,.96) and (1.98,.98) .. (J)
      -- (F);
  \draw[purple!60!black,densely dashed,line width=.7pt]
      (4.4,0)--(2.7,1.7)
      node[pos=.48,above,sloped,font=\footnotesize] {$v=V$};
  \fill (B) circle (1.3pt);
  \fill (IZ) circle (1.3pt);
  \fill (IP) circle (1.3pt);
  \fill[blue!65!black] (S) circle (1.6pt);
  \fill[red!65!black] (G) circle (1.6pt);
  \fill[red!65!black] (J) circle (1.4pt);
  \node[below left] at (B) {$\mathcal B$};
  \node[right] at (IZ) {$i^0$};
  \node[above] at (IP) {$i^+$};
  \node[rotate=45] at (1.55,1.87) {$\HH$};
  \node[rotate=-45] at (4.04,1.25) {$\scri$};
  \node[left,blue!65!black,font=\footnotesize] at (S) {$S_{\mathcal H^+}$};
  \node[blue!65!black] at (1.7,.36) {$\Sigma_0$};
  \node[left,red!65!black] at (G) {$\Gamma_\tau$};
  \node[red!65!black] at (1.78,1.34) {$\mathcal N_\tau$};
  \node[red!65!black] at (1.80,.77) {$\mathcal C_\tau$};
  \node[red!65!black,rotate=45,font=\footnotesize]
      at (2.67,1.83) {$u=\tau$};
  \node[align=center] at (2.5,-.70)
      {(b) Future-horizon endpoint};
 \end{scope}
\end{tikzpicture}
\caption{Initial and future hypersurfaces. }
\label{fig:future-hypersurfaces}
\end{figure}

\paragraph{Outgoing and incoming null hypersurfaces.}
We use $\{u=u_0\}$ for the outgoing hypersurface containing the null
portion of $\mathcal N_{u_0}$, and $\{v=V\}$ for an incoming null
hypersurface. For $\tau_1\le\tau_2$, we denote its portion between
$\mathcal N_{\tau_1}$ and $\mathcal N_{\tau_2}$ by
\[
 I_V(\tau_1,\tau_2)=\{v=V,\ \tau_1\le\tau\le\tau_2\}.
\]
These incoming hypersurfaces also truncate $\mathcal S$ and
$\mathcal N_{u_0}$ in the flux identity of Section~\ref{sec:flux}.
Figure~\ref{fig:future-hypersurfaces} shows the initial hypersurfaces
and the portions of $\mathcal N_\tau$.

\subsection{The spherically symmetric wave equation}
\label{sec:radial-wave}

For a spherically symmetric solution $\psi$ of \eqref{wave}, we define
\begin{equation}
 \phi=r\psi,\qquad Q=\partial_v\phi.
 \label{eq:phi}
\end{equation}
We define the black hole potential by
\begin{equation}
 V(r_*)=\frac{D(r(r_*))D'(r(r_*))}{r(r_*)}.
 \label{eq:potential-definition}
\end{equation}
Here $r=r(r_*)$ is the inverse tortoise-coordinate map, and the prime
on $D$ denotes differentiation with respect to the area radius $r$.
The reduced field satisfies
\begin{equation}
 (\partial_t^2-\partial_{r_*}^2+V)\phi=0,
 \qquad \partial_u\partial_v\phi=-\frac{DD'}{4r}\phi.
 \label{eq:wave}
\end{equation}
As $r\to\infty$, the metric expansion \eqref{eq:D} gives
\begin{equation}
 r_*=r+2M\log r+O(1),\qquad
 V(r_*)=2Mr_*^{-3}+O(r_*^{-4}\log r_*),
 \label{eq:potential}
\end{equation}
with the corresponding differentiated bounds.

\paragraph{The integral formula for the reduced wave equation.}
We regard \eqref{eq:wave} as the one-dimensional equation
$(\partial_t^2-\partial_{r_*}^2)\phi=-V\phi$. We fix a point $P$
with coordinates $(t,r_*)$, where
\[
 t\ge0,\qquad r_*-t\ge r_*(R_0).
\]
Its null coordinates are $u=t-r_*$ and $v=t+r_*$. The past
characteristic triangle between $P$ and the static end of the
initial hypersurface is
\begin{equation}
 \mathcal T_{u,v}
 =\bigl\{(\lambda,\rho_*):0\le\lambda\le t,\ 
      r_*-(t-\lambda)\le\rho_*\le r_*+(t-\lambda)\bigr\}.
 \label{eq:duhamel-domain}
\end{equation}
Its base on $t=0$ is the interval $[r_*-t,r_*+t]$ in the tortoise
coordinate. 

The d'Alembert formula with source $-V\phi$ gives
\begin{align}
 \phi(t,r_*)={}&\frac12\bigl[\phi(0,r_*-t)+\phi(0,r_*+t)\bigr]
       +\frac12\int_{r_*-t}^{r_*+t}T\phi(0,\rho_*)\,\dd\rho_*
       \notag\\
 &-\frac12\int_0^t
       \int_{r_*-(t-\lambda)}^{r_*+(t-\lambda)}
       V(\rho_*)\phi(\lambda,\rho_*)\,\dd\rho_*\dd\lambda.
 \label{eq:duhamel}
\end{align}
All field arguments in \eqref{eq:duhamel} are in the $(t,r_*)$
coordinates; $\lambda$ and $\rho_*$ are the time and tortoise-radius
integration variables. The potential $V$ is given by
\eqref{eq:potential-definition}. We apply this formula in
Section~\ref{sec:exterior} to propagate the initial spatial asymptotics.

\subsection{Initial data and energy assumptions}
\label{sec:data}
\label{sec:analytic}

\paragraph{The initial data.}
We denote by $n_{\mathcal S}$ the future-directed unit normal to $\mathcal S$.
We prescribe the initial data $(f,\pi)$ on $\mathcal S$ and
write $h$ for the stationary time derivative:
\[
 f=\psi|_{\mathcal S},\qquad \pi=n_{\mathcal S}\psi|_{\mathcal S},
 \qquad h=T\psi|_{\mathcal S}.
\]
The pair $(f,\pi)$ determines $h$, and on the static outer end
$h=\sqrt D\,\pi$. Given $p\in\mathbb R$, $\delta>0$, and an integer
$m\ge0$, we consider
\begin{equation}
 f=a r^{-p}+\Os[m+1](r^{-p-\delta}),\qquad
 h=b r^{-p-1}+\Os[m](r^{-p-1-\delta}).
 \label{eq:classification-data}
\end{equation}
Thus the remainders of $f$ and $h$ have $m+1$ and $m$ controlled
radial derivatives, respectively. Section~\ref{sec:main-data-norms}
specifies the count used in each quantitative estimate; the main
theorem takes $m=m(p)$ from \eqref{eq:classification-regularity}.
The combination of leading coefficients used in the propagation estimates is
\begin{equation}
 C_p=b+(1-p)a.
 \label{eq:frac-incoming-coefficient}
\end{equation}

At a bifurcation-sphere endpoint, smoothness is measured in regular
geometric coordinates for the pair $(f,\pi)$. On the static hypersurface $\{t=0\}$,
proper distance $\rho$ satisfies $\rho\asymp\sqrt{r-r_h}$, and
\[
 n_{\{t=0\}}=D^{-1/2}T,
 \qquad \dd\mu_{\{t=0\}}=D^{-1/2}r^2\dd r\dd\omega,
 \qquad \frac{r^2}{D}h=\frac{r^2}{\sqrt D}\pi.
\]
The last expression is locally integrable at $r_h$.
Thus the regularity condition at this endpoint is smoothness in
$\rho$ and in regular angular coordinates. Smooth data on the
one-sided hypersurface admit local smooth extensions across its
endpoint, and uniqueness fixes the exterior development.

For $p\ge3$, we define the convergent geometric functional
\begin{equation}
 B_{\mathcal S}=\frac1{4\pi}\left(
       \int_{S_{\mathcal H^+}}f\,\dd S+
       \int_{\mathcal S}\pi\,\dd\mu_{\mathcal S}\right).
 \label{eq:geometricB}
\end{equation}
Here $\dd S$ is the area measure on the horizon sphere and
$\dd\mu_{\mathcal S}$ is the induced volume measure on $\mathcal S$.
The horizon contribution equals $r_h^2f(S_{\mathcal H^+})$. On the static
outer end, the radial integrand is
$r^2D^{-1}h=O(r^{1-p})$, which is integrable in the stated range.
The compact portion is integrable by geometric smoothness.

\paragraph{Energy assumptions for the future evolution.}
For the data \eqref{eq:classification-data}, the derivative energy
density at infinity has leading radial term
$(b^2+p^2a^2)r^{-2p}\,\dd r$. When this coefficient is nonzero,
the energy on the static end is finite exactly for $p>\tfrac12$.
The ordinary radiation field has a finite limit for generic data with
$p>1$, as described in Table~\ref{tab:intro-black-hole}.

We retain the energy boundedness and integrated local energy decay
assumptions of \cite[Section~2.4]{AAG1}, on the fixed family
$\mathcal N_\tau$ of Section~\ref{sec:hypersurfaces}. We choose a
stationary future timelike vector field $N$, regular at the horizon and
equal to $T$ for $r\ge R$. The letter $N$ denotes this vector field;
$\mathcal N_\tau$ denotes a hypersurface. For a scalar field $\chi$,
we define its energy-momentum tensor and energy current by
\[
 \mathbb T_{\mu\nu}[\chi]
 =\partial_\mu\chi\partial_\nu\chi
    -\tfrac12g_{\mu\nu}g^{\alpha\beta}
                   \partial_\alpha\chi\partial_\beta\chi,
 \qquad J^N_\mu[\chi]=\mathbb T_{\mu\nu}[\chi]N^\nu.
\]
We denote the flux of $J^N[\chi]$ through $\mathcal N_\tau$ by
$\mathcal E_N[\chi](\tau)$. It is the sum of the spacelike flux on
$\mathcal C_\tau$ and the future null flux on
$\mathcal N_\tau\cap\{r\ge R\}$. In the notation below,
$n_\tau$ is the future unit normal on the spacelike portion and a
future null generator on the null portion, with the corresponding
flux measure $\dd\mu_\tau$ on each portion.

For smooth spherically symmetric solutions of $\Box_g\chi=0$ with finite energy
at the starting time $\tau_1$, we assume the following estimates for
$\tau_2\ge\tau_1$ in the future development:
\begin{align}
 \mathcal E_N[\chi](\tau_2)
   +\mathcal F_N[\chi;I_V(\tau_1,\tau_2)]
 &\le C\mathcal E_N[\chi](\tau_1),
 \label{eq:energy-assumption}\\
 \int_{\tau_1}^{\tau_2}\int_{\mathcal C_\tau}
          J^N[\chi]\cdot n_\tau\,\dd\mu_\tau\dd\tau
 &\le C\bigl(\mathcal E_N[\chi](\tau_1)
             +\mathcal E_N[T\chi](\tau_1)\bigr).
 \label{eq:morawetz-assumption}
\end{align}
Here $\mathcal F_N[\chi;I_V(\tau_1,\tau_2)]$ is the nonnegative
energy flux through the incoming boundary defined in
Section~\ref{sec:hypersurfaces}; the first bound is uniform in $V$.
The second estimate is the integrated local energy decay bound on the
compact portion. For every fixed exterior annulus
$\widetilde R\le r\le\widetilde R+1$, with $\widetilde R\ge R$,
we also assume the local estimate
\begin{equation}
 \int_{\tau_1}^{\tau_2}
  \int_{\mathcal N_\tau\cap\{\widetilde R\le r\le\widetilde R+1\}}
       J^T[\partial^\alpha\chi]\cdot n_\tau\,\dd\mu_\tau\dd\tau
 \le C_{\alpha,\widetilde R}
       \sum_{k\le|\alpha|}\mathcal E_N[T^k\chi](\tau_1).
 \label{eq:localmorawetz-assumption}
\end{equation}
The multi-index $\alpha$ specifies the regular coordinate derivatives
on this fixed annulus. We may choose the flux measures and normalization
as in \cite[eqs.~(2.12)--(2.14)]{AAG1}; fixed equivalent choices change
only the constants. We use these estimates also for the finitely many
time derivatives $T^j\chi$ occurring below. Stationarity permits any
fixed starting member, in particular $\mathcal N_{u_0}$ or
$\mathcal N_0$.

These assumptions hold on Schwarzschild and subextremal
Reissner--Nordstr\"om in the setting of \cite{AAG1}. For a general
function $D$, we assume them together with
\eqref{eq:horizon}--\eqref{eq:D}. They supply the global control used
to bound the late-time remainders.

\section{The main theorem}
\label{sec:global-classification}

We now state the main result on a nondegenerate black-hole exterior.
The data are prescribed on an initial hypersurface $\mathcal S$
chosen from $\{t=0\}$, $\Sigma_0$, and $\Sigma_{\mathcal B}$, as defined
in Section~\ref{sec:hypersurfaces}.
The same statement treats the two horizon configurations and
every real power at spatial infinity.  A single profile describes
the transition between bounded radius and the far exterior; at
$p=3$ its coefficient contains both a geometric integral
over the initial hypersurface and the contribution from infinity.

\subsection{Initial-data norms}
\label{sec:main-data-norms}

We work under the geometric and energy assumptions in
Section~\ref{sec:setting}, with any of these choices of $\mathcal S$
and the power-law data \eqref{eq:classification-data} specified there.
We write $[\psi]$ in each initial-data norm to indicate that it is
evaluated on the data of $\psi$ on $\mathcal S$; the argument $[T\psi]$
indicates the corresponding data of the differentiated solution.
We fix $R_0$ sufficiently large for the geometric construction in
Lemma~\ref{lem:compact-transfer}. For every $p\in\mathbb R$ and
integer $m\ge0$, we define
\begin{align}
 \mathfrak D_{p,m}[\psi]={}&
 \|f\|_{H^{m+1}(\mathcal S\cap\{r\le2R_0\})}
 +\|\pi\|_{H^m(\mathcal S\cap\{r\le2R_0\})}\notag\\
 &+\sum_{j=0}^{m+1}\sup_{r\ge R_0}
 r^{p+j}|\partial_r^j f|
 +\sum_{j=0}^{m}\sup_{r\ge R_0}
 r^{p+1+j}|\partial_r^j h|.
 \label{eq:frac-original-norm}
\end{align}
The compact Sobolev norms use regular coordinates, including at a
bifurcation endpoint. The spatial terms control $m+1$ radial derivatives
of $f$ and $m$ of $h$ with the displayed weights.

For the expansions \eqref{eq:classification-data}, we also define
\begin{align}
 \mathfrak A_{p,m,\delta}[\psi]
 ={}&\mathfrak D_{p,m}[\psi]+|a|+|b|\notag\\
 &+\sum_{j=0}^{m+1}\sup_{r\ge R_0}
 r^{p+\delta+j}|\partial_r^j(f-a r^{-p})|\notag\\
 &+\sum_{j=0}^{m}\sup_{r\ge R_0}
 r^{p+1+\delta+j}|\partial_r^j(h-b r^{-p-1})|.
 \label{eq:frac-asymptotic-data-norm}
\end{align}
Thus $\mathfrak D_{p,m}[\psi]$ controls the full data, while
$\mathfrak A_{p,m,\delta}[\psi]$ also controls the leading coefficients
and the remainders after subtracting their powers. The integer $m$
counts the derivatives of $\pi$ and $h$; one additional derivative
of $f$ is included. We use the same definitions for every real $p$.

We make the sufficient choice
\begin{equation}
 m(p)=30+2\left\lceil\max\{0,-p\}\right\rceil,
 \qquad
 \mathfrak A[\psi]=\mathfrak A_{p,m(p),\delta}[\psi].
 \label{eq:classification-regularity}
\end{equation}
This derivative count supplies all commuted estimates used in the proof.

\paragraph{The data and norms for \texorpdfstring{$p\ge3$}{p >= 3}.}
For $p=3$, we also use a derivative count $m\ge10$ and write the data as
\begin{equation}
 f=a r^{-3}+\Os[m+1](r^{-3-\delta}),\qquad
 h=b r^{-4}+\Os[m](r^{-4-\delta}).
 \label{eq:critical}
\end{equation}
For $p=3$, we use the following more economical
initial-data norm:
\begin{align}
 \mathfrak N_{m,\delta}[\psi]
 ={}&\norm{f}_{H^{m+1}(\mathcal S\cap\{r\le2R_0\})}
       +\norm{\pi}_{H^m(\mathcal S\cap\{r\le2R_0\})}
       +|a|+|b|\notag\\
 &+\sum_{j=0}^{m+1}\sup_{r\ge R_0}
       r^{3+\delta+j}|\partial_r^j(f-a r^{-3})|\notag\\
 &+\sum_{j=0}^{m}\sup_{r\ge R_0}
       r^{4+\delta+j}|\partial_r^j(h-b r^{-4})|.
 \label{eq:originalnorm}
\end{align}
Here $R_0$ is chosen large enough for
Lemma~\ref{lem:compact-transfer}. The compact Sobolev norms use
a fixed smooth regular metric and finite coordinate cover.
The bounded-radius and radiation-field estimates below at $p=3$
are controlled by $\mathfrak N_{10,\delta}[\psi]$.

For every $p\ge3$, we define
\begin{equation}
 a_3:=\lim_{r\to\infty}r^3f(r),\qquad
 b_3:=\lim_{r\to\infty}r^4h(r),\qquad
 (a_3,b_3)=
 \begin{cases}
  (a,b),&p=3,\\
  (0,0),&p>3.
 \end{cases}
 \label{eq:pge3-coefficients}
\end{equation}
Here $a,b$ retain their meaning as the coefficients of $r^{-p}$
and $r^{-p-1}$ in the initial data. We set
\begin{equation}
 \delta_3=
 \begin{cases}
  \delta,&p=3,\\
  \tfrac12\min\{\delta,p-3,1\},&p>3,
 \end{cases}
 \label{eq:pge3-remainder-exponent}
\end{equation}
so that the data
have the common representation through the available derivative orders,
\begin{equation}
 f=a_3r^{-3}+\Os[m+1](r^{-3-\delta_3}),\qquad
 h=b_3r^{-4}+\Os[m](r^{-4-\delta_3}).
 \label{eq:pge3-data}
\end{equation}
For $p>3$, this follows from $\delta_3<p-3$, with the same
finite-order symbol bounds. For every $p\ge3$, the notation
$\mathfrak N_{m,\delta_3}[\psi]$ means \eqref{eq:originalnorm}
evaluated with $(a,b,\delta)=(a_3,b_3,\delta_3)$.

\begin{lemma}[Control by the initial data]
\label{lem:classification-data-norm}
Smooth initial data satisfying the symbol bounds in
\eqref{eq:classification-data} through the orders in
\eqref{eq:frac-asymptotic-data-norm} have finite
$\mathfrak A[\psi]$.  This norm controls the initial-data norms used in
Theorems~\ref{thm:pge3-global}, \ref{thm:frac-spatial-global},
\ref{thm:frac-p1}, and~\ref{thm:slow-global}.  At $p=3$ it controls
$\mathfrak N_{10,\delta}[\psi]$ in \eqref{eq:originalnorm}.  For $p>3$,
we regard the data as satisfying \eqref{eq:critical} with zero
coefficients at orders $r^{-3}$ in $f$ and $r^{-4}$ in $h$,
and with the exponent $\delta_3$ in
\eqref{eq:pge3-remainder-exponent}. Then it controls $\mathfrak N_{10,\delta_3}[\psi]$ for that
representation.  The control constants depend only on the fixed parameters
and geometry.
\end{lemma}

\begin{proof}
The un-subtracted symbol seminorms in
\eqref{eq:frac-original-norm} are bounded by $|a|+|b|$ and the
subtracted seminorms in
\eqref{eq:frac-asymptotic-data-norm}.  Compact smoothness
gives the finite Sobolev terms, and lowering the derivative count
only weakens the required norms.  For $p>3$, multiplication by
$r^{3+\delta_3+j}$ or $r^{4+\delta_3+j}$ leaves a negative
power of $r$ in the leading terms, since $\delta_3<p-3$.
The same is true of the remainders.  These observations give
the claimed comparisons; the time-commutation count in the
slow-power argument is recorded in Theorem~\ref{thm:slow-global}.
\end{proof}

\subsection{Global leading-order asymptotics}
\label{sec:result}

We use the hypersurfaces $\mathcal N_\tau$ and the coordinate $\tau$
fixed in Section~\ref{sec:hypersurfaces}, with $\tau=u$ on their
outgoing null portions. The global estimates hold for every
$r\ge r_h$ on every sufficiently late $\mathcal N_\tau$.
For each real exponent $s$, we define the comparison profile
\begin{equation}
 \begin{aligned}
 \Pi_s(\tau,r)
 &=\frac1{2r}\int_{1+\tau}^{1+\tau+2r}y^{-s}\,\mathrm dy\\
 &=\begin{cases}
 \displaystyle
 \frac{(1+\tau)^{1-s}-(1+\tau+2r)^{1-s}}{2(s-1)r},&s\ne1,\\[6pt]
 \displaystyle
 \frac1{2r}\log\left(1+\frac{2r}{1+\tau}\right),&s=1.
 \end{cases}
 \end{aligned}
 \label{eq:classification-profile}
\end{equation}
In particular, $\Pi_0(\tau,r)=1$ and
$\Pi_2(\tau,r)=[(1+\tau)(1+\tau+2r)]^{-1}$.

For $p\ge3$, we use the geometric functional $B_{\mathcal S}$ in
\eqref{eq:geometricB}, which contains the actual horizon endpoint
of the prescribed hypersurface. We set
\begin{equation}
 \mathcal C_p=
 \begin{cases}
 b+(1-p)a,&p<3,\\
 b-2a-8MB_{\mathcal S},&p=3,\\
 -8MB_{\mathcal S},&p>3,
 \end{cases}
 \qquad s=\min\{p,3\}.
 \label{eq:classification-coefficient}
\end{equation}

\begin{theorem}[Global leading-order classification]
\label{thm:global-classification}
\label{thm:main}
We assume the geometric and energy hypotheses in
Section~\ref{sec:setting}. We consider the spherically symmetric solution $\psi$
arising from smooth initial data on $\mathcal S=\{t=0\}$,
$\mathcal S=\Sigma_0$, or $\mathcal S=\Sigma_{\mathcal B}$, as specified
in Section~\ref{sec:hypersurfaces}.
The data satisfy
\eqref{eq:classification-data} for any fixed $p\in\R$ and
$\delta>0$, with the finite norm
\eqref{eq:classification-regularity}. There are constants
$C>0$, $\eta>0$, and $\tau_*\ge0$, depending on the fixed parameters,
background, and family $\mathcal N_\tau$, such that for every
$\tau\ge\tau_*$ and every $r\ge r_h$, including the future horizon,
\begin{equation}
 \boxed{\quad
 |\psi(\tau,r)-\mathcal C_p\Pi_s(\tau,r)|
 \le C\mathfrak A[\psi]\cdot 
 \begin{cases}
 (1+\tau+r)^{-p-\eta},&p<1,\\[2pt]
 \displaystyle\frac{(1+\tau)^{-\eta}}{1+\tau+r},&p=1,\\[6pt]
 \displaystyle\frac{(1+\tau)^{1-s-\eta}}{1+\tau+r},&p>1.
 \end{cases}
 \quad}
 \label{eq:classification-global}
\end{equation}
The constants are independent of the data.  The following
choices are sufficient:
\begin{align}
 \eta&=\frac1{256}\min\{\delta,1/2,1-\alpha\},
 &\alpha&=p+\left\lceil\max\{0,-p\}\right\rceil\in[0,1),
 &&p<1,\label{eq:classification-eta-slow}\\
 \eta&=\frac1{128}\min\{\delta,1\},&&&&p=1,
 \label{eq:classification-eta-one}\\
 \eta&=\frac1{32}\min\{\delta,3-p,1/2\},&&&&1<p<3.
 \label{eq:classification-eta-middle}
\end{align}
For $p\ge3$, we let $\epsilon>0$ and $\epsilon_0>0$ be the remainder
exponents in \eqref{eq:pge3-localtail} and \eqref{eq:pge3-exterior}.
One admissible choice is
\[
 \eta=\frac12\min\{\epsilon,\epsilon_0,1/2\}.
\]
At $p=3$, those estimates use the initial remainder exponent
$\delta_3=\delta$. At $p>3$, they use
$\delta_3=\tfrac12\min\{\delta,p-3,1\}>0$, as in
Lemma~\ref{lem:classification-data-norm}. These choices fix $\eta$
using only $p$, $\delta$, and the fixed geometric and energy parameters;
the resulting constants $C$ and $\tau_*$ are independent of the data.

For each fixed $R_1>r_h$, the global estimate implies
\begin{equation}
 \sup_{r_h\le r\le R_1}
 |\psi(\tau,r)-\mathcal C_p (1+\tau)^{-s}|
 \le C_{R_1}\mathfrak A[\psi]\cdot  (1+\tau)^{-s-\eta}.
 \label{eq:classification-bounded-radius}
\end{equation}
Thus the same coefficient governs the bounded-radius tail
and the future-horizon tail in regular time.

If $p>1$, the ordinary radiation field
$F(u)=\lim_{r\to\infty}r\psi(u,r)$ exists, and for late $u$
\begin{equation}
 F(u)=\frac{\mathcal C_p}{2(s-1)}(1+u)^{1-s}
       +O\bigl(\mathfrak A[\psi]\cdot (1+u)^{1-s-\eta}\bigr).
 \label{eq:classification-radiation}
\end{equation}
If $p=1$, then $\mathcal C_1=b$.  With a fixed reference
length in the logarithms, the renormalized radiation field
\begin{equation}
 F_{\mathrm{ren}}(u)
 =\lim_{r\to\infty}
    \left[r\psi(u,r)-\frac b2\log(2r)\right]
 =-\frac b2\log(1+u)
       +O\bigl(\mathfrak A[\psi]\cdot (1+u)^{-\eta}\bigr)
 \label{eq:classification-log-radiation}
\end{equation}
exists.  In particular, the ordinary radiation field diverges
when $b\ne0$; when $b=0$, this is an ordinary radiation field
with the stated remainder estimate.

If $p<1$, then, at each fixed late retarded time $u$,
\begin{equation}
 r\psi(u,r)
 =\frac{2^{-p}\mathcal C_p}{1-p}r^{1-p}
       +O_u\bigl(\mathfrak A[\psi]\cdot  r^{1-p-\eta}\bigr),
 \qquad r\longrightarrow\infty.
 \label{eq:classification-power-null}
\end{equation}
Consequently the ordinary radiation field diverges for
$\mathcal C_p\ne0$. The estimates
\eqref{eq:classification-global}--\eqref{eq:classification-power-null}
also hold when $\mathcal C_p=0$, in which case they give the
corresponding improved quantitative upper bounds.

\paragraph{The coefficient and quantitative bounds for \texorpdfstring{$p=3$}{p = 3}.}
At $p=3$, the coefficient in
\eqref{eq:classification-coefficient} is $\mathcal C_3=-8\calI$,
where the time-inverted Newman--Penrose charge is
\begin{equation}
 \boxed{\quad \calI=I_0^{(1)}[\psi]
       =M B_{\mathcal S}+\frac{2a-b}{8}.\quad}
 \label{eq:spatialfunctional}
\end{equation}
The following estimates at $p=3$ also hold under the weaker
requirement of smooth geometric data satisfying
\eqref{eq:critical} through the orders measured by
$\mathfrak N_{10,\delta}[\psi]<\infty$. There is an $\epsilon>0$,
depending on the fixed background, hypersurface and $\delta$,
such that, for sufficiently late times and every fixed $R>r_h$,
\begin{equation}
 \sup_{r_h\le r\le R}
   |\psi(\tau,r)+8\calI(1+\tau)^{-3}|
 \le C_R\mathfrak N_{10,\delta}[\psi]\cdot (1+\tau)^{-3-\epsilon}.
 \label{eq:localtail}
\end{equation}
The radiation field satisfies
\begin{equation}
 |F(u)+2\calI(1+u)^{-2}|
 \le C\mathfrak N_{10,\delta}[\psi]\cdot (1+u)^{-2-\epsilon}.
 \label{eq:nulltail}
\end{equation}
Thus the fixed-radius and horizon limits are
\[
 \lim_{t\to\infty}t^3\psi(t,r)=-8\calI,
 \qquad
 \lim_{v\to\infty}v^3\psi|_{\HH}=-8\calI.
\]
For the static hypersurface $\{t=0\}$ of
Section~\ref{sec:hypersurfaces}, the
coefficient has the explicit initial-data expression
\begin{equation}
 \boxed{\quad
 \calI=M\left[r_h^2f(r_h)
       +\int_{r_h}^\infty\frac{r^2}{D(r)}h(r)\,\dd r\right]
       +\frac{2a-b}{8}.
 \quad}
 \label{eq:maincharge}
\end{equation}
For a hypersurface meeting the future horizon strictly to the
future of the bifurcation sphere, the invariant expression
\eqref{eq:geometricB} gives the corresponding horizon and bulk
contribution. In both configurations, spatial infinity contributes
$(2a-b)/8$. The constants in the estimates are independent of
the data. The estimates include $a=b=0$, $2a-b=0$, and $\calI=0$; in the
last case they give a positive-power improvement over the leading
$\tau^{-3}$ and $u^{-2}$ rates.

\end{theorem}

\begin{proof}[Proof of the global classification]
For $p\ge3$, Theorem~\ref{thm:pge3-global} gives
\eqref{eq:classification-global} and the radiation-field asymptotics.
Proposition~\ref{thm:coefficient} identifies its coefficient:
$-8\calI=b-2a-8MB_{\mathcal S}$ at $p=3$ and
$-8\calI=-8MB_{\mathcal S}$ at $p>3$.
For $1<p<3$, we use Theorem~\ref{thm:frac-spatial-global}.
The case $p=1$, including the existence of the renormalized
limit, is Theorem~\ref{thm:frac-p1}; for every $p<1$, we use
Theorem~\ref{thm:slow-global}. Lemma~\ref{lem:classification-data-norm}
places these assertions under the common initial-data norm
$\mathfrak A[\psi]$.
The additional $p=3$ estimates with the more economical norm
$\mathfrak N_{10,\delta}[\psi]$ are
\eqref{eq:pge3-localtail} and \eqref{eq:pge3-nulltail}, proved in
Section~\ref{sec:pge3-conclusion}.

At bounded radius, the elementary identity
\[
 \Pi_s(\tau,r)=(1+\tau)^{-s}+O_{R_1,s}((1+\tau)^{-s-1})
\]
and the displayed choices $\eta<1$ give
\eqref{eq:classification-bounded-radius}.  For $s>1$,
$r\Pi_s(\tau,r)\to (1+\tau)^{1-s}/[2(s-1)]$; the identity
$\tau=u$ on the outgoing null portion gives the coefficient in
\eqref{eq:classification-radiation}.  Finally, for $p<1$
and fixed $\tau$,
\[
 r\Pi_p(\tau,r)
 =\frac{2^{-p}}{1-p}r^{1-p}
       +O_\tau(r^{-p}+1).
\]
Our exponent obeys $\eta<\min\{1,1-p\}$.  Multiplying
\eqref{eq:classification-global} by $r$ therefore gives
\eqref{eq:classification-power-null}.  The direct spatial-to-null
argument in Proposition~\ref{prop:all-p-null-traces} also
gives the same leading power at each earlier fixed retarded
time in the future development.
\end{proof}

\section{Propagation from the initial hypersurface to \texorpdfstring{$\mathcal N_0$}{N0}}
\label{sec:exterior}

We transfer the spatial data on $\mathcal S$ to the same early
outgoing hypersurface $u=u_0<0$ for every real $p$, and then to
$u=0$, the outgoing end of $\mathcal N_0$. We use the coordinates
$u=t-r_*$, $v=t+r_*$ and the fields $\phi=r\psi$, $Q=\partial_v\phi$
of Section~\ref{sec:radial-wave}. On the static end we write
\[
 F_0(r_*)=r(r_*)f(r(r_*)),\qquad
 H_0(r_*)=r(r_*)h(r(r_*)).
\]
Thus $F_0$ and $H_0$ are the initial values of $\phi$ and $T\phi$,
parametrized by the tortoise radius. The data norms
$\mathfrak D_{p,m}[\psi]$ and $\mathfrak A_{p,m,\delta}[\psi]$ are
defined in Section~\ref{sec:main-data-norms}.

We choose $A\ge\max\{2,r_*(R_0)\}$ sufficiently large, put $u_0=-A$,
and use the far-exterior region
\begin{equation}
 \mathcal W_A=\{u\le u_0,\ v\ge-u\}
             =\{t\ge0,\ r_*-t\ge A\}.
 \label{eq:exterior-wedge}
\end{equation}
Its initial boundary is $\{t=0,r_*\ge A\}\subset\mathcal S$.
For $(u,v)\in\mathcal W_A$, the past characteristic triangle is
\begin{equation}
 \mathcal T_{u,v}
 =\{(u',v'):-v\le u'\le u,\ -u'\le v'\le v\}
 \subset\mathcal W_A,
 \label{eq:exterior-triangle}
\end{equation}
with initial base $\{t=0,-u\le r_*\le v\}$.
Figure~\ref{fig:exterior-wedge} shows this region and its finite
truncations. Lemma~\ref{lem:compact-transfer} makes the geometric
choices so that $\mathcal N_{u_0}$ lies to the future of
$\mathcal S$. All subsequent estimates use this same $u_0$.

\subsection{Estimates for \texorpdfstring{$\phi$ and $\partial_v\phi$}{phi and its outgoing derivative}
in \texorpdfstring{$\mathcal W_A$}{the far exterior} for every real \texorpdfstring{$p$}{p}}

The positive weight
\begin{equation}
 E_p(u,v)=\max\{(-u)^{1-p},v^{1-p}\}
                  +\int_{-u}^{v}\rho_*^{-p}\,\dd\rho_*
 \label{eq:exterior-envelope}
\end{equation}
controls the initial-data terms in the characteristic formula.
For fixed $p$, it satisfies
\begin{equation}
 E_p(u,v)\asymp_p
 \begin{cases}
  (-u)^{1-p},&p>1,\\
  1+\log\bigl(v/(-u)\bigr),&p=1,\\
  v^{1-p},&p<1.
 \end{cases}
 \label{eq:exterior-envelope-powers}
\end{equation}
At $p=1$ the displayed expression is an equality. These three
behaviors are therefore treated by one positive weight.

\begin{lemma}[Propagation throughout $\mathcal W_A$ for every real $p$]
\label{lem:volterra}
There are $A_0\ge\max\{2,r_*(R_0)\}$ and $C>0$, depending only
on the fixed background and static end, with the following property.
For any $p\in\mathbb R$ and $A\ge A_0$, we assume
\[
 |F_0(r_*)|\le C_0r_*^{1-p},\qquad
 |H_0(r_*)|\le C_0r_*^{-p}\qquad(r_*\ge A).
\]
Then
\begin{equation}
 |\phi(u,v)|\le C C_0E_p(u,v),\qquad (u,v)\in\mathcal W_A.
 \label{eq:wedgebound}
\end{equation}
If also $|F_0'(r_*)|\le C_0r_*^{-p}$, then
\begin{equation}
 |Q(u,v)|\le C C_0\left(v^{-p}
       +v^{-3}\int_{-v}^{u}E_p(u',v)\,\dd u'\right).
 \label{eq:Qbound}
\end{equation}
The constants are independent of the data, $C_0$, $p$, $A\ge A_0$,
and the point $(u,v)$. The prime denotes differentiation with
respect to the tortoise-radius argument of $F_0$.
\end{lemma}

\begin{proof}
We define the kernel from the black hole potential
\eqref{eq:potential-definition} by
\[
 k(u',v')=\frac14 V\!\left(\frac{v'-u'}2\right).
\]
The formula \eqref{eq:duhamel}, written on
$\mathcal T_{u,v}$ in null coordinates, becomes
\begin{align}
 \phi(u,v)={}&\frac12\bigl(F_0(-u)+F_0(v)\bigr)
       +\frac12\int_{-u}^{v}H_0(\rho_*)\,\dd\rho_* \notag\\
 &-\int_{-v}^{u}\int_{-u'}^{v}
             k(u',v')\phi(u',v')\,\dd v'\dd u'.
 \label{eq:volterra}
\end{align}
The absolute value of its initial-data terms is at most
$C_0E_p(u,v)$. By \eqref{eq:potential} and smoothness of $V$, we may fix
\[
 C_V:=2\sup_{\rho_*\ge1}\rho_*^3|V(\rho_*)|<\infty.
\]
Since $(v'-u')/2\ge-u'\ge1$ in the triangles under consideration,
\[
 |k(u',v')|\le C_V(v'-u')^{-3}.
\]
This constant depends only on the background and the normalization
of $r_*$. Moreover, on every past triangle,
\begin{equation}
 E_p(u',v')\le E_p(u,v).
 \label{eq:exterior-envelope-monotonicity}
\end{equation}
Indeed, $[-u',v']\subset[-u,v]$: both the maximum of the endpoint
powers and the integral of the positive function $\rho_*^{-p}$
decrease when this interval is shortened.

\smallskip\noindent
\emph{The finite triangle and the bootstrap.}
For $L\ge A$, we set
\begin{equation}
 \mathcal T_{u_0,L}
 =\{-L\le u\le u_0,\ -u\le v\le L\},\qquad
 M_L=\max_{\mathcal T_{u_0,L}}
                  \frac{|\phi(u,v)|}{E_p(u,v)}.
 \label{eq:exterior-bootstrap-norm}
\end{equation}
Its initial base is $\{t=0,A\le r_*\le L\}$: $A$ fixes the
outgoing side $u=u_0$, and $L$ fixes the incoming cutoff $v=L$.
The maximum is finite because the triangle is compact and $E_p>0$.

\begin{figure}[H]
\centering
\begin{tikzpicture}[x=1.40cm,y=1.40cm,
 every node/.style={font=\small,inner sep=1.3pt},
 boundary/.style={line width=.8pt},
 initial/.style={draw=blue!65!black,line width=1.25pt},
 future/.style={draw=red!65!black,line width=1.2pt},
 incoming/.style={draw=orange!75!black,line width=1.1pt}]
 \coordinate (B) at (0,0);
 \coordinate (IZ) at (6,0);
 \coordinate (IP) at (3,3);
 \coordinate (A) at (3.20,0);
 \coordinate (L) at (5.30,0);
 \coordinate (P) at (4.25,1.05);
 \coordinate (FU) at (4.60,1.40);
 \coordinate (JU) at (3.425,.225);
 \fill[green!11] (A)--(IZ)--(FU)--cycle;
 \fill[orange!20] (A)--(L)--(P)--cycle;
 \draw[boundary] (B)--(IP)--(IZ);
 \draw[gray!55] (B)--(IZ);
 \draw[initial] (.28,.28) .. controls (.85,.025) and (1.85,0) ..
      (2.70,0)--(IZ);
 \draw[initial,line width=2pt] (A)--(L);
 \draw[future,dashed,line width=.85pt] (A)--(JU);
 \draw[future] (.48,.48) .. controls (1.40,.38) and (3.20,0) ..
      (JU)--(FU);
 \draw[incoming,dashed] (P)--(L);
 \fill (B) circle(1.15pt) (IZ) circle(1.15pt) (IP) circle(1.15pt);
 \fill[blue!65!black] (.28,.28) circle(1.5pt)
      (A) circle(1.5pt) (L) circle(1.5pt);
 \fill[red!65!black] (.48,.48) circle(1.5pt);
 \fill[orange!75!black] (P) circle(1.5pt);
 \node[below left] at (B) {$\mathcal B$};
 \node[right] at (IZ) {$i^0$};
 \node[above] at (IP) {$i^+$};
 \node[above left] at (1.45,1.45) {$\mathcal H^+$};
 \node[above right] at (4.95,1.05) {$\mathcal I^+$};
 \node[blue!65!black,below=2pt] at (1.45,.02) {$\mathcal S$};
 \node[blue!65!black,below=3pt] at (A) {$r_*=A$};
 \node[blue!65!black,below=3pt] at (L) {$r_*=L$};
 \node[blue!65!black] at (4.25,-.48) {$t=0$};
 \node[red!65!black,above=2pt] at (1.95,.255)
      {$\mathcal N_{u_0}$};
 \node[red!65!black,rotate=45,above=2pt] at (4.02,.82)
      {$u=u_0=-A$};
 \node[orange!65!black,rotate=-45,below=2pt] at (4.80,.50)
      {$v=L$};
 \node at (4.06,.32) {$\mathcal T_{u_0,L}$};
 \node[green!35!black] at (5.04,.61) {$\mathcal W_A$};
\end{tikzpicture}
\caption{The first bootstrap, in the region $\mathcal W_A$ above the
static end of $\mathcal S$. The finite triangle $\mathcal T_{u_0,L}$
has the thick blue initial base $\{t=0,\ A\le r_*\le L\}$,
outgoing boundary $u=u_0=-A$, and incoming cutoff $v=L$.}
\label{fig:unified-wedge-bootstrap}\label{fig:exterior-wedge}\label{fig:bootstrap-triangle}
\end{figure}
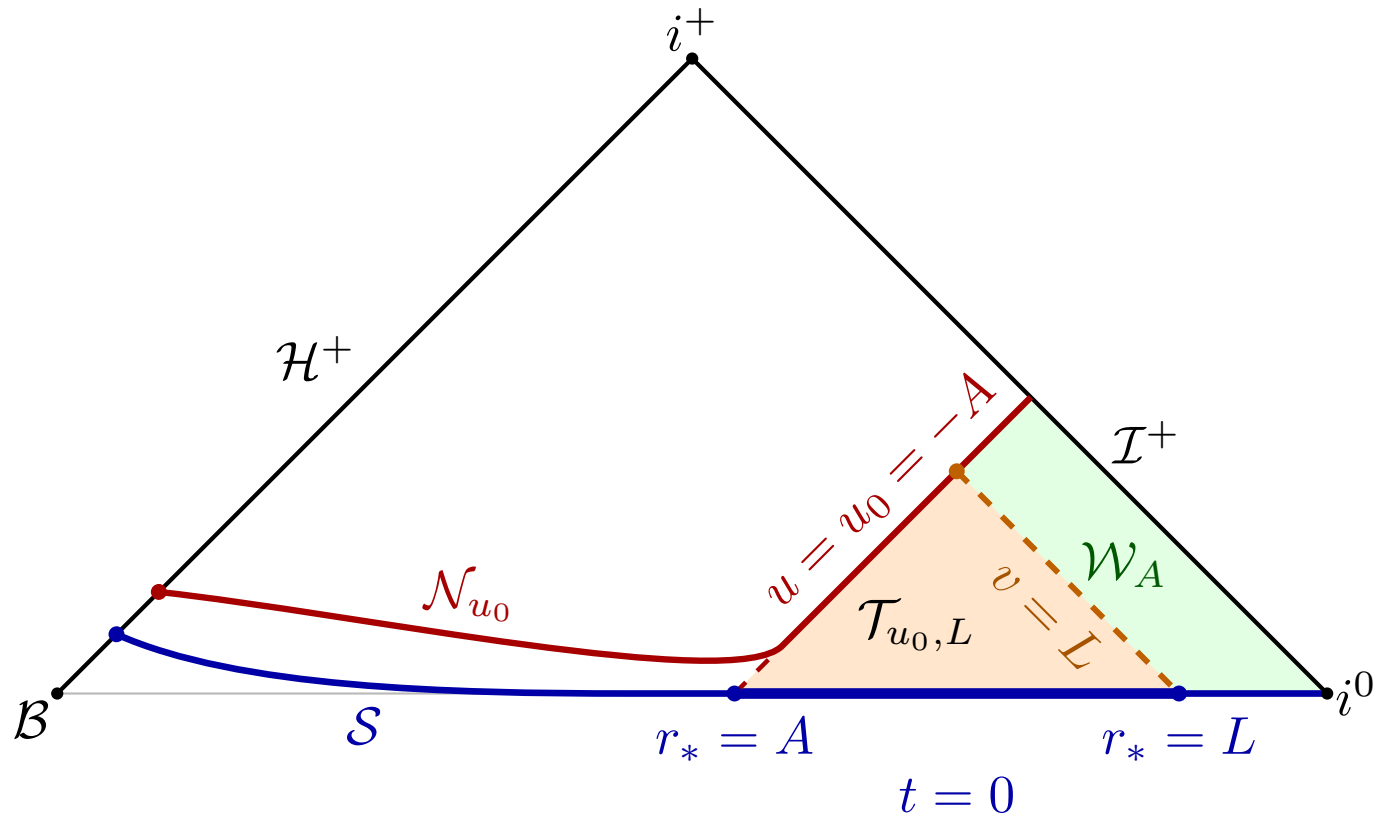

Every past triangle lies in the same finite domain. We denote
the double integral in \eqref{eq:volterra} by $\mathcal K\phi$.
Using \eqref{eq:exterior-envelope-monotonicity}, we obtain
\begin{align}
 \frac{|\mathcal K\phi(u,v)|}{E_p(u,v)}
 &\le C_VM_L\int_{-v}^{u}\int_{-u'}^{v}
                 (v'-u')^{-3}\,\dd v'\dd u'\notag\\
 &\le\frac{C_VM_L}{8}\int_{-\infty}^{u}(-u')^{-2}\,\dd u'
  =\frac{C_VM_L}{8(-u)}
  \le\varepsilon_A M_L,
 \qquad \varepsilon_A:=\frac{C_V}{8A}.
 \label{eq:exterior-source-bootstrap}
\end{align}
The extension of the integration limits in this estimate is applied
to the explicit positive kernel after the solution has been bounded
by the finite maximum $M_L$.
We choose
$A_0=\max\{2,r_*(R_0),C_V/2\}$, so that
$\varepsilon_A\le1/4$ for $A\ge A_0$.
For $C_0>0$, the bootstrap assumption
\begin{equation}
 M_L\le2C_0
 \label{eq:exterior-bootstrap-assumption}
\end{equation}
then improves, by \eqref{eq:volterra}, to
\begin{equation}
 M_L\le C_0+\varepsilon_AM_L\le\frac32C_0<2C_0.
 \label{eq:exterior-bootstrap-improvement}
\end{equation}
At $L=A$ the triangle is an initial point and $M_A\le C_0$.
The maxima vary continuously with $L$; the strict improvement
therefore extends the bootstrap to every finite $L$. The triangles
exhaust $\mathcal W_A$ as $L\to\infty$, proving
\eqref{eq:wedgebound}. The case $C_0=0$ follows by using an
arbitrarily small positive data bound and then taking its limit.

\smallskip\noindent
\emph{The outgoing derivative.}
We differentiate \eqref{eq:volterra} at fixed $u$ to obtain
\begin{equation}
 Q(u,v)=b_0(v)-\int_{-v}^{u}k(u',v)\phi(u',v)\,\dd u',
 \qquad b_0(v)=\frac12\bigl(F_0'(v)+H_0(v)\bigr).
 \label{eq:Qexact}
\end{equation}
We use $|b_0(v)|\le C_0v^{-p}$ and
$|k(u',v)|\le C_Vv^{-3}$, together with the spacetime bound
just proved, to obtain \eqref{eq:Qbound}.
\end{proof}

For later use, evaluating the elementary integral in
\eqref{eq:Qbound} gives, throughout $\mathcal W_A$,
\begin{equation}
 |Q(u,v)-b_0(v)|\le C(p)C_0
 \begin{cases}
  v^{-p-1},&p<2,\\
  v^{-3}\log\bigl(v/(-u)\bigr),&p=2,\\
  v^{-3}(-u)^{2-p},&p>2.
 \end{cases}
 \label{eq:exterior-Q-cases}
\end{equation}
Here $C(p)$ may depend on $p$. In particular, at $p=1$ we use
\[
 \int_{-v}^{u}\left[1+\log\frac{v}{-u'}\right]\,\dd u'
 =2(v+u)+u\log\frac{v}{-u}\le2v,
\]
so the potential contribution to $Q$ is $O(v^{-2})$.

\subsection{Estimates on the compact portions and the incoming null segment}

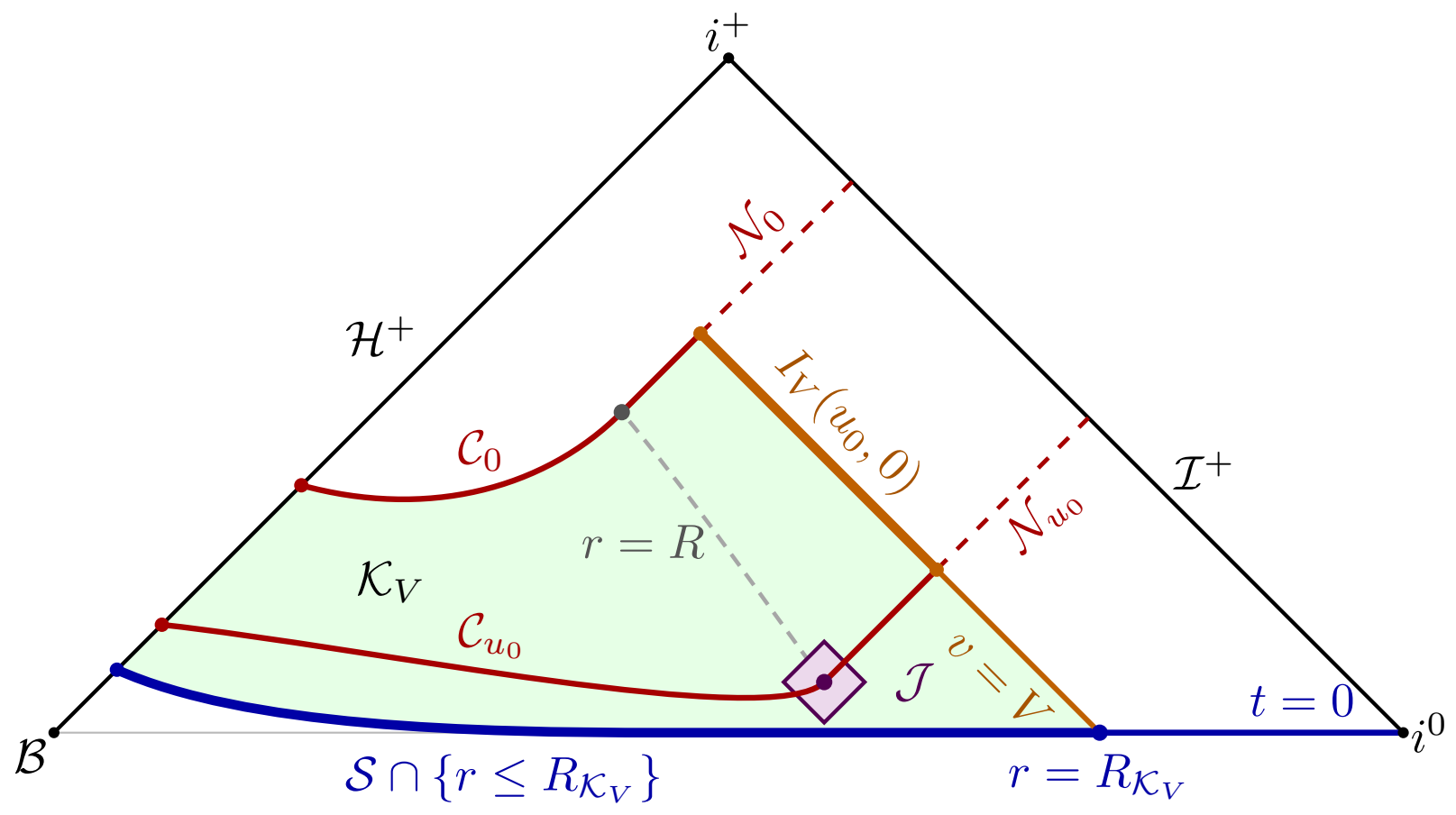
\begin{figure}[H]
\centering
\begin{tikzpicture}[x=1.65cm,y=1.65cm,
 every node/.style={font=\small,inner sep=1.3pt},
 boundary/.style={line width=.8pt},
 initial/.style={draw=blue!65!black,line width=1.25pt},
 future/.style={draw=red!65!black,line width=1.2pt},
 incoming/.style={draw=orange!75!black,line width=1.1pt}]
 \coordinate (B) at (0,0);
 \coordinate (IZ) at (6,0);
 \coordinate (IP) at (3,3);
 \coordinate (SH) at (.28,.28);
 \coordinate (HU) at (.48,.48);
 \coordinate (HZ) at (1.10,1.10);
 \coordinate (JU) at (3.425,.225);
 \coordinate (JZ) at (2.525,1.425);
 \coordinate (FU) at (4.60,1.40);
 \coordinate (FZ) at (3.55,2.45);
 \coordinate (SV) at (4.65,0);
 \coordinate (UV) at (3.925,.725);
 \coordinate (ZV) at (2.875,1.775);

 \fill[green!10]
      (SH) .. controls (.85,.025) and (1.85,0) .. (2.70,0)
      --(SV)--(ZV)--(JZ)
      .. controls (2.175,1.075) and (1.65,.95) .. (HZ)--cycle;
 \draw[boundary] (B)--(IP)--(IZ);
 \draw[gray!55] (B)--(IZ);
 \draw[initial] (SH) .. controls (.85,.025) and (1.85,0) ..
      (2.70,0)--(IZ);
 \draw[initial,line width=2pt]
      (SH) .. controls (.85,.025) and (1.85,0) .. (2.70,0)--(SV);

 \draw[gray!70,densely dashed,line width=.85pt]
      (JU) .. controls (3.15,.60) and (2.80,1.10) .. (JZ);
 \filldraw[fill=violet!15,draw=violet!65!black,line width=.8pt]
      (3.425,.405)--(3.605,.225)--
      (3.425,.045)--(3.245,.225)--cycle;

 \draw[future] (HU) .. controls (1.40,.38) and (3.20,0) ..
      (JU)--(UV);
 \draw[future,dashed,line width=.9pt] (UV)--(FU);
 \draw[future] (HZ) .. controls (1.65,.95) and (2.175,1.075) ..
      (JZ)--(ZV);
 \draw[future,dashed,line width=.9pt] (ZV)--(FZ);

 \draw[incoming] (SV)--(ZV);
 \draw[incoming,line width=2pt] (UV)--(ZV);
 \fill (B) circle(1.15pt) (IZ) circle(1.15pt) (IP) circle(1.15pt);
 \fill[blue!65!black] (SH) circle(1.5pt) (SV) circle(1.7pt);
 \fill[red!65!black] (HU) circle(1.5pt) (HZ) circle(1.5pt);
 \fill[orange!75!black] (UV) circle(1.5pt) (ZV) circle(1.5pt);
 \fill[violet!65!black] (JU) circle(1.7pt);
 \fill[gray!65!black] (JZ) circle(1.7pt);

 \node[below left] at (B) {$\mathcal B$};
 \node[right] at (IZ) {$i^0$};
 \node[above] at (IP) {$i^+$};
 \node[above left] at (1.65,1.65) {$\mathcal H^+$};
 \node[above right] at (4.95,1.05) {$\mathcal I^+$};
 \node at (1.50,.67) {$\mathcal K_V$};
 \node[blue!65!black,below=3pt] at (2.00,0)
      {$\mathcal S\cap\{r\le R_{\mathcal K_V}\}$};
 \node[blue!65!black,below=3pt] at (SV) {$r=R_{\mathcal K_V}$};
 \node[blue!65!black,above=2pt] at (5.55,0) {$t=0$};
 \node[red!65!black,above=2pt] at (1.95,.255)
      {$\mathcal C_{u_0}$};
 \node[red!65!black,above=2pt] at (1.90,1.09)
      {$\mathcal C_0$};
 \node[red!65!black,rotate=45,below=2pt] at (4.28,1.08)
      {$\mathcal N_{u_0}$};
 \node[red!65!black,rotate=45,above=2pt] at (3.23,2.13)
      {$\mathcal N_0$};
 \node[orange!65!black,rotate=-45,below=2pt] at (4.31,.34)
      {$v=V$};
 \node[orange!65!black,rotate=-45,above=2pt] at (3.40,1.25)
      {$I_V(u_0,0)$};
 \node[gray!65!black,left=3pt] at (2.99,.85) {$r=R$};
 \node[violet!65!black,right=2pt] at (3.66,.225) {$\mathcal J$};
\end{tikzpicture}
\caption{The shaded compact region $\mathcal K_V$ lies between
$\mathcal S$ and $\mathcal N_0$ and is truncated by $v=V$.
The purple diamond represents the neighborhood $\mathcal J$
defined in Lemma~\ref{lem:compact-transfer} around the sphere
$\mathcal N_{u_0}\cap\{r=R\}$.}
\label{fig:compact-propagation-region}
\end{figure}

We control the compact portions $\mathcal C_{u_0}$ and
$\mathcal C_0$, together with the fixed incoming segment
$I_V(u_0,0)$ that supplies one boundary of the continuation
argument in Section~\ref{sec:finite-u-continuation}.
Figure~\ref{fig:compact-propagation-region} shows these sets inside
a compact spacetime region and identifies the part of the initial
hypersurface on which its values depend.

\begin{lemma}[Estimates on compact spacetime regions]
\label{lem:compact-transfer}
We may choose the fixed radii and the family
of Section~\ref{sec:hypersurfaces} so that $\mathcal N_{u_0}$,
$u_0=-A$, lies in the future development of $\mathcal S$ and its
outgoing null portion lies in $\mathcal W_A$.
The compact spacelike portion $\mathcal C_{u_0}$ meets the outgoing
null portion at the two-sphere $\mathcal N_{u_0}\cap\{r=R\}$.
Its double-null coordinates are $u=u_0$ and
$v=v_R:=u_0+2r_*(R)$. We fix a sufficiently small $\eta>0$ and
define the closed spacetime neighborhood
\[
 \mathcal J
 =\bigl\{(u,v,\omega):
      |u-u_0|\le\eta,\quad |v-v_R|\le\eta,\quad
      \omega\in\mathbb S^2\bigr\}.
\]
These coordinates are regular near $r=R$. We choose $R_0$
sufficiently large and $\eta$ sufficiently small so that
$\mathcal J$ lies in the future development of $\mathcal S$
and the causal past of $\mathcal C_{u_0}\cup\mathcal J$ on
$\mathcal S$ is contained in $r\le2R_0$. Then
\begin{equation}
 \sup_{\mathcal C_{u_0}\cup\mathcal J}
       \sum_{|\alpha|\le9}|\nabla^\alpha\psi|
 \le C\left(
       \norm{f}_{H^{11}(\mathcal S\cap\{r\le2R_0\})}
       +\norm{\pi}_{H^{10}(\mathcal S\cap\{r\le2R_0\})}\right).
 \label{eq:compact-bound}
\end{equation}
The derivatives and norms use a fixed regular coordinate cover.
The compact energy fluxes of $T^j\psi$, $0\le j\le8$, and of
$N\psi,N^2\psi$ are bounded by the square of this right-hand side.
More generally, for every finite integer $j\ge0$,
\begin{equation}
 \sup_{\mathcal C_{u_0}\cup\mathcal J}
       \sum_{|\alpha|\le j}|\nabla^\alpha\psi|
 \le C_j\left(
       \norm{f}_{H^{j+2}(\mathcal S\cap\{r\le2R_0\})}
       +\norm{\pi}_{H^{j+1}(\mathcal S\cap\{r\le2R_0\})}\right).
 \label{eq:compact-bound-all-orders}
\end{equation}
For any fixed compact spacetime region $\mathcal K$ in the future
development, there is $R_{\mathcal K}$ containing its initial
causal past such that
\begin{equation}
 \sup_{\mathcal K}\sum_{|\alpha|\le j}|\nabla^\alpha\psi|
 \le C_{\mathcal K,j}\left(
       \norm{f}_{H^{j+2}(\mathcal S\cap\{r\le R_{\mathcal K}\})}
       +\norm{\pi}_{H^{j+1}(\mathcal S\cap\{r\le R_{\mathcal K}\})}\right).
 \label{eq:compact-propagation-general}
\end{equation}
This includes compact portions of $\mathcal N_0$ and finite
incoming null segments. Its right-hand side is controlled by
$\mathfrak D_{p,m}[\psi]$ when $m\ge j+1$, for every fixed real $p$.
\end{lemma}

To illustrate the region in \eqref{eq:compact-propagation-general},
we fix $V$ large enough that $v=V$ meets both
$\mathcal N_{u_0}$ and $\mathcal N_0$ on their outgoing null
portions. We denote by $\mathcal K_V$ the closed exterior region
between $\mathcal S$ and $\mathcal N_0$, cut off by $v\le V$.
It contains $\mathcal C_{u_0}$, $\mathcal C_0$, and the finite
incoming segment
\[
 I_V(u_0,0)=\{v=V,\ u_0\le u\le0\}.
\]
We decrease $\eta$ if necessary so that $\mathcal J\subset\mathcal K_V$.
The initial causal past of $\mathcal K_V$ is contained in
$\mathcal S\cap\{r\le R_{\mathcal K_V}\}$, where we may take
$r_*(R_{\mathcal K_V})=V$. All these sets are compact in the
regular coordinates used at the horizon, including at a
bifurcation-sphere endpoint. The estimate here uses fixed $V$;
Section~\ref{sec:finite-u-continuation} then controls the region
$u_0\le u\le0$, $v\ge V$ uniformly towards null infinity.

The role of $\mathcal J$ is to control the spacetime derivatives
$\nabla^\alpha\psi$ near the sphere $\mathcal N_{u_0}\cap\{r=R\}$.
Restricting \eqref{eq:compact-bound-all-orders} to
$\mathcal N_{u_0}\cap\mathcal J$ gives these bounds on a short
portion of $\mathcal N_{u_0}$ on each side of $r=R$; they enter
the energy estimates in Section~\ref{sec:application}.
The choice of $\mathcal J$ ensures that these bounds use only
the compact initial-data norms on $\mathcal S\cap\{r\le2R_0\}$.
The estimate on the larger region $\mathcal K_V$ also controls
$\mathcal C_0$ and $I_V(u_0,0)$, using initial data out to
$R_{\mathcal K_V}$.

\begin{proof}
We first choose $R_0$ large enough to contain the nonstatic part
of $\mathcal S$ and to have $r_*(R_0)\ge\max\{2,C_V/2\}$.
We may take $A=r_*(R_0)$ and a joining radius $R>R_0$ with
$2r_*(R)-A<r_*(2R_0)$. By making this fixed choice sufficiently
far out, we can join the outgoing ray $u=u_0$ at $r=R$ to a
smooth spacelike curve lying to the future of the compact part
of $\mathcal S$ and meeting the future horizon at finite advanced
time $v$. Near $r=R$, we use the function $H(r)$ in
Section~\ref{sec:hypersurfaces} to obtain a smooth hypersurface
that equals $u=u_0$ for $r\ge R$; stationary translates give the
family $\mathcal N_\tau$. The displayed inequality allows us to
choose $\eta>0$ with $v_R+\eta<r_*(2R_0)$.
A past-directed causal curve from $\mathcal J$ that meets the
static end does so at $r_*\le v_R+\eta$; hence the initial causal
past of $\mathcal J$ lies in $r<2R_0$.

For the later-horizon configuration we use regular ingoing
coordinates. At a bifurcation-sphere endpoint we use a regular
double-null chart, or a regular spacelike time coordinate in its
extension. In these charts, each compact region under consideration
has compact causal past on $\mathcal S$. We extend the one-sided
initial data smoothly across their endpoint by a bounded Sobolev
extension operator and insert a cutoff outside this compact past.
Finite propagation identifies the resulting solution on the
specified spacetime region.

On a finite regular coordinate cover, the energy identity for a
uniformly timelike multiplier, commuted through order $j+1$,
and Gronwall's inequality control $H^{j+2}\times H^{j+1}$ over
a fixed finite interval of regular time. Sobolev embedding on
the spacelike slices bounds the displayed spacetime derivatives;
the wave equation supplies the time derivatives. Taking $j=9$
gives \eqref{eq:compact-bound} and its flux bounds, and general
$j$ gives \eqref{eq:compact-bound-all-orders}.
On any additional fixed bounded interval of the static end,
the weighted terms in $\mathfrak D_{p,m}[\psi]$ control the
required Sobolev norms. This proves the final assertion for
arbitrary fixed compact regions.
\end{proof}

\subsection{Continuation from \texorpdfstring{$u=u_0$}{the early outgoing hypersurface}
to \texorpdfstring{$\mathcal N_0$}{N0}}
\label{sec:finite-u-continuation}

We give the continuation on $u_0\le u\le u_2$ for any fixed
$u_2>u_0$. The choice $u_2=0$ reaches the outgoing end of
$\mathcal N_0$ and is shown in Figure~\ref{fig:slow-bootstrap-regions}.

\begin{lemma}[Continuation over a bounded interval of retarded time]
\label{lem:finite-u-transfer}
We fix $p\in\mathbb R$, $m\ge3$, and assume the size bounds
$f=\Os[m+1](r^{-p})$, $h=\Os[m](r^{-p-1})$ on the static end.
For every fixed $u_2>u_0$, there is $V\ge A$ such that
\begin{equation}
 |\phi(u,v)|\le C_{u_2}\mathfrak D_{p,m}[\psi]\cdot G_p(v),
 \qquad G_p(v):=E_p(u_0,v),
 \quad u_0\le u\le u_2,\quad v\ge V.
 \label{eq:finite-u-field-bound}
\end{equation}
Moreover,
\begin{equation}
 Q(u,v)-Q(u_0,v)
 =O\!\left(\mathfrak D_{p,m}[\psi]\cdot v^{-3}G_p(v)\right)
 \label{eq:finite-u-Q-transfer}
\end{equation}
uniformly in this region. For fixed $A,p$,
$G_p(v)$ is comparable to $1$ when $p>1$, to $1+\log v$ when
$p=1$, and to $v^{1-p}$ when $p<1$.
The constants may depend on $p,m,u_2$ and the fixed geometry
and hypersurfaces, and are independent of the solution.
\end{lemma}

\begin{proof}
On the static end, $r_*\asymp r$, so the data-size assumptions
of Lemma~\ref{lem:volterra} hold with
$C_0\le C\mathfrak D_{p,m}[\psi]$.
We choose $V$ sufficiently large that the strip is in the far
exterior and in the future development of $\mathcal S$.
There $|k(u,v)|\le K v^{-3}$, with $K$ depending only on the
background and the fixed interval $[u_0,u_2]$. We also require
\[
 \varepsilon_V:=\frac{K(u_2-u_0)}{2V^2}\le\frac14.
\]
For $L\ge V$, we use the finite rectangles and maxima
\[
 \mathcal R_{u_2,L}=[u_0,u_2]\times[V,L],\qquad
 N_L=\max_{\mathcal R_{u_2,L}}\frac{|\phi(u,v)|}{G_p(v)}.
\]
The two past boundary segments are $u=u_0$ and $v=V$.

\begin{figure}[H]
\centering
\begin{tikzpicture}[x=1.40cm,y=1.40cm,
 every node/.style={font=\small,inner sep=1.3pt},
 boundary/.style={line width=.8pt},
 initial/.style={draw=blue!65!black,line width=1.25pt},
 future/.style={draw=red!65!black,line width=1.2pt},
 incoming/.style={draw=orange!75!black,line width=1.1pt}]
 \coordinate (B) at (0,0);
 \coordinate (IZ) at (6,0);
 \coordinate (IP) at (3,3);
 \coordinate (JU) at (3.425,.225);
 \coordinate (JZ) at (2.525,1.425);
 \coordinate (FU) at (4.60,1.40);
 \coordinate (FZ) at (3.55,2.45);
 \coordinate (UV) at (3.65,.45);
 \coordinate (UL) at (4.25,1.05);
 \coordinate (ZV) at (2.60,1.50);
 \coordinate (ZL) at (3.20,2.10);
 \fill[orange!20] (UV)--(UL)--(ZL)--(ZV)--cycle;
 \draw[boundary] (B)--(IP)--(IZ);
 \draw[gray!55] (B)--(IZ);
 \draw[initial] (.28,.28) .. controls (.85,.025) and (1.85,0) ..
      (2.70,0)--(IZ);
 \draw[future] (.48,.48) .. controls (1.40,.38) and (3.20,0) ..
      (JU)--(FU);
 \draw[future] (1.10,1.10) .. controls (1.65,.95) and (2.175,1.075) ..
      (JZ)--(FZ);
 \draw[incoming] (ZV)--(UV);
 \draw[incoming,dashed] (ZL)--(UL);
 \fill (B) circle(1.15pt) (IZ) circle(1.15pt) (IP) circle(1.15pt);
 \fill[blue!65!black] (.28,.28) circle(1.5pt);
 \fill[red!65!black] (.48,.48) circle(1.5pt)
      (1.10,1.10) circle(1.5pt);
 \fill[orange!75!black] (UV) circle(1.4pt) (UL) circle(1.4pt)
      (ZV) circle(1.4pt) (ZL) circle(1.4pt);
 \node[below left] at (B) {$\mathcal B$};
 \node[right] at (IZ) {$i^0$};
 \node[above] at (IP) {$i^+$};
 \node[above left] at (1.55,1.55) {$\mathcal H^+$};
 \node[above right] at (4.95,1.05) {$\mathcal I^+$};
 \node[blue!65!black,below=2pt] at (1.45,.02) {$\mathcal S$};
 \node[blue!65!black,below=3pt] at (4.80,0) {$t=0$};
 \node[red!65!black,above=2pt] at (1.95,.255)
      {$\mathcal N_{u_0}$};
 \node[red!65!black,above=2pt] at (1.90,1.09)
      {$\mathcal N_0$};
 \node[red!65!black,rotate=45,below=2pt] at (4.08,.88)
      {$u=u_0$};
 \node[red!65!black,rotate=45,above=2pt] at (3.04,1.94)
      {$u=0$};
 \node[orange!65!black,rotate=-45,below=2pt] at (3.15,.95)
      {$v=V$};
 \node[orange!65!black,rotate=-45,above=2pt] at (3.73,1.57)
      {$v=L$};
 \node at (3.425,1.275) {$\mathcal R_{0,L}$};
\end{tikzpicture}
\caption{The shaded null rectangle is
$\mathcal R_{0,L}=[u_0,0]\times[V,L]$.
We choose $V$ beyond the transitions to the outgoing null portions
of both future hypersurfaces. The first bootstrap controls the
boundary $u=u_0$, and finite-time propagation from $\mathcal S$
controls the compact incoming segment $v=V$.}
\label{fig:unified-strip-bootstrap}\label{fig:slow-bootstrap-regions}
\end{figure}
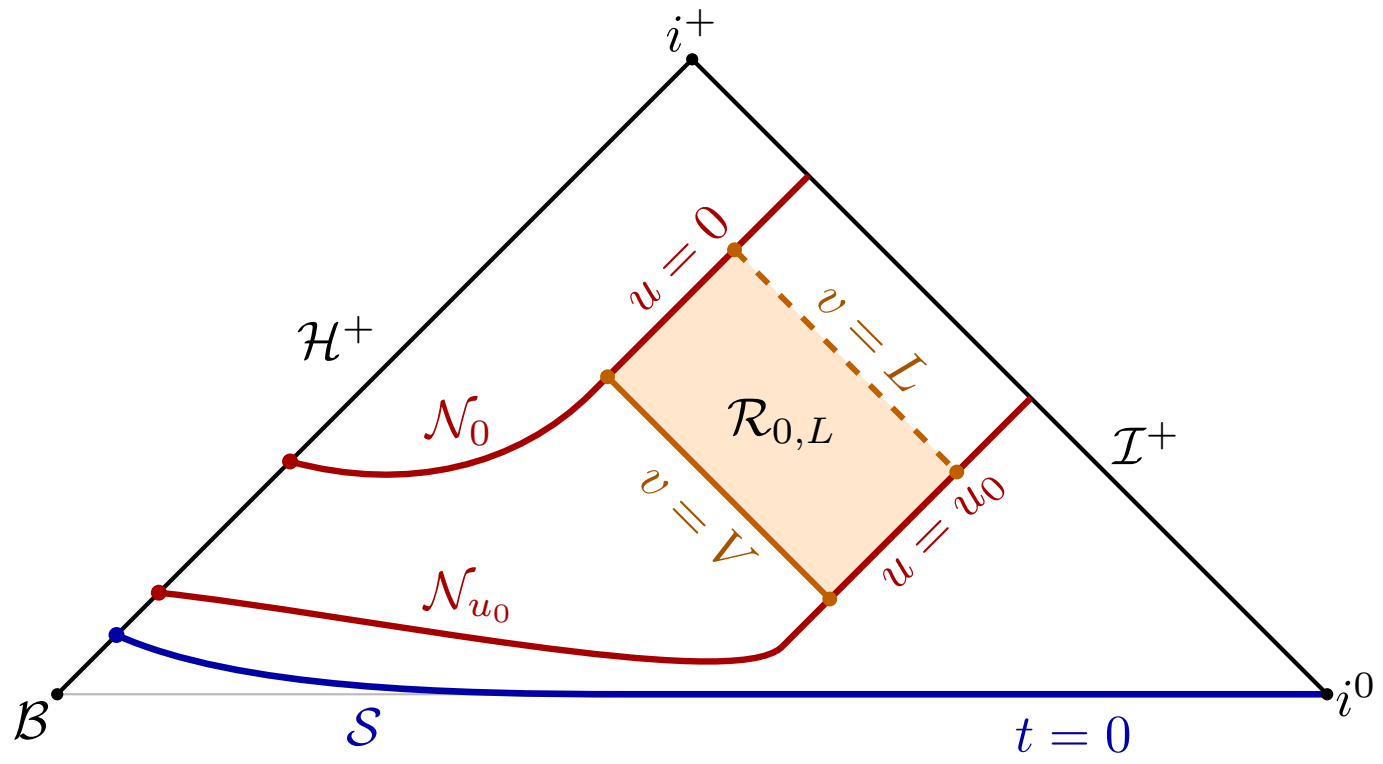

Integrating $\partial_u\partial_v\phi=-k\phi$ on the rectangle gives
\begin{align}
 \phi(u,v)={}&\phi(u_0,v)+\phi(u,V)-\phi(u_0,V)\notag\\
 &-\int_{u_0}^{u}\int_V^v
                 k(u',v')\phi(u',v')\,\dd v'\dd u'.
 \label{eq:characteristic-rectangle}
\end{align}
Lemma~\ref{lem:volterra} bounds the first boundary segment.
Lemma~\ref{lem:compact-transfer} bounds the compact segment
$v=V$. Since $G_p$ is positive and nondecreasing, the boundary
sum is at most $BG_p(v)$ in absolute value, with
$B=C_{u_2}\mathfrak D_{p,m}[\psi]$ independent of $L$.
The integration rectangle lies inside $\mathcal R_{u_2,L}$, and
$G_p(v')\le G_p(v)$ there. Consequently,
\[
 N_L\le B+K N_L(u_2-u_0)\int_V^\infty(v')^{-3}\,\dd v'
       =B+\varepsilon_VN_L.
\]
For nonzero data the bootstrap $N_L\le2B$ improves to
$N_L\le3B/2$. At $L=V$ it starts from the compact boundary data.
The finite maxima are continuous in $L$, so the same continuity
argument as above closes the bootstrap for every finite $L$.
Exhaustion of the strip gives \eqref{eq:finite-u-field-bound};
the zero-data case follows by uniqueness.
Finally,
\[
 Q(u,v)-Q(u_0,v)
       =-\int_{u_0}^{u}k(u',v)\phi(u',v)\,\dd u'
\]
proves \eqref{eq:finite-u-Q-transfer}.
\end{proof}

\subsection{Quantitative asymptotic expansions towards null infinity}
\label{sec:all-p-null-traces}

The following proposition gives the asymptotics as $r\to\infty$
with $u$ in any fixed bounded interval $J$. In particular, it
applies at $u=0$ on the outgoing end of $\mathcal N_0$.
When the second argument is $r$, we express $\phi=r\psi$ in
the coordinates $(u,r)$, related by $v=u+2r_*(r)$.
We use the data coefficient $C_p=b+(1-p)a$ from
\eqref{eq:frac-incoming-coefficient} and abbreviate
$\mathfrak A[\psi]=\mathfrak A_{p,m,\delta}[\psi]$.

\begin{proposition}[Spatial-to-null asymptotics for every real $p$]
\label{prop:all-p-null-traces}
\label{prop:trace}
\label{prop:frac-null-divergence}
We assume \eqref{eq:classification-data} with $m\ge3$.
For every bounded interval $J\subset\mathbb R$, the following
expansions hold in the future development of $\mathcal S$, uniformly
for $u\in J$ and $r\ge R_J$, where $R_J$ is sufficiently large.
All constants depend only on the fixed geometry and hypersurfaces,
$J$, and the displayed exponents.

\smallskip\noindent
\emph{The outgoing derivative for $p<3$.}
For every $0<\beta<\min\{\delta,1,3-p\}$,
\begin{equation}
 Q(u,v)=\frac{C_p}{2}v^{-p}
       +O_J\!\left(\mathfrak A[\psi]\cdot v^{-p-\beta}\right),
 \qquad p<3.
 \label{eq:frac-original-trace}
\end{equation}
Since $\partial_r|_u=2D^{-1}\partial_v$, this gives
\begin{equation}
 \lim_{r\to\infty}r^p\partial_r\phi\big|_u=2^{-p}C_p.
 \label{eq:frac-generalized-charge}
\end{equation}

\smallskip\noindent
\emph{The finite radiation field for $p>1$.}
The limit $F(u)=\lim_{v\to\infty}\phi(u,v)$ exists, uniformly
on $J$, and $|F(u)|\le C\mathfrak A[\psi]\cdot(-u)^{1-p}$ for
$u\le u_0$. When $1<p<3$, the same choices of $\beta$ give
\begin{equation}
 \phi(u,v)=F(u)-\frac{C_p}{2(p-1)}v^{1-p}
       +O_J\!\left(\mathfrak A[\psi]\cdot v^{1-p-\beta}\right).
 \label{eq:finite-radiation-trace}
\end{equation}

\smallskip\noindent
\emph{The logarithmic term for $p=1$.}
With the fixed length normalization in the logarithm, the limit
\begin{equation}
 F_{\rm ren}(u):=\lim_{r\to\infty}
       \left[r\psi(u,r)-\frac b2\log(2r)\right]
 \label{eq:frac-p1-renormalization}
\end{equation}
exists, and for $0<\beta<\min\{\delta,1\}$,
\begin{equation}
 \phi(u,v)=\frac b2\log v+F_{\rm ren}(u)
       +O_J\!\left(\mathfrak A[\psi]\cdot v^{-\beta}\right).
 \label{eq:fixed-u-p1-trace}
\end{equation}
The quantities $F(u)$ and $F_{\rm ren}(u)$ in their respective
ranges are bounded on $J$ by $C_J\mathfrak A[\psi]$.

\smallskip\noindent
\emph{The growing product $r\psi$ for $p<1$.}
For every $0<\gamma<\min\{\delta,1,1-p\}$,
\begin{equation}
 \phi(u,r)=r\psi(u,r)
   =\frac{2^{-p}C_p}{1-p}r^{1-p}
       +O_J\!\left(\mathfrak A[\psi]\cdot r^{1-p-\gamma}\right),
 \label{eq:frac-slow-null-growth}
\end{equation}
and
\begin{equation}
 r^2\partial_r(r\psi)\big|_u
   =2^{-p}C_pr^{2-p}
       +O_J\!\left(\mathfrak A[\psi]\cdot r^{2-p-\gamma}\right).
 \label{eq:frac-slow-NP-growth}
\end{equation}
For $C_p\ne0$, these formulas give divergence in absolute value
of $r\psi$ and of the quantity defining the ordinary
Newman--Penrose constant. They also give quantitative remainder
bounds when $C_p=0$.

\smallskip\noindent
\emph{The coefficient at order $v^{-3}$ for $p\ge3$.}
We use $a_3,b_3,\delta_3$ and the common representation
\eqref{eq:pge3-coefficients}--\eqref{eq:pge3-data}, and set
\begin{equation}
 K_3=\frac{b_3}{2}-a_3.
 \label{eq:K}
\end{equation}
Then
\begin{align}
 v^3Q(u,v)
 ={}&K_3-4M\int_{-\infty}^{u}F(u')\,\dd u'
       +O_J\!\left(\mathfrak A[\psi]\cdot
                    (v^{-\delta_3}+v^{-1}\log v)\right),
 \label{eq:tracev}\\
 L(u):=\lim_{v\to\infty}r(u,v)^3Q(u,v)
 ={}&\frac{K_3}{8}-\frac M2\int_{-\infty}^{u}F(u')\,\dd u'.
 \label{eq:traceL}
\end{align}
The radiation integral converges absolutely, and $I_0[\psi]=0$.
For $p=3$, $K_3=b/2-a$; for $p>3$, $K_3=0$.
The remainder in \eqref{eq:tracev}, the bound for $L(u)$ in
\eqref{eq:traceL}, and the estimate \eqref{eq:raderror} below
are also controlled by $\mathfrak N_{m,\delta_3}[\psi]$ for the
common representation \eqref{eq:pge3-data}.

\smallskip\noindent
\emph{The values on $\mathcal N_0$ for $p<1$.}
On the outgoing null portion of $\mathcal N_0$ we have $u=0$,
so
\begin{equation}
 \psi|_{\mathcal N_0}(r)
 =\frac{2^{-p}C_p}{1-p}r^{-p}
       +O\!\left(\mathfrak A[\psi]\cdot r^{-p-\gamma}\right)
       \qquad(r\to\infty).
 \label{eq:slow-initial-trace}
\end{equation}
On every fixed compact portion of $\mathcal N_0$,
\[
 \sup_{r_h\le r\le R_1}|\psi|_{\mathcal N_0}(r)|
       \le C_{R_1}\mathfrak A[\psi].
\]
Consequently,
\begin{equation}
 \begin{cases}
 |\psi|_{\mathcal N_0}(r)|
       \le C\mathfrak A[\psi]\cdot(1+r)^{-p},
       \quad r\ge r_h,&0<p<1,\\
 \psi|_{\mathcal N_0}(r)\longrightarrow a+b
       \quad\text{as }r\to\infty,&p=0.
 \end{cases}
 \label{eq:slow-initial-trace-limits}
\end{equation}
\end{proposition}

\begin{proof}
The initial-data norm controls the constants in
Lemmas~\ref{lem:volterra}--\ref{lem:finite-u-transfer}.
Direct differentiation on $t=0$ gives
\begin{equation}
 b_0(v)=\frac12\left[rh+D(f+rf')\right]_{r_*=v}
 =\frac{C_p}{2}v^{-p}
   +O\!\left(\mathfrak A[\psi]\cdot
                (v^{-p-\delta}+v^{-p-1}\log v)\right).
 \label{eq:frac-spatial-seed}
\end{equation}
Here $f'$ is the derivative with respect to $r$, and
$r_*=r+2M\log r+O(1)$ accounts for the logarithm.
For $p<3$, combining this formula with \eqref{eq:exterior-Q-cases} proves
\eqref{eq:frac-original-trace} on the early ray $u=u_0$.
The continuation error \eqref{eq:finite-u-Q-transfer} is
$O(v^{-3})$ for $p>1$, $O(v^{-3}\log v)$ for $p=1$, and
$O(v^{-p-2})$ for $p<1$, with the indicated data-norm factors.
It therefore gives the same expansion on every fixed later
interval. The wedge estimate covers the part of a bounded $J$
lying to the past of $u_0$.
The relation $v=2r+O_J(\log r)$ and
$\partial_r|_u=2D^{-1}\partial_v$ give
\eqref{eq:frac-generalized-charge}.

For $p>1$, the bounds for $Q$ are integrable in $v$ and give the
finite limit $F(u)$, uniformly on bounded $u$-intervals. Its
early-$u$ bound follows by taking $v\to\infty$ in
\eqref{eq:wedgebound}. Integrating
\eqref{eq:frac-original-trace} from $v$ to infinity gives
\eqref{eq:finite-radiation-trace} for $1<p<3$.
For $p=1$, the difference $Q-b/(2v)$ is integrable; integration
from $v$ to infinity gives \eqref{eq:fixed-u-p1-trace}.
Since $\log v-\log(2r)\to0$ at fixed $u$, its constant is
precisely \eqref{eq:frac-p1-renormalization}.
Finite compact-region bounds and these convergent integrals
also bound $F$ and $F_{\rm ren}$ on $J$ by the stated data norm.

For $p<1$, we choose
$\gamma<\beta<\min\{\delta,1\}$ and integrate
\eqref{eq:frac-original-trace} from a fixed sufficiently large $V_J$ to
$v$ at fixed $u\in J$. We obtain
\[
 \phi(u,v)=\frac{C_p}{2(1-p)}v^{1-p}
       +O_J\!\left(\mathfrak A[\psi]\cdot v^{1-p-\gamma}\right).
\]
The condition $\gamma<1-p$ absorbs the finite integration
constant, while $\gamma<\beta$ absorbs the integrated error,
including the logarithmic case. Conversion using
$v=2r+O_J(\log r)$ gives \eqref{eq:frac-slow-null-growth}.
Applying $\partial_r|_u=2D^{-1}\partial_v$ directly to the
estimate for $Q$ gives \eqref{eq:frac-slow-NP-growth}.
Restriction to $u=0$, division by $r$, and the compact-region
estimate of Lemma~\ref{lem:compact-transfer} prove
\eqref{eq:slow-initial-trace}--\eqref{eq:slow-initial-trace-limits}.

It remains to compute the coefficient for $p\ge3$.
We use the representation \eqref{eq:pge3-data}; its norm is
controlled through the required derivative orders by
$\mathfrak A[\psi]$. For $p>3$, this follows directly from
$\delta_3<p-3$.
The $p=3$ instance of \eqref{eq:Qbound} gives
\begin{equation}
 |\phi(u,v)-F(u)|
 \le C\mathfrak A[\psi]\cdot
          \bigl(v^{-2}+v^{-2}(-u)^{-1}\bigr),
 \qquad u\le u_0,\quad v\ge-u,
 \label{eq:raderror}
\end{equation}
and $|F(u)|\le C\mathfrak A[\psi]\cdot(-u)^{-2}$ there.
The initial term satisfies
\begin{equation}
 v^3b_0(v)=K_3
       +O\!\left(\mathfrak A[\psi]\cdot
                  (v^{-\delta_3}+v^{-1}\log v)\right).
 \label{eq:seed}
\end{equation}
For $-v\le u'\le u_0$, the potential expansion gives
\begin{equation}
 v^3k(u',v)
 =4M\left(\frac v{v-u'}\right)^3+O(v^{-1}\log v)
 =4M+O\!\left(\frac{-u'}v+\frac{\log v}{v}\right).
 \label{eq:kexpand}
\end{equation}
We insert these formulas in \eqref{eq:Qexact}.
Replacing $\phi(u',v)$ by $F(u')$ costs $O(\mathfrak A[\psi]\cdot v^{-1})$
after multiplication by $v^3$, by \eqref{eq:raderror}.
The kernel error integrated against $F$ costs
$O(\mathfrak A[\psi]\cdot v^{-1}\log v)$, since
$|F(u')|\le C\mathfrak A[\psi]\cdot(-u')^{-2}$.
Extending the lower endpoint from $-v$ to $-\infty$ costs
$O(\mathfrak A[\psi]\cdot v^{-1})$.
This proves \eqref{eq:tracev} for $u\le u_0$.

On a bounded later interval, \eqref{eq:finite-u-Q-transfer}
gives $Q=O(v^{-3})$ and hence
$\phi(u,v)-F(u)=O(v^{-2})$, uniformly in $u$ and with constants
controlled by the data norm. There
$v^3k(u,v)=4M+O(v^{-1}\log v)$. Integrating
$\partial_uQ=-k\phi$ therefore gives
\[
 v^3\bigl(Q(u,v)-Q(u_0,v)\bigr)
 =-4M\int_{u_0}^{u}F(u')\,\dd u'
       +O_J\!\left(\mathfrak A[\psi]\cdot v^{-1}\log v\right).
\]
This proves \eqref{eq:tracev} on the whole bounded interval.
Finally, $r/v\to1/2$ gives \eqref{eq:traceL}, and
$r^2\partial_r\phi=2D^{-1}r^2Q\to0$ gives $I_0[\psi]=0$.
\end{proof}

\subsection{Initial data for \texorpdfstring{$T\psi$}{T psi} on
\texorpdfstring{$\mathcal S$}{S} for every real \texorpdfstring{$p$}{p}}
\label{sec:commuted}

We compute the initial data of $w=T\psi$ on $\mathcal S$ for
every real $p$. The next lemma tracks the change $p\mapsto p+1$
in the spatial exponent, the leading coefficients, and the
initial-data norms. These bounds are used when the later
asymptotic estimates are applied to $T\psi$.

\begin{lemma}[Initial data for $T\psi$]
\label{lem:slow-T-data}
We fix $p\in\mathbb R$ and an integer $N\ge2$, and set $w=T\psi$.
Stationarity gives $\Box_gw=T(\Box_g\psi)=0$. On the static end,
the position datum of $w$ and its stationary time derivative are
\begin{equation}
 w|_{\mathcal S}=h,\qquad
 Tw|_{\mathcal S}=\mathcal L_D f,\qquad
 \mathcal L_D f:=\frac D{r^2}\partial_r(r^2D\partial_r f).
 \label{eq:frac-T-spatial-data}
\end{equation}
For general symbol data $f=\Os[N+1](r^{-p})$ and
$h=\Os[N](r^{-p-1})$, this implies
\begin{equation}
 \mathfrak D_{p+1,k}[T\psi]
 \le C\mathfrak D_{p,k+2}[\psi],\qquad 0\le k\le N-2.
 \label{eq:time-derivative-symbol-norm}
\end{equation}
If the data have the more precise expansions
\eqref{eq:classification-data} through these orders, we set
$\delta_*:=\min\{\delta,1\}$ and obtain
\begin{align}
 w|_{\mathcal S}
 &=b r^{-p-1}+\Os[N](r^{-p-1-\delta_*}),
 \label{eq:slow-T-position}\\
 Tw|_{\mathcal S}
 &=p(p-1)a r^{-p-2}+\Os[N-1](r^{-p-2-\delta_*}).
 \label{eq:slow-T-velocity}
\end{align}
Thus $w$ has spatial exponent $p+1$ and coefficients
$a_w=b$, $b_w=p(p-1)a$. In particular,
\begin{equation}
 \begin{aligned}
 C_{p+1}[T\psi]&=-pC_p[\psi],\\
 \mathfrak A_{p+1,k,\delta_*}[T\psi]
 &\le C\mathfrak A_{p,k+2,\delta}[\psi],
 \qquad 0\le k\le N-2.
 \end{aligned}
 \label{eq:slow-T-norm}
\end{equation}
Here $k$ is an integer norm index, and the brackets indicate the
solution whose initial data are used in the coefficient or norm. The norm on the left
includes the geometric data
$(w|_{\mathcal S},n_{\mathcal S} w|_{\mathcal S})$ on the compact part of
$\mathcal S$. Its bound holds whether $\mathcal S$ meets the
bifurcation sphere or a later cross-section of the future horizon.
\end{lemma}

\begin{proof}
The wave equation on the static end gives
\eqref{eq:frac-T-spatial-data}, with
\[
 \mathcal L_D f
 =D^2f''+\left(\frac{2D^2}{r}+DD'\right)f'.
\]
Since $D=1+\Os[\infty](r^{-1})$, applying this operator to $a r^{-p}$
gives $p(p-1)a r^{-p-2}$ with an error of order $r^{-p-3}$.
Applying it to the remainder of $f$ gives an error of order
$r^{-p-2-\delta}$. The radial derivative bounds in the data
give the corresponding derivative bounds for both errors.
This proves \eqref{eq:slow-T-position} and
\eqref{eq:slow-T-velocity} with $\delta_*=\min\{\delta,1\}$.
The coefficient relation follows from
\[
 b_w+\bigl(1-(p+1)\bigr)a_w
 =p(p-1)a-pb=-p\bigl(b+(1-p)a\bigr).
\]

On the compact part of $\mathcal S$, the field $T$ is smooth in
the regular horizon coordinates. The wave equation expresses
second normal derivatives of $\psi$ through the geometric
initial data $(f,\pi)$ and their tangential derivatives.
The resulting Sobolev bounds for
$(T\psi|_{\mathcal S},n_{\mathcal S}T\psi|_{\mathcal S})$
use at most two additional orders of the initial-data norm.
Together with the symbol estimates, they prove
\eqref{eq:time-derivative-symbol-norm} and \eqref{eq:slow-T-norm}. The same argument in regular coordinates applies at the
bifurcation-sphere endpoint.
\end{proof}

\subsection{Radial derivative estimates on outgoing null hypersurfaces
for \texorpdfstring{$p>1$}{p > 1}}
\label{sec:frac-spatial-transfer}

We estimate radial derivatives on the outgoing null portions of
$\mathcal N_u$, for $r\ge R$ and $u_0\le u\le0$, with constants
uniform in $u$. Throughout this subsection, $\phi=r\psi$ and
$\partial_r$ keeps $u$ fixed. We treat $1<p<3$ first and then
$p\ge3$, including the corresponding estimates for the time
derivatives $T^j\phi$ in the latter case.

\paragraph{Radial derivatives for $1<p<3$.}

\begin{lemma}[Radial derivative bounds on outgoing null hypersurfaces for $1<p<3$]
\label{lem:frac-spatial-transfer}
We fix $1<p<3$ and an integer $m\ge3$, and assume
$f=\Os[m+1](r^{-p})$ and $h=\Os[m](r^{-p-1})$ on the static
end of $\mathcal S$. On the outgoing null portions of the
hypersurfaces $\mathcal N_u$, with $u_0\le u\le0$, we have
\begin{equation}
 \phi(u,r)=F(u)+g_u(r),\qquad
 |\partial_r^j g_u(r)|\le C\mathfrak D_{p,m}[\psi]\cdot r^{1-p-j},
 \quad r\ge R,\quad 0\le j\le m-2.
 \label{eq:frac-transferred-symbol}
\end{equation}
Here $F(u)=\lim_{r\to\infty}\phi(u,r)$, the radial derivatives
keep $u$ fixed, and the constant is uniform for $u\in[u_0,0]$.
Lemma~\ref{lem:compact-transfer} controls the data on the compact
spacelike portions in terms of the same initial-data norm.
\end{lemma}

\begin{proof}
Lemma~\ref{lem:volterra} gives
$|\phi(u,v)|\le C\mathfrak D_{p,m}[\psi]\cdot(-u)^{1-p}$
throughout $\mathcal W_A$. The initial-data term in
\eqref{eq:Qexact} satisfies
$|\partial_v^n b_0(v)|\le C_n\mathfrak D_{p,m}[\psi]\cdot v^{-p-n}$
at the derivative orders used below. We define
\[
 J_p(u,v)=\int_{-v}^{u}(-u')^{1-p}\,du'.
\]
The identity \eqref{eq:Qexact} and the potential bound give
\[
 |Q(u,v)|\le C\mathfrak D_{p,m}[\psi]\cdot
       \bigl(v^{-p}+v^{-3}J_p(u,v)\bigr).
\]
We retain this estimate throughout the wedge when differentiating,
since the integral samples all $-v\le u'\le u$. For
$0\le n\le m-3$, we prove by induction that
\begin{equation}
 |\partial_v^n Q(u,v)|
 \le C_n\mathfrak D_{p,m}[\psi]\cdot
       \left(v^{-p-n}+v^{-3-n}J_p(u,v)\right),
 \qquad u\le u_0=-A,\quad v\ge-u.
 \label{eq:frac-uniform-wedge-derivatives}
\end{equation}
We have just proved the case $n=0$. Differentiating
\eqref{eq:Qexact} $n$ times gives lower-endpoint terms and integrals
\[
 \int_{-v}^u(\partial_v^a k)(u',v)
                 \partial_v^{n-a}\phi(u',v)\,du',
 \qquad 0\le a\le n.
\]
For $a=n$, we use the bound for $\phi$. For $a<n$, we write
$\partial_v^{n-a}\phi=\partial_v^{n-a-1}Q$ and use the induction
hypothesis together with
\[
 |\partial_v^a k(u',v)|\le C_a v^{-3-a},\qquad
 \int_{-v}^u J_p(u',v)\,du'\le vJ_p(u,v).
\]
The terms with $a<n$ are bounded by the right-hand side of
\eqref{eq:frac-uniform-wedge-derivatives}, with an additional
inverse power of $v$. The lower-endpoint terms are
$O(\mathfrak D_{p,m}[\psi]\cdot v^{-p-n-1})$: at $u'=-v$,
the derivatives $\partial_u^j\partial_v^k\phi$ are expressed
through derivatives in $(t,r_*)$ on $t=0$, controlled by the
data and the wave equation. The bound for $\partial_v^n b_0$
therefore closes the induction.

On $u=u_0$, we have
\[
 J_p(u_0,v)\le C(p)
 \begin{cases}
 v^{2-p},&1<p<2,\\
 1+\log v,&p=2,\\
 1,&2<p<3.
 \end{cases}
\]
Consequently
$|\partial_v^nQ(u_0,v)|\le
C\mathfrak D_{p,m}[\psi]\cdot v^{-p-n}$.
We now propagate these derivative bounds across the fixed interval
$u_0\le u\le0$. Lemma~\ref{lem:finite-u-transfer} gives
$|\phi(u,v)|\le C\mathfrak D_{p,m}[\psi]$ for sufficiently large
$v$. Differentiating
\[
 Q(u,v)=Q(u_0,v)-\int_{u_0}^{u}k(u',v)\phi(u',v)\,du'
\]
and using induction in $n$, together with
$|\partial_v^a k|\le C_a v^{-3-a}$ on this strip, gives
\[
 |\partial_v^nQ(u,v)-\partial_v^nQ(u_0,v)|
 \le C\mathfrak D_{p,m}[\psi]\cdot v^{-3-n},
 \qquad 0\le n\le m-3.
\]
Since $p<3$, we obtain the same $v^{-p-n}$ bound uniformly
for $u\in[u_0,0]$. Integrating $Q$ to infinity gives $F(u)$
and the $j=0$ bound in \eqref{eq:frac-transferred-symbol}.
The identity $\partial_r|_u=2D^{-1}\partial_v$ gives its remaining
radial derivative bounds. Lemma~\ref{lem:compact-transfer}
controls the fixed finite-radius portions, completing the estimate
for all $r\ge R$.
\end{proof}

\paragraph{Time and radial derivatives for $p\ge3$.}

We assume the common representation \eqref{eq:pge3-data} with
$m\ge10$. We first estimate $T^j\phi$, $\partial_vT^j\phi$, and
$\partial_v^2T^j\phi$ throughout $\mathcal W_A$. Their restrictions
to $\mathcal N_{u_0}$ supply the weighted energies in
Section~\ref{sec:application}.
We use the potential $V$ in \eqref{eq:potential-definition} to form
the radial operator $\mathcal L_{r_*}=\partial_{r_*}^2-V(r_*)$.
We let $F_j(r_*)$ be the restriction of $T^j\phi$ to $t=0$, expressed in the coordinate
$r_*$. Thus $F_0=rf$ and $F_1=rh$, with $r=r(r_*)$.
Stationarity gives, for each nonnegative integer $\ell$,
\begin{equation}
 F_{2\ell}=\mathcal L_{r_*}^{\ell}(rf),\qquad
 F_{2\ell+1}=\mathcal L_{r_*}^{\ell}(rh).
 \label{eq:commuteddata}
\end{equation}
If the initial bounds for $f$ and $h$ hold through $m+1$ and $m$
radial derivatives, respectively, then
$\partial_{r_*}^nF_j=O(r_*^{-2-j-n})$ for nonnegative integers
$j,n$ with $j+n\le m+1$. In the estimates below we take
$0\le j\le m-1$, so that two radial derivatives are available.
We apply Lemma~\ref{lem:volterra} to $T^j\phi$ with spatial
exponent $3+j$, and use \eqref{eq:exterior-Q-cases}, to obtain
\begin{align}
 |T^j\phi(u,v)|&\le C_j\mathfrak N_{m,\delta_3}[\psi]\cdot(-u)^{-2-j}, \label{eq:Tbound}\\
 |\partial_vT^j\phi(u,v)|
   &\le C_j\mathfrak N_{m,\delta_3}[\psi]\cdot\big(v^{-3-j}+v^{-3}(-u)^{-1-j}\big).
 \label{eq:TQbound}
\end{align}
Both bounds hold for $u\le-A$ and $v\ge-u$.
We apply the differentiated characteristic identity
\eqref{eq:Qexact} to $T^j\phi$ and differentiate once more in $v$.
With $b_j=(F_j'+F_{j+1})/2$, we obtain
\begin{align}
 \partial_v^2T^j\phi(u,v)
 ={}&b_j'(v)-k(-v,v)F_j(v) \notag\\
 &-\int_{-v}^{u}\big((\partial_vk)(u',v)T^j\phi(u',v)
                  +k(u',v)\partial_vT^j\phi(u',v)\big)\,\dd u'.
 \label{eq:secondv}
\end{align}
We use $|\partial_vk(u',v)|\le C(v-u')^{-4}$ to obtain
\begin{equation}
 |\partial_v^2T^j\phi(u,v)|
 \le C_j\mathfrak N_{m,\delta_3}[\psi]\cdot\big(v^{-4-j}+v^{-4}(-u)^{-1-j}\big),
 \qquad u\le-A,\quad v\ge-u.
 \label{eq:TQQbound}
\end{equation}
For $j=0$, the possible intermediate term $v^{-6}\log(v/(-u))$ is bounded
by the right-hand side.

On the fixed outgoing hypersurface $u=u_0$, these estimates imply
\begin{equation}
 T^j\phi=O(1),\qquad
 \partial_rT^j\phi=O(r^{-3}),\qquad
 \partial_r^2T^j\phi=O(r^{-4}).
 \label{eq:radialbounds}
\end{equation}
Here the radial derivatives keep $u$ fixed, and the constants are
bounded by $C_j\mathfrak N_{m,\delta_3}[\psi]$.
These same asymptotic radial bounds hold uniformly for
$u\in[u_0,0]$. To see this, we apply the finite-strip bootstrap to
$w=T^j\phi$. Its incoming boundary values require derivatives
only through order $j$, controlled by
\eqref{eq:compact-propagation-general} since $j\le m-1$.
Thus $w=O(1)$ on the strip. The identities
\[
 \partial_u(\partial_vw)=-kw,\qquad
 \partial_u(\partial_v^2w)=-(\partial_vk)w-k\partial_vw
\]
then propagate $\partial_vw=O(v^{-3})$ and
$\partial_v^2w=O(v^{-4})$ from $u=u_0$ to $u=0$.
The constants have the same initial-data bound.
Time commutation improves
the early-retarded-time power in
\eqref{eq:Tbound}. The backscattering term in \eqref{eq:TQbound}
accounts for the common $v^{-3}$ weight on a fixed null ray.

\section{The time-inverted Newman--Penrose constant for \texorpdfstring{$p\ge3$}{p >= 3}}
\label{sec:flux}

We evaluate $I_0^{(1)}[\psi]$ from the data
\eqref{eq:classification-data} on the initial hypersurface $\mathcal S$
for every $p\ge3$. We use the common data representation
\eqref{eq:pge3-coefficients}--\eqref{eq:pge3-data} from
Section~\ref{sec:main-data-norms} and the coefficient $K_3$ in
\eqref{eq:K}.

The identity to be proved is
\[
 I_0^{(1)}[\psi]=M B_{\mathcal S}+\frac{2a_3-b_3}{8},
\]
where $B_{\mathcal S}$ is the horizon-plus-bulk expression in
\eqref{eq:geometricB}. The calculation applies to each initial
hypersurface $\{t=0\}$, $\Sigma_0$, or $\Sigma_{\mathcal B}$ defined
in Section~\ref{sec:hypersurfaces}.
In particular, on the static hypersurface $\{t=0\}$ of
Section~\ref{sec:hypersurfaces}, it gives
\[
 I_0^{(1)}[\psi]
 =M\left(r_h^2f(r_h)+\int_{r_h}^{\infty}
              \frac{r^2}{D(r)}h(r)\,\dd r\right)+\frac{2a_3-b_3}{8}.
\]
A closed current first relates $\mathcal S$ to a future hypersurface
$\mathcal N_{u_0}$ with an outgoing null end, as defined in
Section~\ref{sec:hypersurfaces}. We then combine
this identity with the null trace from Section~\ref{sec:exterior}
to evaluate the Newman--Penrose constant of the time integral.

\subsection{The closed wave current}

Per unit solid angle, we define
\begin{equation}
 j[\psi]=r^2(\partial_v\psi\,\dd v-\partial_u\psi\,\dd u).
 \label{eq:current}
\end{equation}
The spherically symmetric wave equation gives $\dd j[\psi]=0$. On the radial
projection of $\mathcal S$, oriented from the horizon towards infinity,
its integral is $(4\pi)^{-1}\int_{\mathcal S}n_{\mathcal S}\psi\,\dd\mu_{\mathcal S}$.
In particular, on a
static end,
\begin{equation}
 j[\psi]=\frac{r^2}{D}h\,\dd r.
 \label{eq:currentstatic}
\end{equation}
For a radial curve $\gamma$ starting on a horizon sphere $S_\gamma$,
and a point $P$ on that curve, we define
\begin{equation}
 \mathcal Q_\gamma(P)=r_h^2\psi(S_\gamma)
                         +\int_{S_\gamma}^{P}j[\psi],
 \qquad
 \mathcal P_\gamma(P)=\mathcal Q_\gamma(P)+r(P)\phi(P).
 \label{eq:Qsigma}
\end{equation}
The horizon term absorbs changes of the inner endpoint because the
horizon restriction of $j$ is $r_h^2\partial_v\psi\,\dd v$.
At the bifurcation sphere this assertion is read in regular Kruskal
coordinates $\mathsf U=-e^{-\kappa u}$, $\mathsf V=e^{\kappa v}$, where
$\kappa=D'(r_h)/2$. In these coordinates
\[
 j=r^2(\psi_{\mathsf V}\,\dd\mathsf V
               -\psi_{\mathsf U}\,\dd\mathsf U),
\]
which is smooth for regular geometric data. Its restriction to $\mathsf U=0$
is $r_h^2\,\dd\psi$, including the endpoint $\mathsf U=\mathsf V=0$.

We choose the member $\mathcal N_{u_0}$ with $u_0=-A$ in the future
of $\mathcal S$, as in Lemma~\ref{lem:compact-transfer}. We truncate
$\mathcal S$ and $\mathcal N_{u_0}$ at the same incoming hypersurface
$v=V$. On the static end the endpoint is $u=-V$, whereas on the
outgoing null end of $\mathcal N_{u_0}$ it is $u=u_0$.

\begin{lemma}[Current identity between the initial and future hypersurfaces]
\label{lem:capidentity}
With the above orientations,
\begin{equation}
 \mathcal P_{\mathcal N_{u_0}}(u_0,V)
       -\mathcal P_{\mathcal S}(-V,V)
       =-\int_{-V}^{u_0}D(r(u,V))\phi(u,V)\,\dd u.
 \label{eq:capidentity}
\end{equation}
Consequently the limit
$C_\infty:=\lim_{V\to\infty}\mathcal P_{\mathcal N_{u_0}}(u_0,V)$
exists, and
\begin{equation}
 C_\infty=B_{\mathcal S}-\int_{-\infty}^{u_0}F(u)\,\dd u.
 \label{eq:current-limit}
\end{equation}
\end{lemma}

\begin{proof}
We apply Stokes' theorem and include the horizon contribution to obtain
\[
 \mathcal Q_{\mathcal N_{u_0}}-\mathcal Q_{\mathcal S}
   =\int_{-V}^{u_0}j_u\,\dd u.
\]
Since $\psi=\phi/r$ and $\partial_u r=-D/2$,
\[
 j_u=-r\partial_u\phi-\frac D2\phi,
 \qquad
 \partial_u(r\phi)=r\partial_u\phi-\frac D2\phi,
\]
so $j_u+\partial_u(r\phi)=-D\phi$. This proves
\eqref{eq:capidentity}.

On the static end of $\mathcal S$, $r\phi=r^2f\to0$ and
$\mathcal Q_{\mathcal S}\to B_{\mathcal S}$. On the incoming segment $v=V$,
we have $-V\le u\le u_0=-A$. Lemma~\ref{lem:volterra} gives
$|D\phi(u,V)|\le C(-u)^{-2}$. For each fixed $u\le u_0$,
$D(r(u,V))\to1$ and $\phi(u,V)\to F(u)$.
We extend the integrand by zero for $u<-V$ and apply dominated
convergence to obtain \eqref{eq:current-limit}. In particular, the radiation integral
is absolutely convergent.
\end{proof}

\subsection{Identification with the time-inverted Newman--Penrose constant}

For any spherically symmetric wave $\chi$, the ordinary Newman--Penrose constant has
the normalization
\[
 I_0[\chi]=\lim_{r\to\infty}r^2\partial_r(r\chi)\big|_u.
\]
Proposition~\ref{prop:trace} gives $I_0[\psi]=0$ and a finite limit
of $r^3Q$ on the outgoing null portion of $\mathcal N_{u_0}$. The time-inversion construction of
\cite[Proposition~9.1]{AAG1} therefore supplies a regular spherically symmetric wave
$\Psi$ on its future domain satisfying
\begin{equation}
 T\Psi=\psi,\qquad \Boxg\Psi=0,
 \qquad \lim_{r\to\infty}\Psi|_{\mathcal N_{u_0}}=0.
 \label{eq:timeprimitive}
\end{equation}
The construction gives a regular time integral $\Psi$ on $\mathcal C_{u_0}$,
where $T$ is nonzero, and on the future domain of $\mathcal N_{u_0}$.

\begin{lemma}[The closed current and the time-inverted charge]
\label{lem:primitive-charge}
We write $L=L(u_0)=\lim_{v\to\infty}r^3Q(u_0,v)$. Then
\begin{equation}
 I_0[\Psi]=MC_\infty-2L.
 \label{eq:primitive-charge}
\end{equation}
\end{lemma}

\begin{proof}
We set $Z=r^2(\partial_v\Psi-\partial_u\Psi)$. The wave equation gives
\begin{equation}
 \dd Z=j[T\Psi]=j[\psi].
 \label{eq:primitive-current}
\end{equation}
Indeed, we differentiate $Z$ and use
$2r^2\Psi_{uv}=rD(\Psi_v-\Psi_u)$ to verify both components.
Regularity in ingoing coordinates gives $\Psi_u=0$ at the future
horizon and $\Psi_v=T\Psi=\psi$ there. Thus
$Z|_{\HH}=r_h^2\psi$ and $\mathcal Q_{\mathcal N_{u_0}}=Z$.
On its outgoing portion,
\begin{equation}
 \mathcal P_{\mathcal N_{u_0}}=Z+r^2\psi
                 =2r^2\partial_v\Psi
                 =Dr^2\partial_r\Psi\big|_u.
 \label{eq:Pprimitive}
\end{equation}

The radial equation for $\Psi$ is
\[
 \partial_r(Dr^2\partial_r\Psi)=2r\partial_r\phi.
\]
By Proposition~\ref{prop:trace} and its commuted estimates, for any fixed
$0<\beta<\min\{\delta_3,1\}$,
\[
 \partial_r\phi=2Lr^{-3}+O(r^{-3-\beta}),
 \qquad
 Dr^2\partial_r\Psi
   =C_\infty-\int_r^\infty2s\partial_s\phi\,\dd s
   =C_\infty-\frac{4L}{r}+O(r^{-1-\beta}).
\]
We use $D^{-1}=1+2M/r+O(r^{-2})$ and the normalization in
\eqref{eq:timeprimitive}, and integrate to obtain
\begin{equation}
 r\Psi=-C_\infty-\frac{MC_\infty-2L}{r}
                         +O(r^{-1-\beta}).
 \label{eq:primitive-expansion}
\end{equation}
The equation gives the differentiated expansion, so the defining limit
of $I_0[\Psi]$ is $MC_\infty-2L$.
\end{proof}

\begin{proposition}[Coefficient from the initial data for $p\ge3$]
\label{thm:coefficient}
For the initial data \eqref{eq:classification-data} with $p\ge3$,
we use the notation \eqref{eq:pge3-coefficients}--\eqref{eq:pge3-data}.
The time-inverted charge is
\begin{equation}
 I_0^{(1)}[\psi]=I_0[\Psi]
 =M B_{\mathcal S}-\frac{K_3}{4}
 =M B_{\mathcal S}+\frac{2a_3-b_3}{8}.
 \label{eq:coefficient-proof}
\end{equation}
Equivalently, in terms of the initial coefficients $a,b$,
\begin{equation}
 I_0^{(1)}[\psi]=
 \begin{cases}
  M B_{\mathcal S}+\dfrac{2a-b}{8},&p=3,\\[4pt]
  M B_{\mathcal S},&p>3.
 \end{cases}
 \label{eq:pge3-charge-cases}
\end{equation}
\end{proposition}

\begin{proof}
We set $J=\int_{-\infty}^{u_0}F(u)\,\dd u$. The two independently
computed limits, with the data representation
\eqref{eq:pge3-data}, are
\[
 C_\infty=B_{\mathcal S}-J,\qquad
 L=\frac{K_3}{8}-\frac M2J.
\]
We substitute these limits in Lemma~\ref{lem:primitive-charge} to obtain
the exact cancellation
\[
 I_0[\Psi]
 =M(B_{\mathcal S}-J)-2\left(\frac{K_3}{8}-\frac M2J\right)
 =M B_{\mathcal S}-\frac{K_3}{4}.
\]
This expresses the coefficient entirely in terms of the initial data on $\mathcal S$. At $p=3$, $(a_3,b_3)=(a,b)$ gives the
first line of \eqref{eq:pge3-charge-cases}. At $p>3$,
$a_3=b_3=K_3=0$, which gives its second line.
The static specialization follows from
\eqref{eq:currentstatic}. The horizon trace in
\eqref{eq:Qsigma} already accounts for either permitted horizon endpoint.
\end{proof}

\begin{remark}[Independence of the future hypersurface]
If we use another future hypersurface $\mathcal N_{u_1}$, the current
identity changes $C_\infty$ and $L$ so that $MC_\infty-2L$ has the same
value. The horizon term in \eqref{eq:Qsigma} accounts for the change
of the horizon sphere from $\Gamma_{u_0}$ to $\Gamma_{u_1}$.
\end{remark}

\section{Global leading-order asymptotics for \texorpdfstring{$p\ge3$}{p >= 3}}
\label{sec:application}

For every $p\ge3$, we prove the global leading-order expansion
with a quantitative remainder and the coefficient
$\calI=I_0^{(1)}[\psi]$ computed in
Proposition~\ref{thm:coefficient}. We first obtain the bounded-radius,
horizon, radiation-field, and full exterior estimates, and then
combine them into the uniform global profile in
Theorem~\ref{thm:pge3-global}. Throughout this section,
$\mathfrak N_{10,\delta_3}[\psi]$ denotes the expression
\eqref{eq:originalnorm} evaluated with
\[
 (a,b,\delta)\quad\hbox{replaced by}\quad(a_3,b_3,\delta_3),
\]
where \eqref{eq:pge3-coefficients} and
\eqref{eq:pge3-remainder-exponent} define these quantities.
Thus the norm measures the common representation
\eqref{eq:pge3-data} of the initial data on $\mathcal S$.
Lemma~\ref{lem:classification-data-norm} controls it by the
initial-data norm for the prescribed exponent $p$.

Section~\ref{sec:exterior}, applied to \eqref{eq:pge3-data}, supplies
the estimates on both the compact spacelike portion
$\mathcal C_{u_0}$ and the outgoing null portion of
$\mathcal N_{u_0}$. Here we use these bounds to control the required
weighted energies and the remainder in the asymptotic expansion
of the time integral.
These estimates verify the hypotheses of
\cite[Theorem~1.1]{AAG1}, with all constants controlled by the
initial data on $\mathcal S$. Section~\ref{sec:pge3-conclusion}
records these estimates for $p=3$ and $p>3$ together, and
Section~\ref{sec:pge3-global-profile} gives their single global form.

\subsection{Weighted energy bounds for the late-time asymptotics}

We use the flux $\mathcal E_N[\chi](u_0)$ defined in
Section~\ref{sec:analytic} and abbreviate it here by $\EN{\chi}$.
It is the sum of the spacelike flux on $\mathcal C_{u_0}$ and the null flux
on $\mathcal N_{u_0}\cap\{r\ge R\}$. For $0<\varepsilon<1$, we define
\begin{align}
 \mathcal A_\varepsilon[\psi]
 ={}&\sum_{j=0}^{8}\EN{T^j\psi}
      +\EN{N\psi}+\EN{N^2\psi}\notag\\
 &+\sum_{j=0}^{8}\int_{\mathcal N_{u_0}\cap\{r\ge R\}}
     \left(r^{5-\varepsilon}|\partial_rT^j\phi|^2
          +r^{7-\varepsilon}|\partial_r^2T^j\phi|^2\right)\,\dd r.
 \label{eq:majorenergy}
\end{align}
This deliberately stronger finite collection controls the spherically symmetric
energy required in \cite[Theorem~1.1]{AAG1}. More explicitly, the
definition of $E_{0,I_0=0;1}^{\varepsilon}$ in its Appendix A has ordinary
energies through $T^8\psi$, first radial derivatives with weights at most
$r^{5-\varepsilon}$, and second radial derivatives with weights at most
$r^{7-\varepsilon}$. \cite[Proposition~7.5]{AAG1} adds the needed $N$-commuted
energies; the theorem also assumes the $N^2$ energy. The angular-mode conditions are automatically satisfied for
spherically symmetric solutions.

By \eqref{eq:radialbounds}, the largest weighted integrands are
$O(r^{-1-\varepsilon})$, and hence integrable. For the ordinary null
energies,
\[
 \partial_vT^j\psi
  =r^{-1}\partial_vT^j\phi-\frac D{2r^2}T^j\phi=O(r^{-2}),
\]
so the flux integral with weight $r^2$ is finite. The field $N$ equals
$T$ on this portion. Lemma~\ref{lem:compact-transfer} controls the
compact energies. Consequently
\begin{equation}
 \mathcal A_\varepsilon[\psi]^{1/2}
     \le C_\varepsilon\mathfrak N_{10,\delta_3}[\psi].
 \label{eq:energy-norm-bound}
\end{equation}
The positive loss $\varepsilon$ makes the largest weighted
integrands integrable.

Proposition~\ref{prop:trace} also gives the exact noncompact trace
hypothesis of \cite[Theorem~1.1]{AAG1}:
\begin{equation}
 v^3\partial_v\phi(u_0,v)=C_0+O(v^{-\beta}),
 \qquad 0<\beta<\min\{\delta_3,1\},
 \label{eq:AAG-trace}
\end{equation}
where $|C_0|$ and the remainder constant are bounded by
$C\mathfrak N_{10,\delta_3}[\psi]$.

\subsection{Remainder estimates for the time integral}

The estimates in the proof of \cite[Theorem~1.1]{AAG1} include
a weighted remainder for the time integral. Equation~\eqref{eq:primitive-expansion} and its radial equation
give
\begin{align}
 \partial_r(r\Psi)&=\calI r^{-2}+O(r^{-2-\beta}),\notag\\
 \partial_r^2(r\Psi)&=-2\calI r^{-3}+O(r^{-3-\beta}).
 \label{eq:primitive-derivatives}
\end{align}
These estimates follow from the proved expansion of
$\partial_r\phi$ and the radial equation for $\Psi$.
We convert to $v$ on $u=u_0$ and decrease $\beta$ if necessary to
absorb $\log v/v$, obtaining
\begin{equation}
 \left|\partial_v\left(
       \partial_v(r\Psi)-\frac{2\calI}{v^2}\right)\right|
       \le C\mathfrak N_{10,\delta_3}[\psi]\cdot v^{-3-\beta}.
 \label{eq:primitive-P}
\end{equation}
For this norm alone, we use the shifted advanced parameter
$\bar v=v+c\ge1$ on $\mathcal N_{u_0}$, with $c$ fixed. The same
estimate holds with $\bar v$ in place of $v$: the leading limit is
unchanged and the remainder absorbs the shift because $\beta<1$.
Equation~\eqref{eq:primitive-P} controls the infinite-end part of
$P_{I_0,\beta;1}[\Psi]$ in \cite[eq.~(8.9)]{AAG1}.

We also control its compact part. On $\mathcal C_{u_0}$, we write
$v=s(r):=u_0+H(r)$ and use $\rho=r$ as the radial parameter. We define
\[
 C(r)=r_h^2\psi(\Gamma_{u_0})
           +\int_{\Gamma_{u_0}}^{r}j[\psi].
\]
The current identity \eqref{eq:primitive-current} and $T\Psi=\psi$ give
\begin{equation}
 Dr^2\partial_\rho\Psi
   =C(r)+(Ds'(r)-1)r^2\psi=:S(r),\qquad S(r_h)=0.
 \label{eq:compact-primitive}
\end{equation}
All coefficients here are smooth in $r$ on this cap. Since
$D=(r-r_h)d(r)$ with $d(r_h)>0$, the apparent division by $D$ is
regular:
\[
 \partial_\rho\Psi
  =\frac{\displaystyle\int_0^1
      S'(r_h+\theta(r-r_h))\,\dd\theta}{r^2d(r)}.
\]
This formula and its derivative bound the first two tangential
derivatives of $\Psi$ by finitely many cap derivatives of $\psi$.
The value of $\Psi$ at the joining radius is bounded by integrating
\eqref{eq:Pprimitive} from infinity; that integral converges and is
controlled by $C_\infty$ and the already proved bounds for
$\partial_r\phi$. Integration along the cap then bounds $\Psi$
itself, and $T\Psi=\psi$ controls its time derivatives. Thus the compact
part of $P_{I_0,\beta;1}[\Psi]$ is bounded by
$C\mathfrak N_{10,\delta_3}[\psi]$. The time-inverse energy estimates of
\cite[Section~9.3]{AAG1} bound the remaining time-inverse energies by
the energies controlled in \eqref{eq:energy-norm-bound}.

For completeness, $C_\infty$ is itself bounded by the initial data:
\eqref{eq:current-limit}, the trace bound $|F(u)|\le C|u|^{-2}$, and
\eqref{eq:geometricB} give
$|C_\infty|+|\calI|\le C\mathfrak N_{10,\delta_3}[\psi]$.
We combine these estimates to obtain
\begin{align}
 \mathcal A_\varepsilon[\psi]^{1/2}
 +|C_0|+\sup_{v\ge v_R}v^\beta|v^3Q-C_0|
 +|\calI|+P_{I_0,\beta;1}[\Psi]
 \le C\mathfrak N_{10,\delta_3}[\psi].
 \label{eq:quantitative}
\end{align}

\subsection{Bounded-radius, radiation-field, and exterior estimates}
\label{sec:pge3-conclusion}

We apply \cite[Theorem~1.1]{AAG1} in the future of $\mathcal N_{u_0}$.
For every $p\ge3$, the ordinary charge is zero, the outgoing trace
has the positive-power remainder \eqref{eq:AAG-trace}, and all
spherically symmetric weighted inputs are controlled by \eqref{eq:quantitative}.
Proposition~\ref{thm:coefficient} identifies the time-inverted charge
with $\calI$ in \eqref{eq:pge3-charge-cases}. We choose $\epsilon>0$
smaller than both $\beta$ and the geometric decay margin in that
theorem. For every fixed $R_1>r_h$ and sufficiently late times, we obtain
\begin{align}
 \sup_{r_h\le r\le R_1}
 \left|\psi(\tau,r)+8\calI(1+\tau)^{-3}\right|
 &\le C_{R_1}\mathfrak N_{10,\delta_3}[\psi]\cdot (1+\tau)^{-3-\epsilon},
 \label{eq:pge3-localtail}\\
 \left|F(u)+2\calI(1+u)^{-2}\right|
 &\le C\mathfrak N_{10,\delta_3}[\psi]\cdot (1+u)^{-2-\epsilon}.
 \label{eq:pge3-nulltail}
\end{align}
We obtain the second estimate by taking the radiation-field limit
in the exterior estimate of
\cite[Theorem~1.1 and Proposition~10.1]{AAG1}.
Regularity includes the future horizon. At fixed exterior radius
$\tau=t+\text{constant}$, and on the horizon
$\tau=v+\text{constant}$; these shifts preserve the leading coefficients.
At $p=3$, \eqref{eq:pge3-localtail}--\eqref{eq:pge3-nulltail}
are the estimates \eqref{eq:localtail}--\eqref{eq:nulltail}.

We also record the full exterior estimate from
\cite[Theorem~1.1, eq.~(1.34)]{AAG1}. For a fixed sufficiently
large $R$ and sufficiently late retarded time $u\ge u_1$, it gives
\begin{equation}
 \left|\psi(u,v)+
 \frac{4\calI}{(1+u)^2v}\left(1+\frac uv\right)\right|
 \le C\mathfrak N_{10,\delta_3}[\psi]\cdot (1+u)^{-2-\epsilon_0}v^{-1},
 \qquad r\ge R,
 \label{eq:pge3-exterior}
\end{equation}
for some $\epsilon_0>0$. Together with
\eqref{eq:pge3-localtail}, this controls the solution throughout
the late exterior. In Section~\ref{sec:pge3-global-profile}, we
express these estimates in the profile $\Pi_3(\tau,r)$.

For $p>3$, the representation \eqref{eq:pge3-data} has
$a_3=b_3=0$ and $\delta_3<p-3$. In the weighted norm, the
leading spatial terms are multiplied by $r^{3+\delta_3-p}$,
which is bounded at infinity; the faster remainder terms obey
the same bounds. Lemma~\ref{lem:classification-data-norm} gives
the resulting control by the initial-data norm. All preceding
estimates therefore apply with $\calI=MB_{\mathcal S}$. In particular,
the bounded-radius coefficient is $-8MB_{\mathcal S}$ and the
radiation-field coefficient is $-2MB_{\mathcal S}$, with the
quantitative remainders in
\eqref{eq:pge3-localtail}--\eqref{eq:pge3-exterior}.
The positive remainder exponents may depend on the decay margin
$p-3$ through $\delta_3$.

The finite norm in the statement can be checked directly. The largest
ordinary energy involves one derivative of $T^8\psi$. The stronger
radial terms in \eqref{eq:majorenergy} include
$\partial_r^2T^8\phi$. By \eqref{eq:commuteddata} and
\eqref{eq:secondv}, these use spatial-end derivatives of $f$ through
order ten and of $h$ through order nine. The definition
\eqref{eq:originalnorm} with $m=10$ contains these seminorms and also
the regular compact $H^{11}\times H^{10}$ norms needed for the energy
and derivative bounds on $\mathcal N_{u_0}\cap\mathcal J$
provided by Lemma~\ref{lem:compact-transfer}. This proves the
stated initial-data dependence of the estimates used below.

\subsection{The global expansion from the initial data for \texorpdfstring{$p\ge3$}{p >= 3}}
\label{sec:pge3-global-profile}

We now express the bounded-radius and exterior estimates in the
single profile $\Pi_3(\tau,r)$ defined in
\eqref{eq:classification-profile}. This gives the global conclusion
for every $p\ge3$ directly from the initial data on $\mathcal S$.

\begin{theorem}[Global expansion for $p\ge3$]
\label{thm:pge3-global}
We assume the geometric and energy hypotheses of
Section~\ref{sec:setting}. We choose an initial hypersurface
$\mathcal S$ as in Section~\ref{sec:hypersurfaces} and prescribe
smooth data satisfying, for fixed $p\ge3$ and $\delta>0$,
\[
 f=a r^{-p}+\Os[31](r^{-p-\delta}),\qquad
 h=b r^{-p-1}+\Os[30](r^{-p-1-\delta}),\qquad
 \mathfrak A[\psi]=\mathfrak A_{p,30,\delta}[\psi]<\infty.
\]
We set $\calI=I_0^{(1)}[\psi]$, with the initial-data expression
proved in Proposition~\ref{thm:coefficient}. Thus
\begin{equation}
 \mathcal C_p=-8\calI=
 \begin{cases}
 b-2a-8MB_{\mathcal S},&p=3,\\
 -8MB_{\mathcal S},&p>3,
 \end{cases}
 \label{eq:pge3-global-coefficient}
\end{equation}
where $B_{\mathcal S}$ is defined in \eqref{eq:geometricB}.
There are constants $C>0$, $\eta>0$, and $\tau_0\ge0$ such that
\begin{equation}
 \left|\psi(\tau,r)+8\calI\Pi_3(\tau,r)\right|
 \le C\mathfrak A[\psi]\cdot
       \frac{(1+\tau)^{-2-\eta}}{1+\tau+r},
 \qquad \tau\ge\tau_0,\quad r\ge r_h.
 \label{eq:pge3-global-profile}
\end{equation}
The estimate includes the future horizon, and the constants depend
only on the fixed parameters, background, and hypersurfaces.
Writing $\epsilon,\epsilon_0>0$ for the exponents in
\eqref{eq:pge3-localtail} and \eqref{eq:pge3-exterior}, we may take
\[
 \eta=\frac12\min\{\epsilon,\epsilon_0,1/2\}.
\]
The bounded-radius and radiation-field conclusions are
\eqref{eq:pge3-localtail} and \eqref{eq:pge3-nulltail}, with their
initial-data norm controlled by $\mathfrak A[\psi]$. When
$\calI=0$, \eqref{eq:pge3-global-profile} gives the corresponding
improved global remainder bound.
\end{theorem}

\begin{proof}
Lemma~\ref{lem:classification-data-norm} bounds
$\mathfrak N_{10,\delta_3}[\psi]$ by $C\mathfrak A[\psi]$.
Equation~\eqref{eq:quantitative} also gives
$|\calI|\le C\mathfrak A[\psi]$. We therefore use
\eqref{eq:pge3-localtail} and \eqref{eq:pge3-exterior} with
this same initial-data norm.

For $r\ge R$, we have $\tau=u$ and
$v=u+2r+O(\log r)$, so $v$ is comparable to $1+\tau+r$ at
late times. The exact identity
\[
 8\calI\Pi_3(\tau,r)
 =\frac{4\calI}{(1+\tau)^2}
   \left[\frac1{1+\tau+2r}
        +\frac{1+\tau}{(1+\tau+2r)^2}\right]
\]
allows us to compare the exterior model with this profile:
\begin{align*}
 \left|\frac{4\calI}{(1+\tau)^2v}
              \left(1+\frac{\tau}{v}\right)
          -8\calI\Pi_3(\tau,r)\right|
 &\le C|\calI|\frac{(1+\tau)^{-2}(1+\log(2+r))}
                         {(1+\tau+r)^2}\\
 &\le C_\eta\mathfrak A[\psi]\cdot
             \frac{(1+\tau)^{-2-\eta}}{1+\tau+r}.
\end{align*}
In the second inequality we use, for every fixed $0<\eta<1$,
\[
 \frac{1+\log(2+r)}{1+\tau+r}
       \le C_\eta(1+\tau)^{-\eta}.
\]
Together with \eqref{eq:pge3-exterior}, this proves
\eqref{eq:pge3-global-profile} for $r\ge R$.

On $r_h\le r\le R$, we use
$\Pi_3(\tau,r)=(1+\tau)^{-3}+O_R((1+\tau)^{-4})$
and \eqref{eq:pge3-localtail}. Since $1+\tau+r$ is comparable
to $1+\tau$ on this region and $\eta<\min\{\epsilon,1\}$,
we obtain the same bound through the future horizon. This
completes the global estimate for every $p\ge3$.
\end{proof}

\begin{remark}[Time and radial derivative estimates for $p\ge3$]
\label{rem:pge3-derivatives}
For each fixed $K\in\mathbb N_0$, we take a sufficiently large finite
derivative count $m_K$ for the commuted weighted estimates of
\cite{AAG1}. We assume the common representation
\eqref{eq:pge3-data} through these orders and
$\mathfrak N_{m_K,\delta_3}[\psi]<\infty$.
The derivative estimates and compact estimates in
Section~\ref{sec:exterior} control the corresponding weighted
initial-data norms. We use $Y$ from \eqref{eq:future-derivatives}
and the energy fluxes of Section~\ref{sec:analytic}.
The comparison argument in Section~\ref{sec:frac-decay-23}
transfers the cited spacelike derivative estimates to
$\mathcal N_\tau$ and includes the additional time derivative
needed to compare their energy fluxes.
For $0\le k\le K$, sufficiently small $\varepsilon>0$, sufficiently
late $\tau$, and every $r\ge r_h$, the cited estimates give
\begin{align*}
 |T^k\psi|(\tau,r)
 &\le C\mathfrak N_{m_K,\delta_3}[\psi]\cdot
       \frac{(1+\tau)^{-2-k+\varepsilon}}{1+\tau+r},\\
 |YT^k\psi|(\tau,r)+|NT^k\psi|(\tau,r)
 &\le C\mathfrak N_{m_K,\delta_3}[\psi]\cdot
       (1+\tau)^{-4-k+\varepsilon},\\
 \sqrt{1+r}\bigl(|YT^k\psi|+|NT^k\psi|\bigr)(\tau,r)
 &\le C\mathfrak N_{m_K,\delta_3}[\psi]\cdot
       (1+\tau)^{-7/2-k+\varepsilon},\\
 \mathcal E_N[T^k\psi](\tau)
 &\le C\mathfrak N_{m_K,\delta_3}[\psi]^2\cdot
       (1+\tau)^{-5-2k+2\varepsilon},\\
 \mathcal E_N[YT^k\psi](\tau)+\mathcal E_N[NT^k\psi](\tau)
 &\le C\mathfrak N_{m_K,\delta_3}[\psi]^2\cdot
       (1+\tau)^{-7-2k+2\varepsilon}.
\end{align*}
Here the derivative count is chosen for the specified $K$; the
leading-order expansion in Theorem~\ref{thm:pge3-global} uses
the fixed count $m=30$.
\end{remark}

\Needspace{12\baselineskip}
\section{Interpretation and examples for \texorpdfstring{$p=3$}{p = 3}}
\label{sec:interpretation}
\label{sec:examples}

\paragraph{The case \texorpdfstring{$p>3$}{p > 3}.}
For data satisfying $f=\Os[11](r^{-3-\varepsilon})$ and
$h=\Os[10](r^{-4-\varepsilon})$ for some $\varepsilon>0$,
Sections~\ref{sec:flux} and~\ref{sec:application} give the asymptotic
coefficient and its quantitative late-time estimates, with
\[
 \calI=MB_{\mathcal S}.
\]
This includes the noncompact classes $p>3$ and compactly supported data.

The remainder of this section concerns the $p=3$ data
\eqref{eq:critical}, with $m\ge10$. Here $a,b$ are the
coefficients of $r^{-3}$ in $f$ and $r^{-4}$ in $h$, respectively,
and the formula established in Section~\ref{sec:flux} becomes
\begin{equation}
 \calI=M B_{\mathcal S}+\mathcal I_\infty,
 \qquad \mathcal I_\infty=\frac{2a-b}{8}.
 \label{eq:charge-split}
\end{equation}
We interpret $\mathcal I_\infty$ through the incoming part of the
spatial tail and give a formal frequency-domain comparison with its
late-time coefficients. The final example explains why compactly
supported approximations can fail to capture this contribution.
The coefficient formula and quantitative asymptotics used here are
proved in Sections~\ref{sec:flux} and~\ref{sec:application}.

\subsection{The incoming and outgoing parts of the spatial tail}
\label{sec:incoming-profile}

We explain why the infinity contribution involves the combination
$2a-b$. On the static end of the initial hypersurface, the data
\eqref{eq:critical} give
\[
 \phi(0,r_*)=a r_*^{-2}+o(r_*^{-2}),\qquad
 T\phi(0,r_*)=b r_*^{-3}+o(r_*^{-3}),\qquad \phi=r\psi,
\]
since $r_*/r\to1$. To distinguish the two propagation directions,
we use the free equation
$(\partial_t^2-\partial_{r_*}^2)\phi_{\mathrm{free}}=0$, whose solutions have the form
\[
 \phi_{\mathrm{free}}=F_{\mathrm{out}}(u)+G_{\mathrm{in}}(v),
 \qquad u=t-r_*,\quad v=t+r_*.
\]
The first profile travels towards larger radii and the second towards
smaller radii. For the same data on the static end, we normalize the profiles to vanish at
their respective ends. Their leading terms are
\[
 \begin{aligned}
 F_{\mathrm{out}}(u)&=C_{\mathrm{out}}(-u)^{-2}+o((-u)^{-2})
       &&(u\to-\infty),\\
 G_{\mathrm{in}}(v)&=C_{\mathrm{in}}v^{-2}+o(v^{-2})
       &&(v\to\infty),
 \end{aligned}
\]
with differentiated remainders inherited from \eqref{eq:critical}.
At $t=0$, we have $-u=v=r_*$. Comparing the position and
time-derivative data gives
\[
 a=C_{\mathrm{out}}+C_{\mathrm{in}},\qquad
 b=2(C_{\mathrm{out}}-C_{\mathrm{in}}).
\]
Consequently,
\begin{equation}
 C_{\mathrm{in}}=\frac{2a-b}{4},\qquad
 C_{\mathrm{out}}=\frac{2a+b}{4},\qquad
 \boxed{\mathcal I_\infty=\frac12 C_{\mathrm{in}}.}
 \label{eq:charge-incoming}
\end{equation}
Thus the additional infinity term selects the leading incoming
component of the initial tail.

An incoming null ray starting at $r_*=R_*$ on $t=0$ has $v=R_*$.
It reaches an observer at $r_*=r_{*,\mathrm{obs}}$ at time
$t=R_*-r_{*,\mathrm{obs}}$. Incoming data initially farther away
therefore arrive later, with arbitrarily late arrivals as $R_*\to\infty$.
The outgoing ray from the same initial point has $u=-R_*$ and reaches
null infinity at an increasingly early retarded time. This explains
the causal role of the incoming component. During the black hole
evolution, the potential scatters these components; the full
coefficient \eqref{eq:charge-split} also includes $MB_{\mathcal S}$.

\subsection{The zero-frequency normalization on the black hole exterior}
\label{sec:zero-frequency}

We compute the leading response to a small-frequency incoming mode
and give its formal relation to the two late-time coefficients
associated with $\mathcal I_\infty$.

\begin{proposition}[Zero-frequency response of a spherically symmetric mode]
\label{prop:zero-frequency}
For $\omega>0$, we consider solutions $w_\pm(r_*,\omega)$ of
$w''+(\omega^2-V)w=0$, where primes denote $r_*$-derivatives and
$V$ is defined in \eqref{eq:potential-definition}. We normalize them by
\[
 w_-(r_*,\omega)\sim e^{-i\omega r_*}\quad(r_*\to-\infty),
 \qquad
 w_+(r_*,\omega)\sim e^{i\omega r_*}\quad(r_*\to\infty),
\]
including the corresponding first-derivative asymptotics. The mode
with unit incoming amplitude at infinity is
\begin{equation}
 z_\omega=\overline{w_+}+\mathcal R(\omega)w_+
              =\mathcal T(\omega)w_-,
 \label{eq:mode-scattering}
\end{equation}
where $\mathcal R$ and $\mathcal T$ are the reflection and transmission
coefficients. As $\omega\to0^+$,
\begin{equation}
 \mathcal R(\omega)\to-1,\qquad
 \mathcal T(\omega)=-2i\omega r_h+o(\omega),\qquad
 \frac{z_\omega(r)}r=-2i\omega+o(\omega).
 \label{eq:zero-frequency-response}
\end{equation}
The last estimate is uniform on compact subsets of $r_h<r<\infty$.
\end{proposition}

\begin{proof}
The potential bounds in Section~\ref{sec:setting} give
$\int_{\mathbb R}(1+|r_*|)|V(r_*)|\,\dd r_*<\infty$.
The integral equations for the normalized modes have sine kernels
$\sin(\omega s)/\omega$, bounded by $|s|$. Successive iteration gives
uniformly convergent integral series near $\omega=0$, and hence
continuous limits for $w_\pm$ and their first $r_*$-derivatives on
compact intervals.
Since $V=r''/r$, these limits are explicitly
\begin{equation}
 w_-(r_*,0)=\frac r{r_h},\qquad
 w_+(r_*,0)=r\int_r^\infty\frac{\dd s}{D(s)s^2}.
 \label{eq:zero-solutions}
\end{equation}
We define $W(f,g)=fg'-f'g$. Direct differentiation gives
$W(w_-,w_+)(0)=-1/r_h$, while
$W(\overline{w_+},w_+)=2i\omega$. Hence
\[
 \mathcal T(\omega)=\frac{2i\omega}{W(w_-,w_+)},\qquad
 \mathcal R(\omega)
 =-\frac{W(w_-,\overline{w_+})}{W(w_-,w_+)}.
\]
Continuity at zero gives the first two limits in
\eqref{eq:zero-frequency-response}; the last follows from
$z_\omega/r=\mathcal T w_-/r$ and $w_-(r_*,0)/r=1/r_h$.
\end{proof}

For the corresponding low-energy absorption, see \cite{DGM}.

\paragraph{Formal relation to the late-time coefficients.}
For the convention $e^{-i\omega t}$, the multiplier $-2i\omega$
corresponds to $2\partial_t$, while the reflection limit $-1$
corresponds to a change of sign. Keeping these leading multipliers
formally associates the incoming profile $G_{\mathrm{in}}$ with
$2G_{\mathrm{in}}'(t)$ at bounded radius and with
$-G_{\mathrm{in}}(u)$ for the radiation field. Using its
differentiated expansion from Section~\ref{sec:incoming-profile}, we obtain
\begin{align}
 2G_{\mathrm{in}}'(t)
 &=-4C_{\mathrm{in}}t^{-3}+o(t^{-3})
   =-8\mathcal I_\infty t^{-3}+o(t^{-3}),
 \label{eq:formal-infinity-timelike}\\
 -G_{\mathrm{in}}(u)
 &=-C_{\mathrm{in}}u^{-2}+o(u^{-2})
   =-2\mathcal I_\infty u^{-2}+o(u^{-2}).
 \label{eq:formal-infinity-radiation}
\end{align}
These are precisely the factors multiplying the additional infinity
term in \eqref{eq:pge3-localtail} and \eqref{eq:pge3-nulltail}.
We use this frequency calculation as a formal interpretation.
A frequency-domain proof would also require matching the prescribed
initial data to the scattering representation and estimating the
inverse transforms of the frequency remainders. The rigorous
identification of the full coefficient $MB_{\mathcal S}+\mathcal I_\infty$
from the initial data, together with the quantitative late-time
remainders, is supplied by Sections~\ref{sec:flux} and~\ref{sec:application}.

\subsection{Failure of compactly supported approximations to capture infinity contribution}
\label{sec:cutoff-example}

The following example isolates the infinity contribution completely. It
also explains why an approximation by compact data can lose the leading
coefficient even when it converges in the ordinary energy.

\begin{proposition}[Compact truncations with a different late-time coefficient]
\label{prop:cutoff}
On the static hypersurface $\{t=0\}$ of
Section~\ref{sec:hypersurfaces}, we choose a smooth function $\chi$
which vanishes near $r_h$ and equals one for $r\ge R_0$. We fix $a\ne0$,
and we prescribe
\begin{equation}
 f(r)=a\chi(r)r^{-3},\qquad h(r)=0.
 \label{eq:critical-example}
\end{equation}
We choose a smooth $\vartheta:[0,+\infty)\rightarrow \mathbb{R}$ equal to one on $[0,1]$ and zero on $[2,\infty)$.
For $R>R_0$, we denote by $\psi_R$ the solution with data
$f_R(r)=\vartheta(r/R)f(r)$ and $h_R=0$, and by $\psi$ the solution with data
\eqref{eq:critical-example}. Then
\begin{equation}
 \mathcal I[\psi]=\frac a4,
 \qquad
 \mathcal I[\psi_R]=0\quad\hbox{for every }R.
 \label{eq:cutoff-charges}
\end{equation}
For every fixed $r>r_h$,
\begin{equation}
 \boxed{
 \lim_{R\to\infty}\lim_{t\to\infty}t^3\psi_R(t,r)=0,
 \qquad
 \lim_{t\to\infty}\lim_{R\to\infty}t^3\psi_R(t,r)=-2a.
 }
 \label{eq:noncommuting-limits}
\end{equation}
The initial-data difference satisfies
\begin{equation}
 E_T[f-f_R,0]=O(a^2R^{-5})
 \label{eq:cutoff-energy}
\end{equation}
in the ordinary Killing energy $E_T$ on $\{t=0\}$, defined by
the integral in the proof below.
\end{proposition}

\begin{proof}
The horizon value $f(r_h)$ and the velocity datum $h$ vanish, so
$B_{\{t=0\}}=0$.
The coefficients at orders $r^{-3}$ and $r^{-4}$ are $(a,0)$ for
the untruncated initial datum and $(0,0)$ for every compact truncation.
Proposition~\ref{thm:coefficient}
gives \eqref{eq:cutoff-charges}, and \eqref{eq:pge3-localtail} gives
\[
 \lim_{t\to\infty}t^3\psi(t,r)=-2a,
 \qquad
 \lim_{t\to\infty}t^3\psi_R(t,r)=0.
\]
On every fixed compact spacetime set in the future of the initial slice,
finite propagation gives $\psi_R=\psi$ for all sufficiently large $R$.
For a fixed observation radius, the two solutions agree at least until
$t<r_*(R)-r_*(r)$, when $R$ is large enough. This proves the second
iterated limit in \eqref{eq:noncommuting-limits}; the first follows by
taking the late-time limit at each fixed $R$.

Finally, $f-f_R$ is supported in $r\ge R$, with
$|\partial_r(f-f_R)|\le C|a|r^{-4}$. Since $D\asymp1$ there,
\[
 E_T[f-f_R,0]
 =2\pi\int_R^\infty Dr^2|\partial_r(f-f_R)|^2\,\dd r
 \le Ca^2\int_R^\infty r^{-6}\,\dd r
 \le Ca^2R^{-5}.
\]
\end{proof}

For the untruncated initial datum, \eqref{eq:pge3-nulltail} gives
$\lim_{u\to\infty}u^2F(u)=-a/2$ for the radiation field. Thus its leading late-time signal
is entirely supplied by the infinity term. Each compact truncation has vanishing $t^{-3}$ coefficient.

The limit $\lim_{r\to\infty}r^3f(r)$ is $a$ for the untruncated initial datum
and zero for every compact truncation. A useful approximation within
the $p=3$ class keeps $a,b$ fixed and truncates only the faster-decaying
remainders. For any fixed $0<\delta'<\delta$, these truncations converge
in $\mathfrak N_{m,\delta'}$ from \eqref{eq:originalnorm}: their error
is $O(R^{-(\delta-\delta')})$ in that norm, since the cutoff derivatives
satisfy $|\partial_r^j\vartheta(r/R)|\le C_jR^{-j}$.
This retains the prescribed spatial tail and controls the remaining
contribution to the asymptotic coefficient.

\Needspace{12\baselineskip}

\section{Global leading-order asymptotics for \texorpdfstring{$1<p<3$}{1 < p < 3}}
\label{sec:frac-global}

We divide the argument into two parts. First, we obtain global bounds
for the field $\psi$ and its derivatives. For $2<p<3$, we split the
decaying part of $(r\psi)|_{\mathcal N_{u_0}}$ into radial annuli and combine
decay estimates with powers $2$ and $3$. For $1<p<2$, we apply that
result to $T\psi$ and integrate in time; at $p=2$, we use the estimate
with power $2$ directly.

We then determine the leading term by a common argument for all
$1<p<3$. Section~\ref{sec:exterior} has already identified
$C_p/2$, with $C_p=b+(1-p)a$, as the coefficient of $v^{-p}$
in $Q(0,v)$, where $Q=\partial_v(r\psi)$, on
$\mathcal N_0$. The global bounds control the potential term in the
equation for $Q$, allowing us to propagate this coefficient to late
times. Radial integration towards the horizon and outwards to
arbitrary finite radii gives one global leading profile with a
uniform remainder. Its limit at null infinity gives the
radiation-field asymptotics.
Section~\ref{sec:frac-spatial-global} states the resulting theorem
in terms of the initial data on $\mathcal S$.

\subsection{Statement of the global decay bounds for
\texorpdfstring{$1<p<3$}{1 < p < 3}}
\label{sec:frac-decay-statement}

We use the future hypersurfaces $\mathcal N_\tau$ and the fields
$T=\partial_\tau|_r$, $Y=\partial_r|_\tau$ fixed in
Section~\ref{sec:hypersurfaces}. On the outgoing null portion,
$\tau=u$ and $Y=\partial_r|_u$. The stationary field $N$ of
Section~\ref{sec:analytic} is regular at the horizon and equals $T$
for large $r$; $\mathcal E_N[w](\tau)$ is its energy flux for $w$
through $\mathcal N_\tau$. The constants below may depend on this
fixed family of hypersurfaces.

The proposition supplies the bounds for $\psi$ and $Y\psi$ used
in Section~\ref{sec:frac-asymptotics}. It also gives estimates
for $T\psi$, higher time derivatives, and energies.

\begin{proposition}[Global decay bounds with a small loss for $1<p<3$]
\label{prop:frac-global}
We assume the nondegenerate geometric and energy assumptions of
Section~\ref{sec:setting}, and consider a spherically symmetric wave arising from
smooth data on the chosen initial hypersurface $\mathcal S$ as above.  In particular, these hypotheses
hold on Schwarzschild.  We fix $K\in\mathbb N_0$, set $m=3K+20$,
and assume $\mathfrak D_{p,m}[\psi]<\infty$, where $1<p<3$.
For every sufficiently small $\varepsilon>0$ there is a constant
$C=C(D,\mathcal S,p,K,\varepsilon)>0$ such that, for $0\le k\le K$,
\begin{align}
 |T^k\psi|(\tau,r)
 &\le C\mathfrak D_{p,m}[\psi]\cdot 
 \frac{(1+\tau)^{1-p-k+\varepsilon}}{1+\tau+r},
 \label{eq:frac-global-field}\\
 |YT^k\psi|(\tau,r)+|NT^k\psi|(\tau,r)
 &\le C\mathfrak D_{p,m}[\psi]\cdot (1+\tau)^{-p-k-1+\varepsilon},
 \label{eq:frac-global-radial}\\
 \sqrt{1+r}\bigl(|YT^k\psi|+|NT^k\psi|\bigr)(\tau,r)
 &\le C\mathfrak D_{p,m}[\psi]\cdot (1+\tau)^{-p-k-1/2+\varepsilon}.
 \label{eq:frac-global-radial-weighted}
\end{align}
These bounds hold for every $r\ge r_h$ on sufficiently late
hypersurfaces $\mathcal N_\tau$,
including the future event horizon. We also have
\begin{align}
 \mathcal E_N[T^k\psi](\tau)
 &\le C\mathfrak D_{p,m}[\psi]^2\cdot  (1+\tau)^{1-2p-2k+2\varepsilon},
 \label{eq:frac-global-energy}\\
 \mathcal E_N[YT^k\psi](\tau)
 +\mathcal E_N[NT^k\psi](\tau)
 &\le C\mathfrak D_{p,m}[\psi]^2\cdot  (1+\tau)^{-1-2p-2k+2\varepsilon}.
 \label{eq:frac-global-radial-energy}
\end{align}
The radiation field $F(u):=\lim_{r\to\infty}r\psi(u,r)$,
with $u$ held fixed in this limit, satisfies
\begin{equation}
 |\partial_u^k F(u)|\le
 C\mathfrak D_{p,m}[\psi]\cdot (1+u)^{1-p-k+\varepsilon}.
 \label{eq:frac-global-radiation}
\end{equation}
For example, it suffices to take
$0<\varepsilon<\min\{(p-1)/4,1/8\}$; a smaller loss can always be
used.  The constants may depend on the fixed exponent $p$.
\end{proposition}

We prove the proposition for $2<p<3$ in
Section~\ref{sec:frac-decay-23} and for $1<p\le2$ in
Section~\ref{sec:frac-decay-12}. The positive loss $\varepsilon$
can be chosen as small as required by the subsequent asymptotic
argument.
\subsection{Proof of the decay bounds for \texorpdfstring{$2<p<3$}{2 < p < 3}}
\label{sec:frac-decay-23}

We work with the evolved data on the early hypersurface
$\mathcal N_{u_0}$ supplied by Section~\ref{sec:exterior}. On its
outgoing null portion, we write
$\phi_0=(r\psi)|_{\mathcal N_{u_0}}$ and
$F(u_0)=\lim_{r\to\infty}\phi_0(r)$. The transferred estimates give
\begin{equation}
 \phi_0(r)=F(u_0)+g(r),\qquad
 |\partial_r^j g(r)|\le A r^{1-p-j}
 \quad(0\le j\le 3K+16),
 \label{eq:frac-null-symbol}
\end{equation}
where $A$ also bounds $|F(u_0)|$ and the required Sobolev and
square-root energy norms on the compact spacelike portion
$\mathcal C_{u_0}$. We decompose the decaying function $g$ into
pieces supported on radial annuli. On an annulus of radius $R$,
each weight $r^\ell$ is comparable to $R^\ell$; this will make
explicit how the decay estimates depend on the location of each
piece of the data. The estimate with power $3$ gives faster time
decay at the cost of larger radial weights, while the estimate
with power $2$ requires weaker weights. We apply both to each
annular solution and choose the smaller bound at the observation time.
Related radial cut-off and interpolation arguments, converting spatial
weights into time decay, are used in
\cite[Section~10.2]{GajicInvSquare}.

\paragraph{Step 1: The decay estimates with powers $2$ and $3$.}
The norms in this step are evaluated on $\mathcal N_{u_0}$;
on its outgoing null portion, $u=u_0$ and $\partial_r$ keeps $u$
fixed. The elapsed time is $\tau-u_0$, and replacing
$1+\tau-u_0$ by $1+\tau$ for $\tau\ge0$ changes only constants
because $u_0$ is fixed.
For a spherically symmetric wave $w$, we put $\phi=rw$.
The index $\sigma=2,3$ specifies the bounded-radius decay power
$(1+\tau)^{-\sigma+\varepsilon}$. We denote
by $E_{\sigma;j}^{\varepsilon}$ respectively the norms denoted by
$E_{0,I_0\ne0;j}^{\varepsilon}$ and
$E_{0,I_0=0;j}^{\varepsilon}$ in \cite[Appendix~A]{AAG1}, and set
\begin{equation}
 \mathbb A_{\sigma,K}^{\varepsilon}[w]
 =\sum_{j=0}^{K+2}E_{\sigma;j}^{\varepsilon}[w]
 +\sum_{j=0}^{K+1}\mathcal E_N[NT^jw](u_0).
 \label{eq:frac-endpoint-norm}
\end{equation}
The subscript $I_0\ne0$ in the first norm is its published name;
the estimates of \cite[Propositions~5.1--5.2]{AAG1} require its
finiteness and apply also when the charge vanishes.

We first explain the transfer of the radial estimates from the
spacelike hypersurfaces $\widehat{\mathcal N}_s$ defined in
Section~\ref{sec:hypersurfaces} to $\mathcal N_\tau$.
We write $\widehat{\mathcal E}_N[\chi](s)$ for the flux through
$\widehat{\mathcal N}_s$. A point $(\tau,r)$ belongs to
$\widehat{\mathcal N}_{\tau+\Delta H(r)}$, with $\Delta H$ bounded.
For some fixed $c>0$, the one-dimensional Sobolev inequality in
$s$, applied at each fixed $r$ to the first derivatives of $\chi$,
therefore gives, for sufficiently late $\tau$,
\begin{equation}
 \mathcal E_N[\chi](\tau)
 \le C\int_{\tau-c}^{\tau+c}
 \left(\widehat{\mathcal E}_N[\chi](s)
       +\widehat{\mathcal E}_N[T\chi](s)\right)\,ds.
 \label{eq:spacelike-null-energy-comparison}
\end{equation}
Here $\chi$ may be any smooth spherically symmetric scalar with
the displayed finite energies. To see the uniformity at infinity,
we suppress the common angular factor. For large $r$, where $N=T$,
the spacelike flux density is
\[
 r^2\left(\frac D2|\widehat Y\chi|^2+
          \left(\zeta-\frac D2\zeta^2\right)|T\chi|^2\right)\,dr,
\]
while the null flux density is $\frac D2r^2|Y\chi|^2\,dr$.
The identity $Y=\widehat Y+\zeta T$ and $\zeta=r^{-2}$ give
the required comparison of these densities. On a fixed compact
region the spacelike nondegenerate energy controls every regular
first derivative. Integrating the time-Sobolev bound in $r$
proves \eqref{eq:spacelike-null-energy-comparison}.

The pointwise bounds transfer using the same bounded time shift
and $Y=\widehat Y+\zeta T$. For the derivative energies we also use
\[
 \widehat{\mathcal E}_N[\zeta T^{k+1}w](s)
 \le C\widehat{\mathcal E}_N[T^{k+1}w](s).
\]
The terms containing $\zeta'$ are controlled by the radial Hardy
inequality, since $T^{k+1}w$ tends to zero at infinity.
The ordinary energy of $T^{k+1}w$ has the same decay power as
the energy of $\widehat YT^kw$. The extra time derivative in
\eqref{eq:spacelike-null-energy-comparison} requires the estimates
for $\widehat YT^{k+1}w$ and $T^{k+2}w$, whose energies decay faster.
The norm index $K+2$ and the initial energies of $NT^jw$ for
$j\le K+1$ in \eqref{eq:frac-endpoint-norm} supply these estimates.
Thus this comparison preserves the rates and the derivative counts
used below.

For $0\le k\le K$, the estimates of
\cite[Propositions~6.1--6.2 and~7.5, Corollary~7.6,
Proposition~7.7]{AAG1} give
\begin{align}
 |T^kw|&\le C\sqrt{\mathbb A_{\sigma,K}^{\varepsilon}[w]}\cdot
 (1+\tau)^{-\sigma-k+\varepsilon},
 \label{eq:frac-endpoint-field}\\
 (1+r)|T^kw|&\le
 C\sqrt{\mathbb A_{\sigma,K}^{\varepsilon}[w]}\cdot
 (1+\tau)^{1-\sigma-k+\varepsilon},
 \label{eq:frac-endpoint-radiation}\\
 |YT^kw|+|NT^kw|&\le
 C\sqrt{\mathbb A_{\sigma,K}^{\varepsilon}[w]}\cdot
 (1+\tau)^{-\sigma-k-1+\varepsilon},
 \label{eq:frac-endpoint-radial}\\
 \sqrt{1+r}(|YT^kw|+|NT^kw|)&\le
 C\sqrt{\mathbb A_{\sigma,K}^{\varepsilon}[w]}\cdot
 (1+\tau)^{-\sigma-k-1/2+\varepsilon}.
 \label{eq:frac-endpoint-radial-weighted}
\end{align}
The square-root fluxes
$\sqrt{\mathcal E_N[T^kw](\tau)}$ and
$\sqrt{\mathcal E_N[YT^kw](\tau)+\mathcal E_N[NT^kw](\tau)}$
obey the same norm bound with powers
$(1+\tau)^{1/2-\sigma-k+\varepsilon}$ and
$(1+\tau)^{-1/2-\sigma-k+\varepsilon}$, respectively.
The $\sigma=3$ estimates require $I_0[w]=0$.
The weighted energy estimates underlying these conclusions are
proved in \cite{AAGDecay}; they extend the $r$-weighted energy
method of Dafermos--Rodnianski \cite{DRrp}.

\paragraph{Step 2: The norms of data supported on one annulus.}
We enlarge $R_0$ beyond the compact spacelike portion
$\mathcal C_{u_0}$, keeping $\mathcal N_{u_0}$ fixed.
The constant $A_R$ below bounds the amplitude of the reduced null
datum $\phi_R=rw_R$ on the annulus $R/2\le r\le2R$ and the
correspondingly scaled radial derivatives.

\begin{lemma}[Norms of data supported on one radial annulus]
\label{lem:frac-annular-norms}
We let $R\ge2R_0$. We prescribe zero data on
$\mathcal C_{u_0}$ and reduced null data $\phi_R=rw_R$
supported in $[R/2,2R]$, with
\begin{equation}
 |\partial_r^j\phi_R|\le A_R R^{-j}
 \qquad(0\le j\le 3K+16).
 \label{eq:frac-annular-data}
\end{equation}
The annulus is separated from $\mathcal C_{u_0}$, so the data agree
smoothly where the spacelike and null portions meet.
Then $I_0[w_R]=0$ and
\begin{equation}
 \sqrt{\mathbb A_{2,K}^{\varepsilon}[w_R]}
 \le C A_R R,
 \qquad
 \sqrt{\mathbb A_{3,K}^{\varepsilon}[w_R]}
 \le C A_R R^2.
 \label{eq:frac-annular-norm-scaling}
\end{equation}
The constant is independent of $R$ and $A_R$.
\end{lemma}

\begin{proof}
We obtain the time derivatives of the null data from the wave
equation. On the outgoing portion of $\mathcal N_{u_0}$ it gives
\begin{equation}
 2\partial_r T^j\phi_R
 =\partial_r(D\partial_rT^{j-1}\phi_R)
 -\frac{D'}r T^{j-1}\phi_R,
 \qquad j\ge1.
 \label{eq:frac-null-recurrence}
\end{equation}
We integrate from the sphere
$\partial\mathcal C_{u_0}\setminus\Gamma_{u_0}$, where all the
integration constants vanish because the data are zero nearby.
Since $D'/r=O(r^{-3})$, induction using
\eqref{eq:frac-annular-data} gives, for the derivative orders
entering the norms,
\begin{align}
 |T^j\phi_R|&\le C_j A_R,
 \label{eq:frac-annular-T-value}\\
 |\partial_r^a T^j\phi_R|
 &\le C_{a,j}A_R\left(
 r^{-a-2}+R^{-a-j}\mathbf1_{[R/2,2R]}\right),
 \qquad a,j\ge1.
 \label{eq:frac-annular-mixed}
\end{align}
Here $\mathbf1_{[R/2,2R]}$ is the indicator of the annulus.
We differentiate \eqref{eq:frac-null-recurrence} $a-1$ times to
obtain the second estimate and integrate its case $a=1$ to obtain
the first. The term $r^{-a-2}$ also controls the derivatives of
$T^j\phi_R$ beyond the annulus. The same estimates hold when the
support is enlarged by a fixed factor. For $j=0$, the radial
derivatives are supported in the annulus and satisfy
\eqref{eq:frac-annular-data}.

The largest radial weights in
$\mathbb A_{\sigma,K}^{\varepsilon}[w_R]$ occur in terms of the forms
\[
 \int_{\mathcal N_{u_0}\setminus\mathcal C_{u_0}}r^{2\sigma-1-\varepsilon}
 |\partial_rT^j\phi_R|^2\,dr,
 \qquad
 \int_{\mathcal N_{u_0}\setminus\mathcal C_{u_0}}r^{2\sigma-1+2a-\varepsilon}
 |\partial_r^{1+a}\phi_R|^2\,dr.
\]
On the annulus, $r\asymp R$ and its radial length is comparable to
$R$, so both integrals are bounded by $C A_R^2 R^{2\sigma-2}$.
The time-commuted terms satisfy the same bound by
\eqref{eq:frac-annular-mixed}. Away from the annulus, that estimate
bounds the largest $\sigma=3$ first-derivative integrand by
$C A_R^2 r^{-1-\varepsilon}$, which is integrable. The other terms
have smaller weights and obey the same bound. The ordinary null
energy is bounded by $C A_R^2$, using
\[
 \partial_r(T^jw_R)
 =r^{-1}\partial_rT^j\phi_R-r^{-2}T^j\phi_R.
\]
All terms on $\mathcal C_{u_0}$, including the redshift
commutations, vanish. Finally, $\partial_r\phi_R$ has compact
support, so its Newman--Penrose limit is zero. Taking square
roots gives \eqref{eq:frac-annular-norm-scaling}.
\end{proof}

\paragraph{Step 3: Decomposing the decaying null datum $g$.}
We choose a smooth partition $\sum_j\chi_j=1$ for large $r$, with
$\chi_j$ supported in $[R_j/2,2R_j]$, where
$R_j=2^jR_0$ for $j=1,2,\ldots$, and
$|\partial_r^a\chi_j|\le C_aR_j^{-a}$.
Thus the successive annuli have radii that double, and overlap
only a bounded number of times.

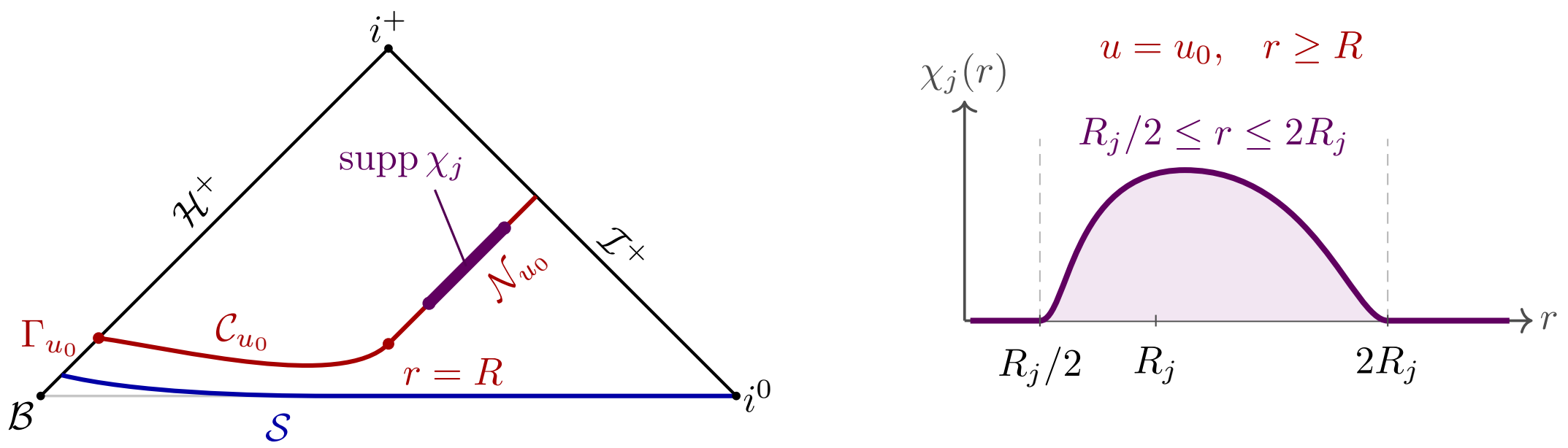
\begin{figure}[H]
\centering
\begin{tikzpicture}[x=1.05cm,y=1.05cm,
 every node/.style={font=\small,inner sep=1.5pt},
 boundary/.style={line width=.8pt},
 initial/.style={draw=blue!65!black,line width=1.15pt},
 future/.style={draw=red!65!black,line width=1.15pt},
 annulus/.style={draw=violet!75!black,line width=3.0pt}]
 \coordinate (B) at (0,0);
 \coordinate (IZ) at (6,0);
 \coordinate (IP) at (3,3);
 \coordinate (SH) at (.18,.18);
 \coordinate (G) at (.50,.50);
 \coordinate (J) at (3.00,.45);
 \coordinate (F) at (4.275,1.725);
 \coordinate (A) at (3.35,.80);
 \coordinate (E) at (4.00,1.45);
 \draw[boundary] (B)--(IP)--(IZ);
 \draw[gray!45,line width=.65pt] (B)--(IZ);
 \draw[initial] (SH)
   .. controls (.80,.03) and (1.80,0) .. (2.50,0)--(IZ);
 \draw[future] (G)
   .. controls (1.20,.40) and (2.60,.05) .. (J)--(F);
 \draw[annulus] (A)--(E);
 \fill (B) circle(1.1pt) (IZ) circle(1.1pt) (IP) circle(1.1pt);
 \fill[red!65!black] (G) circle(1.5pt) (J) circle(1.5pt);
 \fill[violet!75!black] (A) circle(1.7pt) (E) circle(1.7pt);
 \node[below left] at (B) {$\mathcal B$};
 \node[right] at (IZ) {$i^0$};
 \node[above] at (IP) {$i^+$};
 \node[rotate=45] at (1.35,1.70) {$\mathcal H^+$};
 \node[rotate=-45] at (5.04,1.27) {$\mathcal I^+$};
 \node[left=3pt,red!65!black] at (G) {$\Gamma_{u_0}$};
 \node[blue!65!black,below=3pt] at (2.05,0) {$\mathcal S$};
 \node[red!65!black] at (1.75,.55) {$\mathcal C_{u_0}$};
 \node[red!65!black,below right=2pt] at (J) {$r=R$};
 \node[red!65!black,rotate=45,below=4pt] at (3.86,1.31)
      {$\mathcal N_{u_0}$};
 \node[violet!75!black] at (3.12,1.98) {$\operatorname{supp}\chi_j$};
 \draw[violet!65!black,line width=.6pt]
      (3.40,1.78)--(3.67,1.12);
 \node at (3.00,-.72) {(a) An annulus on $\mathcal N_{u_0}$};

 \begin{scope}[xshift=8cm]
  \node[red!65!black] at (2.65,3.00) {$u=u_0$,\quad $r\ge R$};
  \draw[->,black!70,line width=.7pt] (.35,.65)--(5.25,.65)
       node[right] {$r$};
  \draw[->,black!70,line width=.7pt] (.35,.65)--(.35,2.55)
       node[above] {$\chi_j(r)$};
  \draw[gray!55,densely dashed] (1.00,.65)--(1.00,2.22);
  \draw[gray!55,densely dashed] (4.00,.65)--(4.00,2.22);
  \fill[violet!10]
       (1.00,.65)
       .. controls (1.25,.65) and (1.25,1.95) .. (2.25,1.95)
       .. controls (3.35,1.95) and (3.70,.65) .. (4.00,.65)
       --cycle;
  \draw[violet!75!black,line width=1.5pt]
       (.40,.65)--(1.00,.65)
       .. controls (1.25,.65) and (1.25,1.95) .. (2.25,1.95)
       .. controls (3.35,1.95) and (3.70,.65) .. (4.00,.65)
       --(5.05,.65);
  \foreach \x in {1,2,4} {\draw[black!70] (\x,.60)--(\x,.70);}
  \node[below=5pt] at (1.00,.65) {$R_j/2$};
  \node[below=5pt] at (2.00,.65) {$R_j$};
  \node[below=5pt] at (4.00,.65) {$2R_j$};
  \node[violet!75!black] at (2.50,2.25) {$R_j/2\le r\le2R_j$};
  \node at (2.65,-.72) {(b) One smooth radial cutoff};
 \end{scope}
\end{tikzpicture}
\caption{The cutoffs act on the decaying datum
$g(r)=\phi_0(r)-F(u_0)$ on the outgoing null portion of
$\mathcal N_{u_0}$. The purple segment in (a) represents
$\mathcal N_{u_0}\cap\{R_j/2\le r\le2R_j\}$.}
\label{fig:annular-null-data}
\end{figure}

Figure~\ref{fig:annular-null-data} locates the support of one cutoff
on the outgoing null portion of $\mathcal N_{u_0}$.

For each $j$, we let $w_j$ be the solution with zero data on
$\mathcal C_{u_0}$ and reduced null data
\[
 r w_j|_{\mathcal N_{u_0}}=\chi_j g.
\]
The remaining solution $w_{\mathrm b}$ carries the data on
$\mathcal C_{u_0}$ and the reduced null datum
$\phi_0-\sum_j\chi_jg$. This datum equals the constant $F(u_0)$
for all sufficiently large $r$, so $I_0[w_{\mathrm b}]=0$.
Its norms $\mathbb A_{\sigma,K}^{\varepsilon}[w_{\mathrm b}]$
are bounded by $CA^2$. We therefore estimate $w_{\mathrm b}$
using the $\sigma=3$ bounds of Step~1.

For the annular pieces, \eqref{eq:frac-null-symbol} and the
cutoff derivative bounds give $A_{R_j}\le CA R_j^{1-p}$.
Lemma~\ref{lem:frac-annular-norms} consequently gives
\begin{equation}
 \sqrt{\mathbb A_{2,K}^{\varepsilon}[w_j]}
 \le CA R_j^{2-p},
 \qquad
 \sqrt{\mathbb A_{3,K}^{\varepsilon}[w_j]}
 \le CA R_j^{3-p}.
 \label{eq:frac-dyadic-norms}
\end{equation}
Since $\sum_j R_j^{1-p}<\infty$, the series of solutions converges
in the ordinary and finitely commuted energy spaces. Linearity and
uniqueness identify the resulting solution as
$\psi=w_{\mathrm b}+\sum_jw_j$.

\paragraph{Step 4: Choosing the estimate for each annulus and summing.}
Combining \eqref{eq:frac-dyadic-norms} with the two bounds in
\eqref{eq:frac-endpoint-field}, we have
\[
 |T^kw_j|(\tau,r)\le CA\min\left\{
 R_j^{3-p}(1+\tau)^{-3-k+\varepsilon},\,
 R_j^{2-p}(1+\tau)^{-2-k+\varepsilon}\right\}.
\]
The factor $R_j^{3-p}$ grows with $j$, whereas $R_j^{2-p}$
decreases. The two bounds therefore give complementary control
of the smaller and larger annuli. Their ratio is $R_j/(1+\tau)$.
We therefore use the decay power $3$ when $R_j\le1+\tau$ and the
decay power $2$ when $R_j>1+\tau$. For $2<p<3$, the two geometric
series satisfy, with a constant $C(p)>0$ depending only on $p$,
\[
 \sum_{R_j\le1+\tau}R_j^{3-p}\le C(p)(1+\tau)^{3-p},
 \qquad
 \sum_{R_j>1+\tau}R_j^{2-p}\le C(p)(1+\tau)^{2-p}.
\]
Consequently,
\begin{align}
 \sum_j|T^kw_j|
 &\le CA (1+\tau)^{-3-k+\varepsilon}
       \sum_{R_j\le 1+\tau}R_j^{3-p}
   +CA (1+\tau)^{-2-k+\varepsilon}
       \sum_{R_j>1+\tau}R_j^{2-p}\notag\\
 &\le C_{p,\varepsilon,K} A (1+\tau)^{-p-k+\varepsilon}.
 \label{eq:frac-dyadic-sum}
\end{align}
The solution $w_{\mathrm b}$ satisfies the faster power-$3$ bound,
so the same conclusion holds for $T^k\psi$.
Applying the two geometric sums to
\eqref{eq:frac-endpoint-radiation} gives
$(1+r)|T^k\psi|\le CA(1+\tau)^{1-p-k+\varepsilon}$.
Combining this with the unweighted bound proves
\eqref{eq:frac-global-field}.
The same argument applied to
\eqref{eq:frac-endpoint-radial} and
\eqref{eq:frac-endpoint-radial-weighted} gives
\eqref{eq:frac-global-radial} and
\eqref{eq:frac-global-radial-weighted}.
For the energies, we apply the triangle inequality to the
square-root fluxes and then use the same two sums. This proves
\eqref{eq:frac-global-energy} and
\eqref{eq:frac-global-radial-energy}.

Finally, Lemma~\ref{lem:frac-spatial-transfer} supplies
\eqref{eq:frac-null-symbol}, including its radial derivatives,
with $A\le C\mathfrak D_{p,m}[\psi]$. Lemma~\ref{lem:volterra}
bounds $F(u_0)$, and Lemma~\ref{lem:compact-transfer} controls the
norms on $\mathcal C_{u_0}$. Thus every estimate above is controlled
by the initial data on $\mathcal S$. For $2<p<3$, this proof already
works with $m=3K+18$: the transfer loses at most two symbol
derivatives and leaves the $3K+16$ used above. The two additional
derivatives in Proposition~\ref{prop:frac-global} are reserved for
the $1<p<2$ argument, which applies these estimates to $T\psi$.
\subsection{Proof of the decay bounds for
\texorpdfstring{$1<p\le2$}{1 < p <= 2}}
\label{sec:frac-decay-12}

For $1<p<2$, a time derivative raises the spatial exponent into the
range of Section~\ref{sec:frac-decay-23}. We first estimate
$T\psi$, identify the late-time limit of $\psi$, and then integrate
the estimates in time. We treat $p=2$ directly at the end.

\paragraph{Estimating the time derivative when $1<p<2$.}
We set $\mathfrak D[\psi]=\mathfrak D_{p,3K+20}[\psi]$ and apply
the preceding result to the solution $w=T\psi$.
Lemma~\ref{lem:slow-T-data} gives its two initial data in
\eqref{eq:frac-T-spatial-data}; their spatial exponent is
$p+1\in(2,3)$. Equation~\eqref{eq:time-derivative-symbol-norm}
controls the norm of $w$ through order $3K+18$ by the norm of
$\psi$ through order $3K+20$. Thus all the bounds proved in
Section~\ref{sec:frac-decay-23} apply to $T\psi$, with exponent
$p+1$.

\paragraph{The limiting value of $\psi$ and its time integral.}
The estimates for $T\psi$ and $YT\psi$ are locally integrable in
$\tau$. Hence
$L(r):=\lim_{\tau\to\infty}\psi(\tau,r)$ exists locally in $C^1$.
The local integrated energy estimate forces $YL=0$: a nonzero
limiting radial derivative on a compact exterior interval would
give a positive local energy for all sufficiently late times.
Thus $L$ is independent of $r$. Uniform energy boundedness and
the radial Hardy inequality give
\[
 |\psi(\tau,r)|\le C r^{-1/2}
       \sqrt{\mathcal E_N[\psi](u_0)}
\]
for all sufficiently large $r$. Taking first $\tau\to\infty$
and then $r\to\infty$ gives $L=0$. Local convergence in the
regular coordinates gives the same limit at the horizon.
We therefore have the exact identity
\begin{equation}
 \psi(\tau,r)=-\int_\tau^\infty T\psi(s,r)\,ds,
 \label{eq:frac-future-integration}
\end{equation}
where the integral keeps $r$ fixed.

We integrate the bounds for $T\psi$ and $(1+r)T\psi$ separately:
\[
 |\psi|\le C\mathfrak D[\psi]\cdot(1+\tau)^{-p+\varepsilon},
 \qquad
 (1+r)|\psi|\le C\mathfrak D[\psi]\cdot(1+\tau)^{1-p+\varepsilon}.
\]
The second integral converges because $\varepsilon<p-1$.
Combining these estimates gives \eqref{eq:frac-global-field}
for $k=0$. Since $Y$ and $N$ commute with $T$, the integrable
bounds for $YT\psi$ and $NT\psi$ also allow differentiation
under \eqref{eq:frac-future-integration}; this gives the radial
estimates \eqref{eq:frac-global-radial} and
\eqref{eq:frac-global-radial-weighted}. For $k\ge1$, we use the
estimates for $T\psi$ directly with $k-1$ time derivatives.

For the energy estimates, Minkowski's integral inequality in the
stationary energy norms gives
\[
 \sqrt{\mathcal E_N[\psi](\tau)}
 \le\int_\tau^\infty
       \sqrt{\mathcal E_N[T\psi](s)}\,ds
 \le C\mathfrak D[\psi]\cdot(1+\tau)^{1/2-p+\varepsilon}.
\]
Applying the same argument to $Y\psi$ and $N\psi$ gives
\eqref{eq:frac-global-radial-energy}. All these integrals converge
for the stated choice of $\varepsilon$.

\paragraph{The case $p=2$.}
The transferred data on $\mathcal N_{u_0}$ satisfy
$\partial_r(r\psi)=\Os[3K+16](r^{-2})$. The norms
$\mathbb A_{2,K}^{\varepsilon}[\psi]$ from
\eqref{eq:frac-endpoint-norm} are therefore finite: their largest
radial integrands are $O(r^{-1-\varepsilon})$.
We apply \eqref{eq:frac-endpoint-field}--\eqref{eq:frac-endpoint-radial-weighted}
with $\sigma=2$, together with the corresponding energy estimates.
These give every assertion of Proposition~\ref{prop:frac-global}
at $p=2$.

\paragraph{Number of derivatives.}
The largest norm index used above is $K+2$. Its ordinary energy
terms involve at most $T^{3K+11}\psi$ and one further derivative;
the radial terms follow from repeated use of
\eqref{eq:frac-null-recurrence}. The symbol bounds through order
$3K+16$ in Section~\ref{sec:frac-decay-23} control these terms.
Applying the argument to $T\psi$ requires two further initial-data
derivatives, so $m=3K+20$ in \eqref{eq:frac-original-norm} suffices.
This completes the proof of Proposition~\ref{prop:frac-global}.
\subsection{Quantitative global asymptotics for \texorpdfstring{$1<p<3$}{1 < p < 3}}
\label{sec:frac-asymptotics}

We now combine the global bounds for $\psi$ and $Y\psi$ from
Proposition~\ref{prop:frac-global} with the expansion of
$Q=\partial_v(r\psi)$ on $\mathcal N_0$ supplied by
Proposition~\ref{prop:all-p-null-traces}. We first transport the
coefficient of $Q$ through the exterior and recover $\psi$ at
$r=\sqrt{1+\tau}$. Radial integration towards the horizon and
outwards to arbitrary finite radii then gives one global expansion.
The bounded-radius and radiation-field asymptotics follow from this
expansion. These null and radial integrations use the approach of
\cite[Sections~7--8]{AAG1}.

We write $\phi=r\psi$ and use the potential $V$ defined in
\eqref{eq:potential-definition}. The geometry fixed in
Section~\ref{sec:hypersurfaces} gives
\begin{equation}
 \tau=u,\qquad H'(r)=\frac2{D(r)}\quad(r\ge R),
 \label{eq:frac-asymptotic-hyp-slope}
\end{equation}
so that
\begin{equation}
 \left.\partial_r\right|_u=Y\qquad(r\ge R).
 \label{eq:frac-asymptotic-radial-conversion}
\end{equation}
The coordinate $\tau$ and the vector field $Y$ extend regularly
to the future horizon along the compact portion of each
$\mathcal N_\tau$.

\begin{proposition}[Global leading-order asymptotics for $1<p<3$]
\label{prop:frac-asymptotics}
We fix $1<p<3$ and $\delta>0$, and set
\begin{equation}
 \eta=\frac1{16}\min\left\{\delta,3-p,\frac12\right\},
 \qquad \widehat\eta=\frac\eta2.
 \label{eq:frac-asymptotic-eta}
\end{equation}
We assume that the expansion on the outgoing null portion of
$\mathcal N_0$ is
\begin{equation}
 Q(0,v)=A v^{-p}+O(\mathfrak A[\psi]\cdot v^{-p-4\eta}),
 \qquad |A|\le C\mathfrak A[\psi].
 \label{eq:frac-asymptotic-trace}
\end{equation}
We also assume the following bounds supplied by
Proposition~\ref{prop:frac-global}, with the estimates on bounded
time intervals included in the constants: for every sufficiently
small $\epsilon>0$, for $\tau\ge0$ and $r\ge r_h$,
\begin{align}
 |\psi(\tau,r)|
 &\le C\mathfrak A[\psi]\cdot
   \frac{(1+\tau)^{1-p+\epsilon}}{1+\tau+r},
 \label{eq:frac-asymptotic-global-field}\\
 |Y\psi(\tau,r)|
 &\le C\mathfrak A[\psi]\cdot(1+\tau)^{-p-1+\epsilon}.
 \label{eq:frac-asymptotic-global-radial}
\end{align}
Here $A$ is the signed coefficient of $v^{-p}$ and
$\mathfrak A[\psi]$ bounds the size of the data. For the initial
data considered here, $A=C_p/2$ and we may take
$\mathfrak A[\psi]=\mathfrak A_{p,23,\delta}[\psi]$.
The constants may depend on the fixed $p$ and $\epsilon$.

We define
\begin{equation}
 \Pi_p(\tau,r)
 :=\frac{(1+\tau)^{1-p}-(1+\tau+2r)^{1-p}}{2(p-1)r}.
 \label{eq:frac-asymptotic-profile}
\end{equation}
Then, uniformly for $\tau\ge0$ and $r\ge r_h$,
\begin{equation}
 \left|\psi(\tau,r)-2A\Pi_p(\tau,r)\right|
 \le C\mathfrak A[\psi]\cdot
       \frac{(1+\tau)^{1-p-\widehat\eta}}{1+\tau+r}.
 \label{eq:frac-asymptotic-global-remainder}
\end{equation}
In particular, the coefficient of this global profile is
$2A=C_p=b+(1-p)a$.
For each fixed $R_1>r_h$, we also have
\begin{equation}
 \sup_{r_h\le r\le R_1}
 \left|\psi(\tau,r)-2A(1+\tau)^{-p}\right|
 \le C_{R_1}\mathfrak A[\psi]\cdot(1+\tau)^{-p-\eta}.
 \label{eq:frac-asymptotic-timelike}
\end{equation}
The radiation field $F(u)=\lim_{v\to\infty}\phi(u,v)$ exists and
satisfies
\begin{equation}
 \left|F(u)-\frac{A}{p-1}(1+u)^{1-p}\right|
 \le C\mathfrak A[\psi]\cdot(1+u)^{1-p-\eta}.
 \label{eq:frac-asymptotic-radiation}
\end{equation}
All three estimates include $A=0$.
\end{proposition}

\begin{proof}
We fix
\[
 0<\epsilon<\min\left\{\frac\eta4,\frac{p-1}4\right\}.
\]
All constants below may depend on this choice. We first work at
sufficiently large times, so that $\sqrt{1+\tau}\ge R$, and
include bounded time intervals at the end. The field and radial
derivative bounds in the hypotheses control the source terms and
radial integrals below.

\paragraph{Step 1. Propagating $Q$ to late times in the exterior.}
The wave equation \eqref{eq:wave} gives
\begin{equation}
 \partial_u Q=-\frac14V\phi.
 \label{eq:frac-asymptotic-Q-transport}
\end{equation}
For $r\ge R$ we have $r\asymp v-u$ after enlarging the fixed $R$,
$|V|\le Cr^{-3}$, and $\tau=u$. In particular,
\eqref{eq:frac-asymptotic-global-field} implies
\begin{equation}
 |\phi(u,v)|\le C\mathfrak A[\psi]\cdot
 \frac{r}{1+u+r}(1+u)^{1-p+\epsilon}.
 \label{eq:frac-asymptotic-phi-global}
\end{equation}
We integrate \eqref{eq:frac-asymptotic-Q-transport} from $u'=0$
to $u'=u$ at fixed $v$. The bound
\eqref{eq:frac-asymptotic-phi-global} controls the unknown field
$\phi$ in this integral. We split at $u'=v/2$: on the first
part the radius is comparable to $v$, while on the second part we
integrate the decreasing radius directly. Empty parts are omitted.
For $0\le u'\le\min\{u,v/2\}$, we have $r(u',v)\asymp v$, so
\[
 \int_0^{\min\{u,v/2\}}|V\phi|(u',v)\,\mathrm du'
 \le C\mathfrak A[\psi]\cdot v^{-3}H_p(v),
 \qquad
 H_p(v):=\int_0^v(1+s)^{1-p+\epsilon}\,\mathrm ds.
\]
On $v/2\le u'\le u$, if this interval is nonempty,
\eqref{eq:frac-asymptotic-phi-global} gives
$|\phi(u',v)|\le C\mathfrak A[\psi]\cdot r(u',v)v^{-p+\epsilon}$.
Since $\partial_{u'}r=-D/2$ and $D$ is bounded away from zero in
this far-exterior region,
\[
 \int_{v/2}^{u}|V\phi|(u',v)\,\mathrm du'
 \le C\mathfrak A[\psi]\cdot v^{-p+\epsilon}
       \int_{v/2}^{u}r(u',v)^{-2}\,\mathrm du'
 \le C\mathfrak A[\psi]\cdot\frac{v^{-p+\epsilon}}{v-u}.
\]
Consequently,
\begin{equation}
 |Q(u,v)-Q(0,v)|
 \le C\mathfrak A[\psi]\cdot\left[
       v^{-3}H_p(v)+\frac{v^{-p+\epsilon}}{v-u}\right].
 \label{eq:frac-asymptotic-Q-error}
\end{equation}
The elementary integral in this estimate satisfies
\[
 H_p(v)\le C
 \begin{cases}
  (1+v)^{2-p+\epsilon},&p<2+\epsilon,\\
  \log(2+v),&p=2+\epsilon,\\
  1,&p>2+\epsilon.
 \end{cases}
\]
The choices of $\eta$ and $\epsilon$ give
$v^{-3}H_p(v)\le C v^{-p-4\eta}$ in all three cases.
For the second error in \eqref{eq:frac-asymptotic-Q-error}, we
keep a sufficiently large separation between $v$ and $u$.
For any fixed $c>0$, in the region
\[
 \mathcal B_{1/2,c}
 :=\{r\ge R,\ u\ge0,\ v-u\ge c(1+v)^{1/2}\}
\]
the second error also has size
$O(\mathfrak A[\psi]\cdot v^{-p-4\eta})$, since
$\tfrac12-\epsilon>4\eta$. Combining these bounds with
\eqref{eq:frac-asymptotic-trace}, we obtain
\begin{equation}
 Q(u,v)=A v^{-p}+O(\mathfrak A[\psi]\cdot v^{-p-4\eta})
 \quad\hbox{in }\mathcal B_{1/2,c}.
 \label{eq:frac-asymptotic-Q-matching}
\end{equation}
The restriction $p<3$ supplies the positive gain in the first
source contribution $v^{-3}H_p(v)$.

\paragraph{Step 2. Recovering $\psi$ at $r=\sqrt{1+\tau}$.}
We set $r_\gamma(\tau)=\sqrt{1+\tau}$ and denote by $P_\tau$
the point on $\mathcal N_\tau$ with $r(P_\tau)=r_\gamma(\tau)$.
This radius tends to infinity more slowly than $\tau$.
The curve lies in $\mathcal B_{1/2,c}$ for a suitable fixed $c>0$,
because $u(P_\tau)=\tau$ and
\[
 v(P_\tau)=\tau+2r_*(P_\tau)
          =\tau+O((1+\tau)^{1/2}).
\]
The exact identity
\begin{equation}
 \frac{2Q}{D}
 =\left.\partial_r(r\psi)\right|_u
 =\psi+r\left.\partial_r\psi\right|_u
 \label{eq:frac-asymptotic-Q-field}
\end{equation}
expresses $\psi$ in terms of $Q$ and $Y\psi$.
By \eqref{eq:frac-asymptotic-radial-conversion} and
\eqref{eq:frac-asymptotic-global-radial},
\[
 r|Y\psi|(P_\tau)
 \le C\mathfrak A[\psi]\cdot(1+\tau)^{-p-1/2+\epsilon}.
\]
Moreover $D(P_\tau)^{-1}=1+O((1+\tau)^{-1/2})$ and
\[
 v(P_\tau)^{-p}
 =(1+\tau)^{-p}+O((1+\tau)^{-p-1/2}).
\]
Together with \eqref{eq:frac-asymptotic-Q-matching}, these bounds give
\begin{equation}
 \psi(P_\tau)
 =2A(1+\tau)^{-p}
   +O(\mathfrak A[\psi]\cdot(1+\tau)^{-p-4\eta}).
 \label{eq:frac-asymptotic-curve-field}
\end{equation}

\paragraph{Step 3. Integrating towards the future horizon.}
For every $r_h\le r\le r_\gamma(\tau)$, we integrate along
the same hypersurface $\mathcal N_\tau$. Since $Y=\partial_r|_\tau$
is regular at the horizon,
\begin{align*}
 |\psi(\tau,r)-\psi(P_\tau)|
 &\le\int_r^{r_\gamma(\tau)}|Y\psi(\tau,\rho)|\,\mathrm d\rho\\
 &\le C\mathfrak A[\psi]\cdot(1+\tau)^{-p-1/2+\epsilon}.
\end{align*}
Hence \eqref{eq:frac-asymptotic-curve-field} gives, uniformly on
this entire inner region,
\[
 \psi(\tau,r)=2A(1+\tau)^{-p}
       +O(\mathfrak A[\psi]\cdot(1+\tau)^{-p-4\eta}).
\]
Taylor's formula for \eqref{eq:frac-asymptotic-profile}, with
$r/(1+\tau)\le(1+\tau)^{-1/2}$, gives
\[
 \Pi_p(\tau,r)
 =(1+\tau)^{-p}+O\bigl(r(1+\tau)^{-p-1}\bigr).
\]
Since $4\eta<1/2$, we conclude that
\begin{equation}
 |\psi(\tau,r)-2A\Pi_p(\tau,r)|
 \le C\mathfrak A[\psi]\cdot(1+\tau)^{-p-4\eta}
 \quad(r_h\le r\le r_\gamma(\tau)).
 \label{eq:frac-asymptotic-inner-profile}
\end{equation}

\paragraph{Step 4. Integrating outwards to an arbitrary finite radius.}
We now fix a sufficiently large $u=\tau$ and consider every
$r\ge r_\gamma(u)$. This whole outgoing segment lies in
$\mathcal B_{1/2,c}$ for a fixed $c>0$. We have
\[
 D=1+O(r^{-1}),\qquad
 v-(1+u+2r)=2(r_*-r)-1=O(\log(2+r)).
\]
Also $r\ge c(1+v)^{1/2}$ on this segment. Thus
\begin{align*}
 \left|\frac{2A}{D}v^{-p}-2A(1+u+2r)^{-p}\right|
 &\le C|A|\left[r^{-1}v^{-p}+v^{-p-1}\log(2+v)\right]\\
 &\le C\mathfrak A[\psi]\cdot v^{-p-4\eta}.
\end{align*}
Using \eqref{eq:frac-asymptotic-Q-matching} and
$\partial_r\phi|_u=2Q/D$, we therefore write
\begin{equation}
 \left.\partial_r\phi\right|_u
 =2A(1+u+2r)^{-p}+e(u,r),\qquad
 |e(u,r)|\le C\mathfrak A[\psi]\cdot(1+u+r)^{-p-4\eta}.
 \label{eq:frac-asymptotic-radial-expansion}
\end{equation}
Here $e$ is the error in this radial derivative expansion.

We next express the already known value at $r_\gamma=r_\gamma(u)$
in a form that joins directly to the radial integral. Taylor's formula
gives
\[
 \int_0^{r_\gamma}(1+u+2\rho)^{-p}\,\mathrm d\rho
 =r_\gamma(1+u)^{-p}
      +O\bigl(r_\gamma^2(1+u)^{-p-1}\bigr).
\]
Combining this with \eqref{eq:frac-asymptotic-curve-field}, we obtain
\begin{equation}
 \begin{aligned}
 \phi(u,r_\gamma)
 &=2A\int_0^{r_\gamma}(1+u+2\rho)^{-p}\,\mathrm d\rho+E_\gamma(u),\\
 |E_\gamma(u)|
 &\le C\mathfrak A[\psi]\cdot r_\gamma(1+u)^{-p-4\eta}.
 \end{aligned}
 \label{eq:frac-asymptotic-matched-value}
\end{equation}
The integral from $0$ is an elementary integral defining the explicit
profile; the spacetime integration starts at $r=r_\gamma$.
We integrate \eqref{eq:frac-asymptotic-radial-expansion} from
$r_\gamma$ to the finite radius $r$. The leading integrals combine as
\begin{align}
 \phi(u,r)
 &=2A\int_0^r(1+u+2\rho)^{-p}\,\mathrm d\rho+E(u,r)\notag\\
 &=\frac{A}{p-1}
       \left[(1+u)^{1-p}-(1+u+2r)^{1-p}\right]+E(u,r),
 \label{eq:frac-asymptotic-finite-radius}
\end{align}
where
\[
 E(u,r)=E_\gamma(u)+\int_{r_\gamma}^r e(u,\rho)\,\mathrm d\rho.
\]
The estimates for $E_\gamma$ and $e$ imply
\begin{align}
 |E(u,r)|
 &\le C\mathfrak A[\psi]\cdot\left[
      r_\gamma(1+u)^{-p-4\eta}
      +\int_{r_\gamma}^r(1+u+\rho)^{-p-4\eta}\,\mathrm d\rho
                         \right]\notag\\
 &\le C\mathfrak A[\psi]\cdot
       \min\{r,1+u\}(1+u)^{-p-4\eta}.
 \label{eq:frac-asymptotic-integrated-error}
\end{align}
For $r\le1+u$, we bound the integrand by $(1+u)^{-p-4\eta}$;
for $r\ge1+u$, we integrate to infinity, using $p+4\eta>1$.
The boundary error satisfies the same bound since
$r_\gamma\le\min\{r,1+u\}$.

The leading term in \eqref{eq:frac-asymptotic-finite-radius} equals
$2Ar\Pi_p(u,r)$. Dividing by $r$ and using
$\min\{r,1+u\}/r\le2(1+u)/(1+u+r)$ proves the required
outer estimate. Together with \eqref{eq:frac-asymptotic-inner-profile}
and $\tau=u$ on the outgoing null portion, this gives the stronger
late-time bound
\begin{equation}
 |\psi(\tau,r)-2A\Pi_p(\tau,r)|
 \le C\mathfrak A[\psi]\cdot
       \frac{(1+\tau)^{1-p-4\eta}}{1+\tau+r}
 \quad(\tau\ge\tau_0,\ r\ge r_h)
 \label{eq:frac-asymptotic-global-strong}
\end{equation}
for some fixed sufficiently large $\tau_0$.

\paragraph{Step 5. Limiting asymptotics and bounded time intervals.}
At bounded radius, Taylor's formula gives
$\Pi_p(\tau,r)=(1+\tau)^{-p}+O_{R_1}((1+\tau)^{-p-1})$.
Equation~\eqref{eq:frac-asymptotic-global-strong} therefore proves
\eqref{eq:frac-asymptotic-timelike} at late times.

For each sufficiently large $u$, the expansion
\eqref{eq:frac-asymptotic-Q-matching} is integrable in $v$, since
$p>1$. Thus $F(u)=\lim_{v\to\infty}\phi(u,v)$ exists. Multiplying
\eqref{eq:frac-asymptotic-global-strong} by $r$ and taking this limit,
using
\[
 \lim_{r\to\infty}2Ar\Pi_p(u,r)
 =\frac{A}{p-1}(1+u)^{1-p},
\]
gives \eqref{eq:frac-asymptotic-radiation}, with the stronger
gain $4\eta$ at late times.

On the bounded interval $0\le u\le\tau_0$, we have
$r(u',v)\asymp v$ for large $v$ and $0\le u'\le u$.
The field bound gives $|\phi(u',v)|\le C_{\tau_0}\mathfrak A[\psi]$.
Integrating \eqref{eq:frac-asymptotic-Q-transport} from $0$ to $u$
therefore yields
\[
 |Q(u,v)|\le C_{\tau_0}\mathfrak A[\psi]\cdot(v^{-p}+v^{-3})
          \le C_{\tau_0}\mathfrak A[\psi]\cdot v^{-p}.
\]
This proves existence of $F$ on that interval as well, and the
bound on $\phi$ bounds $F$ uniformly there.
Finally, on $0\le\tau\le\tau_0$, the field bound and the explicit
profile give
\[
 |\psi(\tau,r)|+|2A\Pi_p(\tau,r)|
 \le C_{\tau_0}\mathfrak A[\psi]\cdot\frac1{1+r}.
\]
These bounded-time estimates extend all three conclusions to the
stated time ranges after increasing the constants. The global
estimate follows from \eqref{eq:frac-asymptotic-global-strong}
by weakening the late-time gain to $\widehat\eta$.
\end{proof}

\begin{remark}
\label{rem:frac-asymptotic-scope}
The exponents in \eqref{eq:frac-asymptotic-eta} give a positive
power improvement for every fixed $1<p<3$. At $p=2$, the statement
includes $a=b$ and gives the corresponding improved remainder bound.
For $a\ne b$, it agrees with the ordinary-NP normalization
$I_0=(b-a)/4$ and the coefficients $4I_0$ and $2I_0$ in
\cite{AAG1}.
\end{remark}

\subsection{The global expansion from the initial data}
\label{sec:frac-spatial-global}

We now state the expansion entirely in terms of the initial data
on $\mathcal S$. Section~\ref{sec:exterior} supplies the incoming
coefficient $A=C_p/2$ and the required data on $\mathcal N_0$;
Sections~\ref{sec:frac-decay-23}--\ref{sec:frac-decay-12} supply the
global decay bounds; Section~\ref{sec:frac-asymptotics} then gives
the profile and its uniform remainder.

\begin{theorem}[Global expansion for $1<p<3$]
\label{thm:frac-spatial-global}
We assume the nondegenerate geometric and energy assumptions of
Section~\ref{sec:setting}. We choose an initial hypersurface
$\mathcal S$ as in Section~\ref{sec:hypersurfaces} and prescribe smooth
data on it. We fix $1<p<3$, $\delta>0$, take $m=23$, and assume
\[
 f=a r^{-p}+\Os[m+1](r^{-p-\delta}),\qquad
 h=b r^{-p-1}+\Os[m](r^{-p-1-\delta}),\qquad
 \mathfrak A[\psi]=\mathfrak A_{p,23,\delta}[\psi]<\infty.
\]
We set
\begin{equation}
 \mathcal C_p=b+(1-p)a,
 \label{eq:frac-spatial-coefficient}
\end{equation}
and use the profile
\[
 \Pi_p(\tau,r)
 =\frac{(1+\tau)^{1-p}-(1+\tau+2r)^{1-p}}{2(p-1)r}.
\]
There are constants $C>0$ and $\tau_0\ge0$ such that
\begin{equation}
 \left|\psi(\tau,r)-\mathcal C_p\Pi_p(\tau,r)\right|
 \le C\mathfrak A[\psi]\cdot
       \frac{(1+\tau)^{1-p-\eta}}{1+\tau+r},
 \qquad
 \eta=\frac1{32}\min\{\delta,3-p,1/2\}.
 \label{eq:frac-spatial-global}
\end{equation}
The estimate holds for all $\tau\ge\tau_0$ and $r\ge r_h$,
including the future horizon. The constants depend only on the
fixed parameters, background, and hypersurfaces.
Consequently, as $\tau\to\infty$ on each bounded radial interval
and as $u\to\infty$ along future null infinity, respectively,
\begin{align}
 \psi(\tau,r)
 &=\mathcal C_p(1+\tau)^{-p}
       +O_{R_1}(\mathfrak A[\psi]\cdot (1+\tau)^{-p-\eta}),
       &&r_h\le r\le R_1,
 \label{eq:frac-spatial-timelike}\\
 F(u)
 &=\frac{\mathcal C_p}{2(p-1)}(1+u)^{1-p}
       +O(\mathfrak A[\psi]\cdot (1+u)^{1-p-\eta}).
 \label{eq:frac-spatial-radiation}
\end{align}
The estimate also holds when $\mathcal C_p=0$, giving the stated
improved remainder bound.
\end{theorem}

\begin{proof}
Proposition~\ref{prop:all-p-null-traces} gives the expansion of
$Q(0,v)$ with $A=\mathcal C_p/2$.
Proposition~\ref{prop:frac-global}, with $K=1$ and $m=23$,
gives the field and radial derivative bounds. Applying
Proposition~\ref{prop:frac-asymptotics} to these inputs gives
\eqref{eq:frac-spatial-global}.
The norm $\mathfrak A_{p,23,\delta}[\psi]$ controls both these
global decay estimates and the quantitative expansion of $Q(0,v)$,
so the coefficient and all remainder constants are controlled by
the initial data on $\mathcal S$.

Finally,
$\Pi_p(\tau,r)=(1+\tau)^{-p}+O_{R_1}((1+\tau)^{-p-1})$
at bounded radius, while
$r\Pi_p\to(1+u)^{1-p}/[2(p-1)]$ at infinity, with our
normalization $\tau=u$ for $r\ge R$. These limits prove the
bounded-radius and radiation-field expansions.
\end{proof}

In particular, $p=2$ gives the initial-hypersurface charge formula
\begin{equation}
 I_0[\psi]=\frac{b-a}{4},
 \qquad
 \psi=(b-a)\Pi_2(\tau,r)+
 O\!\left(\mathfrak A[\psi]\cdot \frac{(1+\tau)^{-1-\eta}}{1+\tau+r}\right).
 \label{eq:frac-p2-summary}
\end{equation}

\begin{remark}[Cancellation at $p=2$]
\label{rem:frac-p2-log}
For exact exterior data $f=a r^{-2}$ and $h=a r^{-3}$,
$C_2=I_0=0$, but the next null term is generally logarithmic:
\[
 Q(u_0,v)=-4Ma\,v^{-3}\log v+O(v^{-3}).
\]
Indeed, the leading exterior solution has
$\phi(u',v)=a/(-u')$ up to an error integrable in $u'$ for the logarithmic
coefficient, and the kernel is $4M(v-u')^{-3}$ to leading order.
Its integral produces the displayed logarithm.  Correspondingly, the truncated integral contributing to $B_{\mathcal S}$
satisfies $\int_{R_0}^{R}r^2D^{-1}h\,dr=a\log R+O(1)$.
Theorem~\ref{thm:frac-spatial-global} applies with zero leading
$\tau^{-2}$ coefficient.  
\end{remark}

These results give the $1<p<3$ entries of
Table~\ref{tab:intro-black-hole}.
The intermediate-power decay estimates follow from the decay
hierarchies in \cite{AAG1,AAGDecay} by the explicit dyadic argument above.
The relation between the limits
$\lim_{r\to\infty}r^p\partial_r(r\psi)|_u$ and late-time coefficients
is also discussed in
\cite[Section~4.1]{GK22}; here its needed global estimates and
coefficient are established from the initial data on $\mathcal S$.

The time-commuted and radial estimates for $1<p<3$ are those of
Proposition~\ref{prop:frac-global}, with the corresponding
initial-data derivative count.

\section{Global leading-order asymptotics for \texorpdfstring{$p=1$}{p = 1}}
\label{sec:slow-spatial-endpoints}

For $p=1$, the leading coefficient is $C_1=b$.
Section~\ref{sec:p1-global-profile} proves the global expansion and
defines the finite radiation field $F_{\rm ren}(u)$ obtained by
subtracting the logarithmic growth of $r\psi$.
Section~\ref{sec:p1-next-term} refines the case $b=0$ under more
detailed spatial expansions: it computes $I_0[\psi]$ and the next
bounded-radius and radiation-field coefficients.
Table~\ref{tab:p1-radiation-NP} summarizes the radiation field and
Newman--Penrose constant for these cases.
All data are prescribed on the chosen initial hypersurface
$\mathcal S$ of Section~\ref{sec:hypersurfaces}.

\subsection{Global expansion and the renormalized radiation field}
\label{sec:p1-global-profile}

\begin{theorem}[Global asymptotics for $p=1$]
\label{thm:frac-p1}
We assume the geometric and energy hypotheses of
Section~\ref{sec:setting}, and let the initial data satisfy
\begin{equation}
 f=\frac a r+\Os[26](r^{-1-\delta}),\qquad
 h=\frac b{r^2}+\Os[25](r^{-2-\delta}),\qquad \delta>0.
 \label{eq:frac-p1-data}
\end{equation}
We use the time coordinate $\tau$ of
Section~\ref{sec:hypersurfaces}, with $\tau=u$ on the outgoing null
portion of $\mathcal N_\tau$.
For example, we take
$\eta=\min\{\delta,1\}/128$ and
$\mathfrak A[\psi]=\mathfrak A_{1,25,\delta}[\psi]$, with the same expression
\eqref{eq:frac-asymptotic-data-norm} used at $p=1$.
For sufficiently late $\tau$, uniformly for all $r\ge r_h$,
including the future horizon, we have
\begin{equation}
 \left|\psi(\tau,r)
       -\frac b{2r}\log\left(1+\frac{2r}{1+\tau}\right)\right|
 \le C\mathfrak A[\psi]\cdot \,\frac{(1+\tau)^{-\eta}}{1+\tau+r}.
 \label{eq:frac-p1-global}
\end{equation}
In particular, uniformly on $r_h\le r\le R_1$,
\begin{equation}
 \psi(\tau,r)
 =b(1+\tau)^{-1}
   +O_{R_1}\bigl(\mathfrak A[\psi]\cdot (1+\tau)^{-1-\eta}\bigr).
 \label{eq:frac-p1-timelike}
\end{equation}
For $b\ne0$, the product $r\psi$ diverges logarithmically
as $r\to\infty$ at fixed $u$. The renormalized radiation field
$F_{\rm ren}(u)$ defined in \eqref{eq:frac-p1-renormalization}
obeys
\begin{equation}
 F_{\rm ren}(u)
 =-\frac b2\log(1+u)
       +O\bigl(\mathfrak A[\psi]\cdot (1+u)^{-\eta}\bigr).
 \label{eq:frac-p1-renormalized-tail}
\end{equation}
This radiation asymptotic is asserted for late $u\ge0$.
The definition \eqref{eq:frac-p1-renormalization} uses the reference length one.
Replacing $\log(2r)$ by $\log(2r/\ell)$, for $\ell>0$, adds
$(b/2)\log\ell$ to $F_{\rm ren}$.
\end{theorem}

\begin{proof}
For $w=T\psi$, the two initial data are given by
\eqref{eq:frac-T-spatial-data}.  In the present case they have the
expansions
\[
 w|_{\mathcal S}=b r^{-2}+\Os[25](r^{-2-\delta_1}),\qquad
 Tw|_{\mathcal S}=0\cdot r^{-3}+\Os[24](r^{-3-\delta_1}),
 \qquad 0<\delta_1<\min\{\delta,1\}.
\]
Consequently $C_2[w]=-b$, and the $p=2$ global expansion proved
above gives
\begin{equation}
 \begin{aligned}
 T\psi&=-\frac b{(1+\tau)(1+\tau+2r)}+\mathcal R(\tau,r),\\
 |\mathcal R(\tau,r)|
 &\le C\mathfrak A[\psi]\cdot \frac{(1+\tau)^{-1-\eta}}{1+\tau+r}.
 \end{aligned}
 \label{eq:frac-p1-derivative-expansion}
\end{equation}
The bounds for $T\psi$ and $YT\psi$ give a limit
$L(r)=\lim_{\tau\to\infty}\psi(\tau,r)$ on bounded radial intervals.
Local integrated decay makes $L$ constant, and the uniform Hardy
bound in the proof of \eqref{eq:frac-future-integration} gives
$L=0$. Hence
$\psi(\tau,r)=-\int_\tau^\infty T\psi(s,r)\,ds$.
The main integral equals
\[
 \int_{1+\tau}^\infty \frac{ds}{s(s+2r)}
 =\frac1{2r}\log\left(1+\frac{2r}{1+\tau}\right).
\]
Since $s+r\ge 1+\tau+r$ for $s\ge 1+\tau$,
\[
 \int_{1+\tau}^\infty\frac{s^{-1-\eta}}{s+r}\,ds
 \le\frac{(1+\tau)^{-\eta}}{\eta(1+\tau+r)}.
\]
This proves \eqref{eq:frac-p1-global} and its compact-radius
consequence.

The limit $\lim_{r\to\infty}r\mathcal R(s,r)$ at fixed
time $s$ exists, since $u=s$ on the outgoing portion of
$\mathcal N_s$ and $w=T\psi$ has an ordinary radiation field.
The integrand $r\mathcal R(s,r)$ is bounded by
$C\mathfrak A[\psi]\cdot (1+s)^{-1-\eta}$ uniformly in $r$ and is integrable in
$s$.  Dominated convergence in the integrated remainder therefore
proves \eqref{eq:frac-p1-renormalized-tail} for the renormalized
radiation field defined in \eqref{eq:frac-p1-renormalization}.
Here $\tau=u$ on the outgoing null portion of $\mathcal N_\tau$.
\end{proof}

Proposition~\ref{prop:all-p-null-traces} gives the logarithmic
behavior of $\phi=r\psi$ towards null infinity at fixed $u$.
Theorem~\ref{thm:frac-p1} determines the late-time behavior of its
renormalized radiation field $F_{\rm ren}(u)$ as $u\to\infty$.

\subsection{The next asymptotic term when \texorpdfstring{$b=0$}{b = 0}}
\label{sec:p1-next-term}

When $b=0$, the $\tau^{-1}$ coefficient in
Theorem~\ref{thm:frac-p1} vanishes. Under the additional two-term
expansions below, $a_2,b_2$ and the curvature contribution from $a_1$
determine $I_0[\psi]$, hence the $\tau^{-2}$ and $u^{-1}$ coefficients.
The following proposition supplements the theorem with these finer
asymptotics. A concrete example and a summary table follow the proof.

\begin{proposition}[The ordinary Newman--Penrose charge for a $1/r$ position tail]
\label{prop:frac-coulomb-charge}
We assume
\begin{equation}
 f=\frac{a_1}{r}+\frac{a_2}{r^2}
                 +\Os[26](r^{-2-\delta}),\qquad
 h=\frac{b_2}{r^3}+\Os[25](r^{-3-\delta}),\qquad \delta>0.
 \label{eq:frac-coulomb-data}
\end{equation}
Then $F(u):=\lim_{r\to\infty}r\psi(u,r)$ exists and
\begin{equation}
 I_0[\psi]=\frac{b_2-a_2-3Ma_1}{4}.
 \label{eq:frac-coulomb-NP}
\end{equation}
For a finite ordinary Newman--Penrose charge, the estimates of
\cite[Propositions~8.2 and~8.6]{AAG1} give
\[
 \psi=4I_0(1+\tau)^{-2}+O_{R_1}((1+\tau)^{-2-\eta}),
 \qquad
 F(u)=2I_0(1+u)^{-1}+O((1+u)^{-1-\eta}).
\]
The first estimate is uniform for $r_h\le r\le R_1$ as
$\tau\to\infty$; the second holds as $u\to\infty$.
Here $\eta>0$, and the remainders are controlled by finitely many
weighted derivatives of the data in \eqref{eq:frac-coulomb-data}
and compact Sobolev norms of $(f,\pi)$.
These are expansions with possibly zero leading coefficient.
\end{proposition}

\begin{proof}
We use the common outgoing hypersurface $u=u_0=-A$ fixed in
Section~\ref{sec:exterior}.
We apply the finite-triangle argument of Lemma~\ref{lem:volterra}
to $\phi-a_1$. Its initial-data term is $O((-u)^{-1})$, and the
additional source from the constant obeys
\[
 |a_1|\int_{-v}^{u}\int_{-u'}^{v}|k(u',v')|\,\dd v'\dd u'
 \le C|a_1|(-u)^{-1}.
\]
The bootstrap with the $p=2$ weight therefore gives
$\phi(u',v)=a_1+O((-u')^{-1})$ throughout $\mathcal W_A$.
The initial-data term $b_0(v)$ in \eqref{eq:Qexact} is
\[
 b_0(v)=\frac{b_2-a_2}{2v^2}
          +O(v^{-2-\delta}+v^{-3}\log v).
\]
The contribution of $a_1$ to the integral in \eqref{eq:Qexact} is
determined by the potential $V$ in \eqref{eq:potential-definition}.
Its expansion \eqref{eq:potential} gives
\[
 \lim_{v\to\infty}v^2\int_{-v}^{u_0} k(u',v)a_1\,du'
 =4Ma_1\int_0^1(1+z)^{-3}\,dz
 =\frac32Ma_1.
\]
Here we use the substitution $z=-u'/v$ in the integral.
The $O((-u')^{-1})$ part costs $O(v^{-3}\log v)$ before multiplication
by $v^2$.  The asymptotic kernel error has the same or better order.
It follows that
\[
 Q(u_0,v)
 =\frac{b_2-a_2-3Ma_1}{2v^2}
       +O(v^{-2-\beta}),
 \qquad 0<\beta<\min\{\delta,1\}.
\]
The conversion $I_0=\lim 2r^2D^{-1}Q$ proves
\eqref{eq:frac-coulomb-NP}.  Differentiating the same integral
identity supplies the required finite symbol bounds.  In particular,
the largest required radial energy integrands are
$r^{3-\epsilon}|\partial_r\phi|^2$ and
$r^{5-\epsilon}|\partial_r^2\phi|^2$, both
$O(r^{-1-\epsilon})$.  Lemma~\ref{lem:compact-transfer} controls the data on the compact
spacelike portion $\mathcal C_{u_0}$ of $\mathcal N_{u_0}$.  The asymptotic estimates of
\cite[Propositions~8.2 and~8.6]{AAG1} now apply.
\end{proof}

For example, we take $f=a/r$, $h=0$ outside a compact set. Then
\begin{align*}
 I_0&=-\frac{3Ma}{4},\\
 \psi&=-3Ma(1+\tau)^{-2}+O_{R_1}((1+\tau)^{-2-\eta}),\\
 F&=-\frac{3Ma}{2}(1+u)^{-1}+O((1+u)^{-1-\eta}).
\end{align*}
The coefficient tends to zero with $M$.  In Minkowski the same exact exterior pair gives $\psi=0$ once
$u>R_0$.

We summarize the possibilities in Table~\ref{tab:p1-radiation-NP}.
The condition $b=0$ gives a finite ordinary radiation field $F(u)$;
the existence and value of the ordinary Newman--Penrose constant
$I_0[\psi]$ then depend on the subleading initial data.
For $b\ne0$, \eqref{eq:frac-original-trace} and
$\partial_r|_u=2D^{-1}\partial_v$ give the linear divergence
of $r^2\partial_r\phi|_u$ displayed in the table.

\begin{table}[H]
\centering
\small
\setlength{\tabcolsep}{5pt}
\renewcommand{\arraystretch}{1.35}
\begin{tabularx}{\textwidth}{@{}
 >{\raggedright\arraybackslash}p{1.64in}
 >{\raggedright\arraybackslash}X
 >{\raggedright\arraybackslash}X@{}}
\toprule
Initial data at spatial infinity
 & Ordinary radiation field $F(u)$
 & Ordinary Newman--Penrose constant $I_0[\psi]$\\
\midrule
Data \eqref{eq:frac-p1-data}, $b\ne0$
 & $\infty$ (logarithmic):\newline
   $\phi\sim\dfrac b2\log(2r)$
 & $\infty$ (linear):\newline
   $r^2\partial_r\phi\big|_u\sim\dfrac b2 r$\\
\addlinespace[3pt]
Data \eqref{eq:frac-p1-data}, $b=0$
 & Finite; $F(u)=F_{\rm ren}(u)$
 & Depends on the subleading data: the defining expression may
   converge to a finite value, diverge, or have no limit.\\
\addlinespace[3pt]
Two-term data \eqref{eq:frac-coulomb-data}
 & Finite; $F(u)=F_{\rm ren}(u)$
 & $\displaystyle\frac{b_2-a_2-3Ma_1}{4}$\\
\addlinespace[3pt]
$f=a/r$, $h=0$ outside a compact set
 & Finite; $F(u)=F_{\rm ren}(u)$
 & $\displaystyle-\frac{3Ma}{4}$\\
\bottomrule
\end{tabularx}
\caption{Radiation field and Newman--Penrose constant for $p=1$.
All radial limits are taken as $r\to\infty$ at fixed retarded time $u$.}
\label{tab:p1-radiation-NP}
\end{table}

\section{Global leading-order asymptotics for \texorpdfstring{$p<1$}{p < 1}}
\label{sec:slow-global}

We now determine the leading term of the scalar field $\psi$ for
every real spatial exponent $p<1$. The data are prescribed on the
chosen initial hypersurface $\mathcal S$ of Section~\ref{sec:hypersurfaces}, with
the data notation of Section~\ref{sec:data}:
\[
 f=\psi|_{\mathcal S},\qquad \pi=n_{\mathcal S}\psi|_{\mathcal S},\qquad
 h=T\psi|_{\mathcal S}.
\]
Here $n_{\mathcal S}$ is the future unit normal, $T=\partial_t$ is the
stationary Killing field, and $h=\sqrt D\,\pi$ on the static end
of $\mathcal S$. The two coefficients $a,b$ in the
spatial expansions below determine the leading term.

We use the future hypersurfaces $\mathcal N_\tau$, the reference
hypersurface $\mathcal N_0$, and the time coordinate $\tau$ fixed in
Section~\ref{sec:hypersurfaces}. We take the common time origin
sufficiently late that the estimates from
Sections~\ref{sec:frac-global} and~\ref{sec:slow-spatial-endpoints}
used below hold for every $\tau\ge0$.
The initial data remain on $\mathcal S$, while $\psi(0,r)$ denotes
the evolved solution on $\mathcal N_0$. Every time integration
below keeps $r$ fixed, with integrand $T\psi=\partial_\tau\psi|_r$.

\paragraph{The argument and the data norms.}
For $0\le p<1$, we first prove that
\[
 L(r):=\lim_{\tau\to\infty}\psi(\tau,r)
\]
exists uniformly on every bounded radial interval
$[r_h,R_1]$. The wave equation and regularity at the horizon then
show that $L(r)$ is independent of $r$: its value is $0$ for
$0<p<1$ and $a+b$ for $p=0$. This identifies the constant in
\[
 \psi(\tau,r)-L=-\int_\tau^\infty T\psi(s,r)\,ds.
\]
For $p<0$, we instead recover $\psi$ from its values on $\mathcal N_0$
by integrating time derivatives over $[0,\tau]$. Proposition~\ref{prop:all-p-null-traces}
in Section~\ref{sec:exterior} supplies the behavior of
$\psi|_{\mathcal N_0}$ as $r\to\infty$.

We use $\mathfrak D_{p,m}[\psi]$ from \eqref{eq:frac-original-norm} and
$\mathfrak A_{p,m,\delta}[\psi]$ from
\eqref{eq:frac-asymptotic-data-norm}, with the same expressions for
all real $p$. The first norm controls the full data with spatial
weights and their Sobolev norms on a fixed compact part of
$\mathcal S$. The second also controls $|a|+|b|$ and the remainders
after subtracting $a r^{-p}$ and $b r^{-p-1}$. The integer $m$
is the norm index: the spatial terms involve derivatives of $f$
through order $m+1$ and of $h$ through order $m$. The compact
Sobolev norms of $(f,\pi)$ are computed in regular coordinates
at the horizon endpoint.

The superscript in $\Os[N]$ specifies the radial derivative orders
$j=0,1,\ldots,N$, as in \eqref{eq:symbol-notation}.
These weighted bounds allow polynomially growing data.
Finite propagation defines their smooth solution by solving with
compactly supported data that agree with the prescribed data on
successively larger parts of $\mathcal S$. The resulting solutions
agree on each fixed compact spacetime region once the retained
part of $\mathcal S$ contains the intersection of its causal past
with $\mathcal S$. The estimates below
are expressed in the initial-data norm
$\mathfrak A_{p,m,\delta}[\psi]$ and include data with infinite ordinary
derivative energy.

\begin{theorem}[Global asymptotics for every real $p<1$]
\label{thm:slow-global}
We assume the geometric and energy assumptions of
Section~\ref{sec:setting}. We choose an initial hypersurface
$\mathcal S$ as in Section~\ref{sec:hypersurfaces}. We take
$m=m(p)$ from \eqref{eq:classification-regularity} and prescribe
smooth spherically symmetric data with
\begin{equation}
 f=a r^{-p}+\Os[m+1](r^{-p-\delta}),\qquad
 h=b r^{-p-1}+\Os[m](r^{-p-1-\delta}),
 \qquad p<1,\quad\delta>0.
 \label{eq:slow-spatial-data}
\end{equation}
We define the data coefficient $C_p$, the model profile $\Pi_p$,
and the parameters of the estimate by
\begin{align}
 C_p&=b+(1-p)a,\notag\\
 \Pi_p(\tau,r)&=
 \frac{(1+\tau+2r)^{1-p}-(1+\tau)^{1-p}}{2(1-p)r},
 \label{eq:slow-profile}\\
 n(p)&=\left\lceil\max\{0,-p\}\right\rceil,
 \qquad\alpha(p)=p+n(p)\in[0,1),
 \label{eq:slow-induction-index}\\
 m(p)&=30+2n(p),
 \qquad\eta(p,\delta)=
 \frac1{256}\min\{\delta,1/2,1-\alpha(p)\}.
 \label{eq:slow-explicit-parameters}
\end{align}
Here $n(p)$ is the number of time differentiations needed to
raise the spatial exponent to $\alpha(p)\in[0,1)$, $m(p)$
specifies the initial-data norm, and $\eta(p,\delta)>0$ is the
power gained by the remainder.

On the hypersurfaces $\mathcal N_\tau$ fixed above, for all
$\tau\ge0$ and $r_h\le r<\infty$,
\begin{equation}
 \left|\psi(\tau,r)-C_p\Pi_p(\tau,r)\right|
 \le C\mathfrak A_{p,m(p),\delta}[\psi]\cdot 
                  (1+\tau+r)^{-p-\eta(p,\delta)}.
 \label{eq:slow-global-expansion}
\end{equation}
The same constant applies at every radius, including $r=r_h$
on the future event horizon. It depends only on $p$, $\delta$,
the fixed geometry, $\mathcal S$, and the chosen family $\mathcal N_\tau$.
The estimate also applies when $C_p=0$.

For every fixed $R_1>r_h$, it follows that
\begin{equation}
 \psi(\tau,r)
 =C_p(1+\tau)^{-p}
  +O_{R_1}\!\left(
       \mathfrak A_{p,m(p),\delta}[\psi]\cdot 
       (1+\tau)^{-p-\eta(p,\delta)}\right),
 \qquad r_h\le r\le R_1.
 \label{eq:slow-compact-expansion}
\end{equation}
Thus $0<p<1$ gives decay, $p=0$ gives convergence to $a+b$,
and $p<0$ gives polynomial growth when $C_p\ne0$.
\end{theorem}

\paragraph{Uniform convergence to a constant for
\texorpdfstring{$p=0$}{p = 0} and linear growth for
\texorpdfstring{$p=-1$}{p = -1}.}
For $p=0$, the identities $C_0=a+b$ and $\Pi_0(\tau,r)=1$ give
\[
 \left|\psi(\tau,r)-(a+b)\right|
 \le C\mathfrak A_{0,30,\delta}[\psi]\cdot 
             (1+\tau+r)^{-\eta(0,\delta)}.
\]
Thus the scalar field converges to $a+b$ uniformly throughout the
exterior, including the future event horizon, as $\tau\to\infty$.

For $p=-1$, we have $C_{-1}=b+2a$ and
$\Pi_{-1}(\tau,r)=1+\tau+r$, and hence
\[
 \left|\psi(\tau,r)-(b+2a)(1+\tau+r)\right|
 \le C\mathfrak A_{-1,32,\delta}[\psi]\cdot 
             (1+\tau+r)^{1-\eta(-1,\delta)}.
\]
When $b+2a\ne0$, this gives linear growth in $\tau$ at bounded
radius and linear growth in $r$ along each fixed hypersurface
$\mathcal N_\tau$, $\tau\ge0$. The error is smaller than the leading term
by the positive power $\eta(-1,\delta)$, uniformly over the exterior.

For general $p<1$, the profile is positive and satisfies
\begin{equation}
 \begin{gathered}
 \Pi_p(\tau,r)=\frac1{2r}\int_{1+\tau}^{1+\tau+2r}s^{-p}\,ds,\\
 c_p(1+\tau+r)^{-p}\le\Pi_p(\tau,r)\le C_p^{\rm cmp}(1+\tau+r)^{-p},
 \quad p<1.
 \end{gathered}
 \label{eq:slow-profile-comparison}
\end{equation}
Here $c_p,C_p^{\rm cmp}>0$ depend only on $p$; the symbol $C_p$
without a superscript denotes the data coefficient
$b+(1-p)a$. The remainder in \eqref{eq:slow-global-expansion}
therefore gains a fixed positive power relative to the size of
the model profile everywhere in the exterior.

Proposition~\ref{prop:all-p-null-traces} gives the estimates
\eqref{eq:slow-initial-trace}--\eqref{eq:slow-initial-trace-limits}
for $\psi|_{\mathcal N_0}$, together with control on its compact
portion. For $p=0$, the initial position datum tends to $a$ on
$\mathcal S$, while $\psi|_{\mathcal N_0}(r)$ tends to $a+b$.
Along the outgoing end of $\mathcal N_0$, we have
$u=0$ and $t=r_*(r)\sim r$, so the initial velocity tail contributes
$b$ to these evolved values. We use the values on $\mathcal N_0$
to determine the constants arising when integrating in time.

\subsection{The late-time limit of \texorpdfstring{$\psi$}{psi} at bounded radius}

We next identify the limit of $\psi(\tau,r)$ as $\tau\to\infty$
at fixed radius. Time integrability of $T\psi$ first gives a
function $L(r)$. The wave equation and the regularity of that
limit at $r=r_h$ then force $L(r)$ to have the same value at
every radius.

\begin{lemma}[The late-time limit of $\psi$ is independent of $r$]
\label{lem:slow-stationary}
We let $\psi$ be a smooth spherically symmetric solution for $\tau\ge0$,
$r\ge r_h$, regular up to the future horizon. We assume that for
every $R_1>r_h$,
\begin{equation}
 \int_0^\infty
       \sup_{r_h\le r\le R_1}|T\psi(\tau,r)|\,d\tau<\infty.
 \label{eq:slow-local-integrability}
\end{equation}
There is a single constant $L$ such that
\[
 \lim_{\tau\to\infty}
       \sup_{r_h\le r\le R_1}|\psi(\tau,r)-L|=0
 \qquad\text{for every }R_1>r_h.
\]
For each $\tau\ge0$ and $r\ge r_h$,
\begin{equation}
 \psi(\tau,r)-L=-\int_\tau^\infty T\psi(s,r)\,ds.
 \label{eq:slow-future-integral}
\end{equation}
If, in addition, the restriction of $\psi$ to $\mathcal N_0$ and the
integrated time derivative satisfy
\[
 \lim_{r\to\infty}\psi(0,r)=\ell,\qquad
 \lim_{r\to\infty}\int_0^\infty|T\psi(s,r)|\,ds=0,
\]
then $L=\ell$.
\end{lemma}

\begin{proof}
For $\tau_2\ge\tau_1$, integration at fixed $r$ gives
\[
 \sup_{r_h\le r\le R_1}
       |\psi(\tau_2,r)-\psi(\tau_1,r)|
 \le\int_{\tau_1}^{\tau_2}
       \sup_{r_h\le r\le R_1}|T\psi(s,r)|\,ds.
\]
Condition \eqref{eq:slow-local-integrability} makes this bound
tend to zero as $\tau_1\to\infty$. Hence $\psi(\tau,\cdot)$
converges uniformly on $[r_h,R_1]$ to a continuous function
$L(r)$. These limits agree on overlapping intervals.

We justify the equation for $L$ by testing the wave equation
before passing to the limit. For integers $j\ge1$, we set
$\psi_j(s,r,\omega):=\psi(j+s,r,\omega)$.
The convergence just proved gives $\psi_j\to L(r)$ locally
uniformly in $(s,r,\omega)$, with $s$ in any fixed bounded interval.
The metric coefficients are independent of $\tau$, so these
translations preserve $g$ and $\Box_g\psi_j=0$ wherever $j+s\ge0$.
For any smooth test function $\chi$ compactly supported in
$\mathbb R_s\times(r_h,\infty)\times\mathbb S^2$, its support lies
in this domain for all sufficiently large $j$, and integration
by parts and local uniform convergence give
\[
 \int L\,\Box_g\chi\,d\mathrm{Vol}_g
 =\lim_{j\to\infty}\int\psi_j\,\Box_g\chi\,d\mathrm{Vol}_g
 =0,
\]
where $d\mathrm{Vol}_g$ is the stationary spacetime volume element.
All derivatives in these integrals act on the fixed test function
$\chi$. Thus $\Box_gL=0$ in the distributional sense in $r>r_h$.
Since $L$ depends only on $r$, this equation becomes
\[
 \partial_r(r^2D\,\partial_rL)=0
\]
in distributions. Consequently $r^2D\,\partial_rL=c_1$;
as $r^2D>0$ in the open exterior, $L$ is smooth there and,
for any fixed reference radius $r_{\rm ref}>r_h$,
\[
 L(r)=c_0+c_1\int_{r_{\rm ref}}^r
                         \frac{d\rho}{\rho^2D(\rho)}.
\]
Because $D(r_h)=0$ and $D'(r_h)>0$, the integral satisfies
\[
 \int_{r_{\rm ref}}^r\frac{d\rho}{\rho^2D(\rho)}
 =\frac{\log(r-r_h)}{r_h^2D'(r_h)}+O(1)
 \qquad(r\downarrow r_h).
\]
The continuity and boundedness of $L(r)$ at the horizon force
$c_1=0$. Thus $L(r)$ is one constant $L$.

Integrating $T\psi$ from $\tau$ to a finite future time and
passing to the limit proves \eqref{eq:slow-future-integral}.
At $\tau=0$ it gives
\[
 |L-\psi(0,r)|
 \le\int_0^\infty|T\psi(s,r)|\,ds.
\]
The two additional limits in the statement then imply $L=\ell$.
\end{proof}

\subsection{Proof of the global expansion}

We first record the two elementary identities needed for time
integration. For $0<\beta<1$, $\tau\ge0$, and $r\ge r_h$,
\begin{equation}
 \int_{1+\tau}^\infty\frac{s^{-\beta}}{s+r}\,ds
 \le C_\beta(1+\tau+r)^{-\beta}.
 \label{eq:slow-elementary-integral}
\end{equation}
If $r\le 1+\tau$, we compare with $\int_{1+\tau}^\infty s^{-\beta-1}\,ds$.
If $r>1+\tau$, we split at $s=r$: the first integral is at most
$r^{-1}\int_{1+\tau}^r s^{-\beta}\,ds$, and the second is at most
$\int_r^\infty s^{-\beta-1}\,ds$. These give the same
bound because $0<\beta<1$.

The model profiles satisfy
\begin{equation}
 \partial_\tau\Pi_p(\tau,r)=-p\Pi_{p+1}(\tau,r).
 \label{eq:slow-profile-recursion}
\end{equation}
When $p+1=1$, we use the continuous extension
\[
 \Pi_1(\tau,r)=\frac1{2r}\log\left(1+\frac{2r}{1+\tau}\right)
\]
from Theorem~\ref{thm:frac-p1}; in particular,
$\Pi_0=1$ and $\partial_\tau\Pi_0=0$. Thus the model profile
obeys the same coefficient relation under time differentiation
as the actual solution in \eqref{eq:slow-T-norm}.

\begin{proof}[Proof of Theorem~\ref{thm:slow-global}]
We write
\[
 \mathfrak A[\psi]=\mathfrak A_{p,m(p),\delta}[\psi],\qquad
 \eta=\eta(p,\delta).
\]
The hypersurface $\mathcal N_0$ was chosen so that the estimates used
below hold for every $\tau\ge0$. Constants depending on this
fixed hypersurface and on propagation from $\mathcal S$ are
included in the constants multiplying $\mathfrak A[\psi]$.

\smallskip
\noindent\emph{The range $0<p<1$: decay to zero and the global remainder.}
We let $w=T\psi$ and $q=p+1\in(1,2)$. The data computation in
Lemma~\ref{lem:slow-T-data} allows us to apply
Theorem~\ref{thm:frac-spatial-global} to the wave solution $w$. It gives
\begin{equation}
 \begin{aligned}
 w(\tau,r)&=-pC_p\Pi_{p+1}(\tau,r)+R_w(\tau,r),\\
 |R_w|&\le C\mathfrak A[\psi]\cdot \frac{(1+\tau)^{-p-\eta_w}}{1+\tau+r},
 \qquad\eta_w=\frac1{32}\min\{\delta,1/2\}.
 \end{aligned}
 \label{eq:slow-positive-T-expansion}
\end{equation}
Here $R_w$ denotes the error after subtracting the displayed
profile of $T\psi$. The norm required by that theorem has
index $23$. Passing from the initial data of $\psi$ to those
of $T\psi$ increases the required index by at most two, so
$m(p)=30$ is sufficient.

The integral formula for $\Pi_{p+1}$, considered separately
for $r\le 1+\tau$ and $r>1+\tau$, gives
\[
 0<\Pi_{p+1}(\tau,r)
 \le C(p)\frac{(1+\tau)^{-p}}{1+\tau+r},
\]
where $C(p)$ depends only on $p$. Since $|C_p|\le C\mathfrak A[\psi]$,
we obtain
\begin{equation}
 |T\psi(\tau,r)|\le C\mathfrak A[\psi]\cdot \frac{(1+\tau)^{-p}}{1+\tau+r}.
 \label{eq:slow-positive-T-bound}
\end{equation}
For every fixed $R_1$, the supremum of this bound over
$r_h\le r\le R_1$ is integrable in $\tau$. Moreover,
\eqref{eq:slow-elementary-integral} and
Proposition~\ref{prop:all-p-null-traces} give
\[
 \psi(0,r)\longrightarrow0,\qquad
 \int_0^\infty|T\psi(s,r)|\,ds
 \le C\mathfrak A[\psi]\cdot (1+r)^{-p}\longrightarrow0
 \quad(r\to\infty).
\]
Lemma~\ref{lem:slow-stationary} therefore proves that
$\psi(\tau,r)\to0$ uniformly on every bounded radial interval,
and that $\psi(\tau,r)=-\int_\tau^\infty T\psi(s,r)\,ds$.

For fixed $r$, also $\Pi_p(\tau,r)\to0$ as $\tau\to\infty$.
Integrating \eqref{eq:slow-positive-T-expansion} and using
\eqref{eq:slow-profile-recursion} therefore yields
\[
 \psi(\tau,r)-C_p\Pi_p(\tau,r)
       =-\int_\tau^\infty R_w(s,r)\,ds.
\]
Our choice of $\eta$ satisfies $\eta\le\eta_w$ and
$0<p+\eta<1$. We bound the last integrand by
$C\mathfrak A[\psi]\cdot (1+s)^{-p-\eta}/(1+s+r)$, and apply
\eqref{eq:slow-elementary-integral} with $\beta=p+\eta$.
This proves \eqref{eq:slow-global-expansion} for $0<p<1$.

\smallskip
\noindent\emph{The case $p=0$: convergence to $a+b$.}
For $w=T\psi$, Lemma~\ref{lem:slow-T-data} gives spatial
exponent $1$ and leading coefficients $a_w=b$, $b_w=0$.
The coefficient of the logarithmic profile in
Theorem~\ref{thm:frac-p1} is $b_w$, so that theorem gives
\begin{equation}
 |T\psi(\tau,r)|
 \le C\mathfrak A[\psi]\cdot \frac{(1+\tau)^{-\eta_0}}{1+\tau+r},
 \qquad \eta_0=\frac1{128}\min\{\delta,1\}.
 \label{eq:slow-zero-T-bound}
\end{equation}
Its sufficient norm has index $25$ for $w$, and hence requires
at most index $27$ for the initial data of $\psi$.
This is covered by $m(0)=30$.

We apply \eqref{eq:slow-elementary-integral} with
$\beta=\eta_0\in(0,1)$. The integral of $|T\psi|$ converges
uniformly on every bounded radial interval and is at most
$C\mathfrak A[\psi]\cdot (1+r)^{-\eta_0}$ when integrated over $[0,\infty)$.
It therefore tends to zero as $r\to\infty$. Since
$\psi(0,r)\to a+b$ on $\mathcal N_0$ by
\eqref{eq:slow-initial-trace-limits},
Lemma~\ref{lem:slow-stationary} identifies the late-time limit
as $L=a+b=C_0$. Integrating
\eqref{eq:slow-zero-T-bound} over $[\tau,\infty)$ now gives
\[
 |\psi(\tau,r)-(a+b)|
 \le C\mathfrak A[\psi]\cdot (1+\tau+r)^{-\eta_0}
 \le C\mathfrak A[\psi]\cdot (1+\tau+r)^{-\eta(0,\delta)}.
\]
This is \eqref{eq:slow-global-expansion} at $p=0$, since
$\Pi_0=1$.

\smallskip
\noindent\emph{The range $p<0$: integration from $\mathcal N_0$.}
We argue by induction on the nonnegative integer $n(p)$
in \eqref{eq:slow-induction-index}. The cases already proved
give the statement for $n(p)=0$, namely $0\le p<1$.
For $p<0$, the exponent of $w=T\psi$ is $p+1<1$, with
$n(p+1)=n(p)-1$. We set $\delta_*=\min\{\delta,1\}$. The
parameters obey
\begin{equation}
 \alpha(p+1)=\alpha(p),\qquad
 \eta(p+1,\delta_*)=\eta(p,\delta),\qquad
 m(p+1)+2=m(p).
 \label{eq:slow-induction-invariance}
\end{equation}
Thus the induction hypothesis applies to $T\psi$ with the
same positive remainder exponent and a norm controlled by $\mathfrak A[\psi]$.

We define the error in the desired expansion by
\[
 R(\tau,r):=\psi(\tau,r)-C_p\Pi_p(\tau,r).
\]
Combining the induction hypothesis with
\eqref{eq:slow-T-norm} and
\eqref{eq:slow-profile-recursion} gives
\begin{equation}
 |TR(\tau,r)|\le C\mathfrak A[\psi]\cdot (1+\tau+r)^{-p-1-\eta}.
 \label{eq:slow-negative-R-derivative}
\end{equation}
To integrate this estimate from $\tau=0$, we need the values
$R(0,r)$ on $\mathcal N_0$. We choose $\gamma=2\eta$. The definitions give
\[
 0<\eta<\min\{\gamma,-p\},\qquad
 \gamma<\min\{\delta,1,1-p\}.
\]
Indeed, $-p=n(p)-\alpha(p)\ge1-\alpha(p)$ for $p<0$,
and the factor $1/256$ in
\eqref{eq:slow-explicit-parameters} gives the strict inequalities.

The profile on $\mathcal N_0$ satisfies
\[
 \Pi_p(0,r)=\frac{2^{-p}}{1-p}r^{-p}
                    +O(r^{-p-\gamma}),\qquad \gamma<1.
\]
Its leading coefficient agrees with the one in
Proposition~\ref{prop:all-p-null-traces}. Subtracting the two
expansions therefore yields
\begin{equation}
 |R(0,r)|\le C\mathfrak A[\psi]\cdot (1+r)^{-p-\gamma},
 \qquad r\ge r_h.
 \label{eq:slow-negative-initial-error}
\end{equation}
The estimate for large $r$ comes from that subtraction, and
the estimate on the fixed compact part of $\mathcal N_0$ follows from
its bound in the same proposition.

We now integrate at fixed $r$ over the finite interval $[0,\tau]$:
\[
 R(\tau,r)=R(0,r)+\int_0^\tau TR(s,r)\,ds.
\]
We put $d=-p-\eta>0$. Equations
\eqref{eq:slow-negative-R-derivative} and
\eqref{eq:slow-negative-initial-error}, with the change of
variable $s\mapsto1+s$, give
\begin{align*}
 |R(\tau,r)|
 &\le C\mathfrak A[\psi]\cdot (1+r)^{-p-\gamma}
        +C\mathfrak A[\psi]\cdot \int_1^{1+\tau}(s+r)^{d-1}\,ds\\
 &\le C\mathfrak A[\psi]\cdot (1+\tau+r)^d
        +\frac{C}{d}\mathfrak A[\psi]\cdot \bigl((1+\tau+r)^d-(1+r)^d\bigr)\\
 &\le C\mathfrak A[\psi]\cdot (1+\tau+r)^{-p-\eta}.
\end{align*}
Here $(1+r)^{-p-\gamma}\le(1+r)^d\le(1+\tau+r)^d$ because
$\gamma>\eta$ and $d>0$. This completes the induction.
After $n(p)$ differentiations the argument reaches the exponent
$\alpha(p)\in[0,1)$; for negative integer $p$ it reaches $0$.

Finally, for $r_h\le r\le R_1$, the integral formula for
$\Pi_p$ gives
\[
 \Pi_p(\tau,r)=(1+\tau)^{-p}+O_{p,R_1}((1+\tau)^{-p-1}).
\]
Since $\eta<1$, combining this with the global error proves
\eqref{eq:slow-compact-expansion}. To obtain
\eqref{eq:slow-profile-comparison}, we use the same integral
formula and consider $r\le 1+\tau$ and $r>1+\tau$. In the first case
the integration variable is comparable to $(1+\tau)+r$; in the
second case, rescaling it by $r$ bounds the resulting integral
above and below by positive constants depending only on $p$.
\end{proof}

\section*{Acknowledgements}
We acknowledge support from the Natural Sciences and Engineering Research
Council of Canada (NSERC) through grant 502581.


\begin{thebibliography}{99}
\raggedright

\bibitem{AAG1}
Y.~Angelopoulos, S.~Aretakis, and D.~Gajic,
\emph{Late-time asymptotics for the wave equation on spherically symmetric,
stationary spacetimes},
Advances in Mathematics \textbf{323} (2018), 529--621.
\href{https://arxiv.org/abs/1612.01566}{arXiv:1612.01566}.

\bibitem{AAGDecay}
Y.~Angelopoulos, S.~Aretakis, and D.~Gajic,
\emph{A vector field approach to almost-sharp decay for the wave
 equation on spherically symmetric, stationary spacetimes},
Annals of PDE \textbf{4} (2018), no.~2, 15.
\href{https://doi.org/10.1007/s40818-018-0051-2}
{doi:10.1007/s40818-018-0051-2}.
\href{https://arxiv.org/abs/1612.01565}{arXiv:1612.01565}.

\bibitem{AAG2}
Y.~Angelopoulos, S.~Aretakis, and D.~Gajic,
\emph{Asymptotics for scalar perturbations from a neighborhood of the
bifurcation sphere},
Classical and Quantum Gravity \textbf{35} (2018), 155007.
\href{https://doi.org/10.1088/1361-6382/aacc1e}
{doi:10.1088/1361-6382/aacc1e}.
\href{https://arxiv.org/abs/1802.05692}{arXiv:1802.05692}.

\bibitem{BCCPV}
S.~Bhattacharjee, B.~Chakrabarty, D.~D.~K.~Chow, P.~Paul, and A.~Virmani,
\emph{On late time tails in an extreme Reissner--Nordstr\"om black hole:
frequency domain analysis},
Classical and Quantum Gravity \textbf{35} (2018), 205002.
\href{https://doi.org/10.1088/1361-6382/aade59}
{doi:10.1088/1361-6382/aade59}.
\href{https://arxiv.org/abs/1805.10655}{arXiv:1805.10655}.

\bibitem{BSB}
D.~Baskin and A.~S\'a Barreto,
\emph{Radiation fields for semilinear wave equations},
Transactions of the American Mathematical Society
\textbf{367} (2015), no.~6, 3873--3900.
\href{https://arxiv.org/abs/1208.2743}{arXiv:1208.2743}.

\bibitem{DRRedshift}
M.~Dafermos and I.~Rodnianski,
\emph{The red-shift effect and radiation decay on black hole spacetimes},
Communications on Pure and Applied Mathematics
\textbf{62} (2009), no.~7, 859--919.
\href{https://doi.org/10.1002/cpa.20281}{doi:10.1002/cpa.20281}.
\href{https://arxiv.org/abs/gr-qc/0512119}{arXiv:gr-qc/0512119}.

\bibitem{DRrp}
M.~Dafermos and I.~Rodnianski,
\emph{A new physical-space approach to decay for the wave equation
with applications to black hole spacetimes},
in \emph{XVIth International Congress on Mathematical Physics},
P.~Exner (ed.), World Scientific, 2010, pp.~421--432.
\href{https://doi.org/10.1142/9789814304634_0032}
{doi:10.1142/9789814304634\_0032}.
\href{https://arxiv.org/abs/0910.4957}{arXiv:0910.4957}.

\bibitem{DRSDecay}
M.~Dafermos, I.~Rodnianski, and Y.~Shlapentokh-Rothman,
\emph{Decay for solutions of the wave equation on Kerr exterior
spacetimes III: The full subextremal case $|a|<M$},
Annals of Mathematics (2) \textbf{183} (2016), no.~3, 787--913.
\href{https://doi.org/10.4007/annals.2016.183.3.2}
{doi:10.4007/annals.2016.183.3.2}.
\href{https://arxiv.org/abs/1402.7034}{arXiv:1402.7034}.

\bibitem{DRSScattering}
M.~Dafermos, I.~Rodnianski, and Y.~Shlapentokh-Rothman,
\emph{A scattering theory for the wave equation on Kerr black hole exteriors},
Annales scientifiques de l'\'Ecole Normale Sup\'erieure (4)
\textbf{51} (2018), no.~2, 371--486.
\href{https://doi.org/10.24033/asens.2358}{doi:10.24033/asens.2358}.
\href{https://arxiv.org/abs/1412.8379}{arXiv:1412.8379}.

\bibitem{DGM}
S.~R.~Das, G.~W.~Gibbons, and S.~D.~Mathur,
\emph{Universality of low energy absorption cross sections for black holes},
Physical Review Letters \textbf{78} (1997), 417--419.
\href{https://doi.org/10.1103/PhysRevLett.78.417}
{doi:10.1103/PhysRevLett.78.417}.
\href{https://arxiv.org/abs/hep-th/9609052}{arXiv:hep-th/9609052}.

\bibitem{GajicInvSquare}
D.~Gajic,
\emph{Late-time asymptotics for geometric wave equations with
inverse-square potentials},
Journal of Functional Analysis \textbf{285} (2023), no.~7, 110058.
\href{https://doi.org/10.1016/j.jfa.2023.110058}
{doi:10.1016/j.jfa.2023.110058}.
\href{https://arxiv.org/abs/2203.15838}{arXiv:2203.15838}.

\bibitem{GajicExtremalKerr}
D.~Gajic,
\emph{Azimuthal instabilities on extremal Kerr},
\href{https://arxiv.org/abs/2302.06636}{arXiv:2302.06636} (2023).

\bibitem{GajicCharged}
D.~Gajic,
\emph{Charged scalar fields on Reissner--Nordstr\"om spacetimes II:
late-time tails and instabilities},
\href{https://arxiv.org/abs/2603.28861}{arXiv:2603.28861} (2026).

\bibitem{GK22}
D.~Gajic and L.~M.~A.~Kehrberger,
\emph{On the relation between asymptotic charges, the failure of peeling
and late-time tails},
Classical and Quantum Gravity \textbf{39} (2022), 195006.
\href{https://doi.org/10.1088/1361-6382/ac8863}
{doi:10.1088/1361-6382/ac8863}.
Corrigendum: Classical and Quantum Gravity \textbf{41} (2024), 119501.
\href{https://doi.org/10.1088/1361-6382/ad422c}
{doi:10.1088/1361-6382/ad422c}.

\bibitem{GK}
D.~Gajic and L.~Kehrberger,
\emph{Linear and nonlinear late-time tails on dynamical black hole
spacetimes via time integrals}.
\href{https://arxiv.org/abs/2511.23242}{arXiv:2511.23242} (2025).

\bibitem{Hintz}
P.~Hintz,
\emph{A sharp version of Price's law for wave decay on asymptotically flat
spacetimes},
Communications in Mathematical Physics \textbf{389} (2022), 491--542.
\href{https://doi.org/10.1007/s00220-021-04276-8}{doi:10.1007/s00220-021-04276-8}.

\bibitem{John}
F.~John,
\emph{Partial Differential Equations}, fourth edition,
Applied Mathematical Sciences, vol.~1,
Springer-Verlag, New York, 1982.
\href{https://link.springer.com/book/9780387906096}
{ISBN 978-0-387-90609-6}.

\bibitem{LX}
S.~Looi and H.~Xiong,
\emph{Asymptotic expansions for semilinear waves on asymptotically flat
spacetimes},
Analysis and Mathematical Physics \textbf{16} (2026), 119.
\href{https://doi.org/10.1007/s13324-026-01264-w}
{doi:10.1007/s13324-026-01264-w}.
\href{https://arxiv.org/abs/2407.08997}{arXiv:2407.08997}.
 
\bibitem{Price}
R.~H.~Price,
\emph{Nonspherical perturbations of relativistic gravitational collapse.
I. Scalar and gravitational perturbations},
Physical Review D \textbf{5} (1972), 2419--2438.
\href{https://doi.org/10.1103/PhysRevD.5.2419}
{doi:10.1103/PhysRevD.5.2419}.

\bibitem{Tataru}
D.~Tataru,
\emph{Local decay of waves on asymptotically flat stationary space-times},
American Journal of Mathematics \textbf{135} (2013), 361--401.
\href{https://arxiv.org/abs/0910.5290}{arXiv:0910.5290}.

\end{thebibliography}
\end{document}